\documentclass[apj]{emulateapj}
\usepackage{lscape}

\newcommand{\noprint}[1]{}

\def\fnsiz{\footnotesize}
\def\ts{\thinspace}
\newcommand{\am}{\mbox{$'$}}
\newcommand{\as}{\mbox{$''$}}

\newcommand{\dg}{\mbox{$^{\circ}$}}
\newcommand{\ebv}{\mbox{$E(B-V)$}}

\newcommand{\etal}{et~al.\/}

\newcommand{\sigew}{\mbox{$\sigma_{EW}$}}
\newcommand{\ew}{\mbox{$EW$}}
\newcommand{\ewha}{\mbox{$EW(\Halpha)$}}
\newcommand{\eweff}{\mbox{$EW_{50}$}}
\newcommand{\eweffo}{\mbox{$EW_{50,0}$}}

\newcommand{\FHa}{\mbox{$F_{\rm H\alpha}$}}
\newcommand{\FHI}{\mbox{$F_{\rm HI}$}}
\newcommand{\FFIR}{\mbox{$F_{\rm FIR}$}}
\newcommand{\fion}[2]{\mbox{\rm [{#1}\ts{\fnsiz {#2}}]}} 

\newcommand{\Hline}[1]{\mbox{H{\footnotesize {#1}}}}
\newcommand{\Halpha}{\Hline{\mbox{$\alpha$}}}
\newcommand{\Hbeta}{\Hline{\mbox{$\beta$}}}
\newcommand{\HI}{{\sc Hi}}
\newcommand{\HII}{{\sc Hii}}

\newcommand{\HIPASS}{{\sc HiPASS}}
\newcommand{\bhipass}{{\bf H{\small I}PASS}}

\newcommand{\kms}{\mbox{km\thinspace s$^{-1}$}}
\newcommand{\MHI}{\mbox{${\cal M}_{\rm HI}$}}
\newcommand{\MHtwo}{\mbox{${\cal M}_{\rm H_2}$}}
\newcommand{\LHa}{\mbox{$L_{\fnsiz  H{\tiny \alpha}}$}}
\newcommand{\logmhi}{\mbox{$\log({\cal M}_{\rm HI})$}}

\newcommand{\Msun}{\mbox{${\cal M}_\odot$}}
\newcommand{\Mstar}{\mbox{${\cal M}_\star$}}
\newcommand{\pino}{\parindent=0mm}

\newcommand{\reha}{\mbox{$r_e({\rm H\alpha})$}}
\newcommand{\reR}{\mbox{$r_e({\rm R})$}}
\newcommand{\rmax}{\mbox{$r_{\rm max}$}}
\newcommand{\rsky}{\mbox{$r_{\rm sky}$}}
\newcommand{\Seha}{\mbox{$S_e({\rm H\alpha})$}}

\newcommand{\SSFR}{\mbox{$\Sigma_{\rm SFR}$}}

\newcommand{\tgas}{\mbox{$t_{\rm gas}$}}

\newcommand{\vhel}{\mbox{$V_{\rm h}$}}
\newcommand{\fwhm}{\mbox{$W_{50}$}}

\bibpunct{(}{)}{;}{a}{}{}

\shorttitle{SINGG. I. Initial Results}
\shortauthors{Meurer et al.}

\begin{document}

\slugcomment{2006, ApJS in press}

\title{The Survey for Ionization in Neutral Gas Galaxies: I. Description
  and Initial Results}

\author{Gerhardt R.\ Meurer\altaffilmark{1}, 
  D.J.\ Hanish\altaffilmark{1}, H.C.\ Ferguson\altaffilmark{2}, 
  P.M.\ Knezek\altaffilmark{3}, V.A.\ Kilborn\altaffilmark{4,5}, 
  M.E.\ Putman\altaffilmark{6}, R.C.\ Smith\altaffilmark{7}, 
  B.\ Koribalski\altaffilmark{5}, M.\ Meyer\altaffilmark{2}, 
  M.S.\ Oey\altaffilmark{6}, E.V.\ Ryan-Weber\altaffilmark{8}, 
  M.A.\ Zwaan\altaffilmark{9}, T.M.\ Heckman\altaffilmark{1}, 
  R.C.\ Kennicutt, Jr.\altaffilmark{10}, J.C.\ Lee\altaffilmark{10}, 
  R.L.\ Webster\altaffilmark{11}, J.\ Bland-Hawthorn\altaffilmark{12}
  M.A.\ Dopita\altaffilmark{13}, K.C.\ Freeman\altaffilmark{13},
  M.T.\ Doyle\altaffilmark{14}, M.J.\ Drinkwater\altaffilmark{14}, 
  L.\ Staveley-Smith\altaffilmark{5}, and J.\ Werk\altaffilmark{6}}
\altaffiltext{1}{Department of Physics and Astronomy, Johns Hopkins
                 University, 3400 North Charles Street, Baltimore, MD
                 21218}
\altaffiltext{2}{Space Telescope Science Institute, 3700 San Martin
                 Drive, Baltimore, MD 21218}
\altaffiltext{3}{WIYN Consortium, Inc., 950 North Cherry Avenue, Tucson,
                 AZ 85726}
\altaffiltext{4}{Centre for Astrophysics and Supercomputing, Swinburne 
                 University of Technology, Mail 31, PO Box 218, 
                 Hawthorn, VIC 3122, Australia}
\altaffiltext{5}{Australia Telescope National Facility, CSIRO, P.O. Box
                 76, Epping, NSW 1710, Australia}
\altaffiltext{6}{University of Michigan, Department of Astronomy, 830
                 Denison Building, Ann Arbor, MI 48109-1042} 
\altaffiltext{7}{Cerro Tololo Inter-American Observatory (CTIO), Casilla
                 603, La Serena, Chile} 
\altaffiltext{8}{Institute of Astronomy, Madingley Road, Cambridge 
                  CB3 0HA, United Kingdom}
\altaffiltext{9}{European Southern Observatory,
                 Karl-Schwarzschild-Str.\ 2, 85748 Garching b.\ 
                 M\"{u}nchen, Germany}
\altaffiltext{10}{Steward Observatory, University of Arizona, Tucson, AZ
                 85721}
\altaffiltext{11}{School of Physics, University of Melbourne, VIC 3010,
                 Australia} 
\altaffiltext{12}{Anglo-Australian Observatory, Epping NSW 2121, Australia}
\altaffiltext{13}{Research School of Astronomy and Astrophysics, 
                  Australian National University, Cotter Road, Weston 
                  Creek, ACT 2611, Australia}
\altaffiltext{14}{Department of Physics, University of Queensland, 
                  Brisbane, QLD 4072, Australia}

\begin{abstract}
  We introduce the Survey for Ionization in Neutral Gas Galaxies
  (SINGG), a census of star formation in \HI-selected galaxies.  The
  survey consists of \Halpha\ and $R$-band imaging of a sample of 468
  galaxies selected from the \HI\ Parkes All Sky Survey (\HIPASS).  The
  sample spans three decades in \HI\ mass and is free of many of the
  biases that affect other star forming galaxy samples.  We present the
  criteria for sample selection, list the entire sample, discuss our
  observational techniques, and describe the data reduction and
  calibration methods. This paper focuses on 93 SINGG targets whose
  observations have been fully reduced and analyzed to date.  The
  majority of these show a single Emission Line Galaxy (ELG).  We see
  multiple ELGs in 13 fields, with up to four ELGs in a single field.
  All of the targets in this sample are detected in \Halpha\ indicating
  that dormant (non-star-forming) galaxies with $\MHI \gtrsim 3\times
  10^7\, \Msun$ are very rare. A database of the measured global
  properties of the ELGs is presented.  The ELG sample spans four orders
  of magnitude in luminosity (\Halpha\ and $R$-band), and \Halpha\
  surface brightness, nearly three orders of magnitude in $R$ surface
  brightness and nearly two orders of magnitude in \Halpha\ equivalent
  width (EW).  The surface brightness distribution of our sample is
  broader than that of the Sloan Digital Sky Survey (SDSS) spectroscopic
  sample, the EW distribution is broader than prism-selected samples,
  and the morphologies found include all common types of star forming
  galaxies (e.g.\ irregular, spiral, blue compact dwarf, starbursts,
  merging and colliding systems, and even residual star formation in S0
  and Sa spirals).  Thus SINGG presents a superior census of star
  formation in the local universe suitable for further studies ranging
  from the analysis of \HII\ regions to determination of the local
  cosmic star formation rate density.
\end{abstract}


\keywords{galaxies: ISM -- galaxies: evolution -- HII regions -- stars:
formation -- surveys}

\section{Introduction}\label{s:intro}

Selection biases have had a serious influence in our understanding of
the universe.  This is especially true with regards to star formation in
the local universe.  Attempts at a global census of star formation
depend critically on the limitations of the methods used.  For example,
prism-based emission line samples \citep[e.g.][]{gzar95,kiss01} are
biased toward systems with high equivalent widths; ultraviolet (UV)
selected samples \citep[e.g.][]{temdb98} are biased against very dusty
systems; and far-infrared (FIR) selected samples \citep[e.g.][]{sm96}
are biased against low-dust (and perhaps low-metallicity) systems.
Broad-band optical surveys have a well-known bias against low surface
brightness (LSB) systems \citep{d76} that are at least as common as
normal and ``starburst'' galaxies \citep{bim97}.  Conversely, the
techniques used to discover LSB systems tend to discard compact and high
surface brightness galaxies \citep{dsgss97}, as do surveys that
distinguish galaxies from stars by optical structure \citep{dghbcfjp02}.
Broad-band surveys from the optical, UV, and infrared also suffer from
spectroscopic incompleteness.  The missed galaxies are typically faint,
may be at low distances, and hence may make major contributions to the
faint end of the luminosity function.  Large fiber-spectroscopy surveys
such as 2dF \citep{2df} and SDSS \citep{sdss} are affected by the
selection function for placing fibers \citep[e.g.][]{sdssfib}, large
aperture corrections \citep[which are variable even for galaxies of
similar morphology;][; hereafter B04]{bcwtkhb04}, ``fiber collisions''
\citep{bllmyzl03}, and the requirements for classification as ``star
forming'' (B04).  While these effects are mostly small and well studied
(e.g.\ B04), they may still introduce subtle biases in our understanding
of the phenomenology of extra-galactic star formation.  Finally, the
different tracers of star formation (UV, FIR, \Halpha, X-ray and radio
emission) result from different physical processes, and often trace
different masses of stars.  Imprecise knowledge of the physics of these
processes and particularly the Initial Mass Function (IMF) may result in
systematic errors in the star formation rate (SFR).

A more complete census of star formation in the local universe would be
sensitive to all types of star-forming galaxies.  Here we report initial
results from the Survey for Ionization in Neutral Gas Galaxies (SINGG),
which we will show meets this requirement.  SINGG surveys \HI-selected
galaxies in the light of \Halpha\ and the $R$-band continuum.  \Halpha\
traces the presence of the highest mass stars ($\Mstar
\gtrsim 20 \Msun$) through their ability to ionize the interstellar
medium (ISM).  For any metallicity, \Halpha\ (at rest wavelength
$\lambda = 6562.82$\AA) is one of the main emission line coolants in
star forming regions and typically the strongest at optical wavelength.
The modest typical levels of extinction ($A_{\rm H\alpha} \lesssim 1.5$
mag) found in previous \Halpha\ surveys \citep{k83,gzrav96,wsjgm03}
suggest that dust absorption corrections are manageable, perhaps even
in extremely dusty systems \citep{ms01}. The starting point for SINGG is
the recently completed \HI\ Parkes All-Sky Survey
\citep[\HIPASS;][]{hipass1} the largest survey to select galaxies
entirely by their \HI\ 21-cm emission.  \citet{hwbod04} have also
obtained $R$ and \Halpha\ observations (as well as $B$ band data) of a
sample of \HIPASS\ galaxies similar in number to those whose images we
present here.  Since their goals were more oriented toward studying low
surface-brightness galaxies, their sample selection was less
comprehensive than ours.  Our sample is more inclusive, for instance
having no angle of inclination selection, and our observations
generally have higher quality and are deeper.  Because interstellar
hydrogen is the essential fuel for star formation, \HIPASS\ is an ideal
sample to use in star formation surveys.  \HI\ redshifts are available
for all sources thus allowing a consistent measurement of distance.
Furthermore, because it is a radio-selected survey, it is not directly
biased by optical properties such as luminosity, surface brightness, or
Hubble type.  Instead, the distribution of these properties that we find
will be determined by their dependence on the \HI\ selection criteria we
adopt.

This paper describes SINGG and presents initial results for a subsample
of targets.  Section \ref{s:samp} describes the sample selection process
and lists the full SINGG sample.  The rest of the paper concentrates on
the first sub-sample of SINGG data that has been fully reduced and
analyzed.  It consists of 93 SINGG targets observed over four observing
runs.  Since we are releasing these data, with the publication of this
paper, we refer to this data set as SINGG Release 1, or SR1.  Section
\ref{s:data} describes the SR1 data and their reduction and analysis.  A
database of the measured properties is presented in Section
\ref{s:measres} which includes a detailed discussion of data quality and
errors.  Science results are discussed in Sec.\ \ref{s:results}.  Chief
among them is that all targets in SR1 are detected in \Halpha.  These
cover a wide range in \Halpha\ luminosity, surface brightness and
equivalent width, verifying that an \HI-selected sample is well suited
for star formation surveys.  We discuss the implications of this result
and how the relationship between star formation and \HI\ may arise.  The
paper is summarized in Sec.\ \ref{s:conc}.

\section{Sample selection}\label{s:samp}

The full list of SINGG targets was selected from \HIPASS\ source
catalogs.  \HIPASS\ used the 64-m Parkes Radio Telescope with a
multibeam receiver \citep{multibeam96} to map the entire southern sky
for neutral hydrogen emission from -1280 to 12,700 \kms\ in heliocentric
radial velocity (\vhel).  The original survey, and the source catalogs
used for SINGG, extend from $-$90\dg\ to $+$2\dg\ in declination.  The
northern extension of the survey, $+$2\dg\ to $+$25\dg\ in declination,
has recently been cataloged \citep{nhicat05}.  Processing of the
\HIPASS\ data resulted in cubes 8\dg\ $\times$ 8\dg\ in size with a
velocity resolution of 18.0 \kms, a spatial resolution of $\sim15$\am,
and a 3$\sigma$ limiting flux of 40 mJy beam$^{-1}$.  \citep{hicat2}
determined the completeness of the survey using a fake source analysis: 
fake sources were inserted into the \HIPASS\ data cubes and the HIPASS
source finder was used to determine whether the source was detected.
The fake sources had a wide range of peak fluxes, integrated fluxes,
random velocities, and a variety of velocity profile shapes (Gaussian,
double-horn, and flat-top) and FWHM velocity widths ranging from 20 to
650 \kms.  Integrated over all profile shapes and widths, the 95\%\
completeness level for integrated flux is $7.4\, {\rm Jy\, km\, s^{-1}}$
\citep{hicat2} and corresponds to an \HI\ mass limit of $\MHI \approx
1.7 \times 10^6 \Msun\, D^2$, where $D$ is the distance in Mpc.  The
details of the observing and reduction methods of HIPASS are outlined in
\citet{hipass01}.  In this section we describe how the full SINGG sample
was chosen from the HIPASS catalogs, while the rest of the paper focuses
on the targets comprising SR1.

\subsection{Sample size}\label{s:sampsiz}

The primary goal of SINGG is to uniformly survey the star formation
properties of \HI-selected galaxies across the entire \HI\ mass function
sampled by \HIPASS\ in a way that is blind to previously known optical
properties of the sources.  An essential aspect of the project is its
ability to measure not only mean star formation quantities, but also the
distribution about the mean among galaxies of different \HI\ mass
(\MHI), Hubble type, surface brightness, and environment.  Our goal is
to image 180 targets per decade of \MHI.  The available sources found by
\HIPASS\ allow this goal to be obtained over the mass range
$\log(\MHI/\Msun) \approx 8.0$ to 10.6.  A sample this size allows the
width in the \Halpha\ emissivity ($\FHa/\FHI$, where \FHa\ and \FHI\ are
the integrated \Halpha\ and \HI\ fluxes) distribution to be measured to
statistical accuracy better than 10\%\ per decade of \MHI\ and allows
sensitive tests for non-Gaussian distributions.  This is important for
testing models such as the stochastic self-propagating star formation
scenario of \citet{gss80} which predicts a wider range of star-formation
properties with decreasing galaxy mass.  A large sample also makes the
selection of rare systems more likely, including extreme starburst and
dormant systems.

\begin{deluxetable*}{l c c c c l}
  \tablewidth{0pt}
  \tablecaption{Correlation of SINGG sample and \HIPASS\ Catalogs\label{t:catsum}}
  \tablehead{\colhead{~} &
             \multicolumn{2}{c}{Targets in common} & 
             \multicolumn{2}{c}{\HI\ parameters source} & 
             \colhead{~} \\
             \colhead{Catalog} &
             \colhead{SINGG all} &
             \colhead{SR1} &
             \colhead{SINGG all} &
             \colhead{SR1} &
             \colhead{Reference}}
\startdata
HICAT      & 450 &  89 & 449 &  89 & Meyer et al.\ (2004)         \\
BGC        & 269 &  83 &   4 &   3 & Koribalski et al.\ (2004)    \\
SCCC       &  19 &   7 &   0 &   0 & Kilborn et al.\ (2002)       \\
AVCC       &  10 &   6 &   0 &   0 & Putman et al.\ (2002)        \\
Additional &  15 &   1 &  15 &   1 & This study                   \\
Total      & --  & --  & 468 &  93 & ~                            \\
\enddata
\end{deluxetable*}

\subsection{Source catalogs}\label{s:scats}

Our final sample was selected primarily from two catalogs known as HICAT
and BGC.  (1) HICAT, the full \HIPASS\ catalog \citep{hipass1} selects
candidate sources from the \HIPASS\ cubes using two different automated
techniques: a peak flux density threshold algorithm, and a technique of
convolving the spectral data with top-hat filters of various
scales.  Extensive automated and eye quality checks were used to verify
candidates.  HICAT only includes targets with Galactic standard of rest
velocity, $V_{\rm GSR} > 300$ \kms, in order to minimize the
contribution of high velocity clouds (HVCs), and was created totally
blind to the optical properties of the targets. The completeness and
reliability of this catalog are well understood \citep{hicat2}, hence it
was the primary source for our sample selection and all \HI\ parameters.
(2) The \HIPASS\ Bright Galaxy Catalog (BGC) contains the 1000 \HIPASS\ 
targets with the brightest peak flux density \citep{bgc04}.  The BGC
uses the same input data cubes as HICAT; however, it catalogs sources to
lower radial velocities.  Special attention was paid to insure that all
known nearby galaxies were considered for inclusion, irrespective of
velocity and confusion with Galactic \HI.  Care was taken to split the
\HI\ flux from contaminating sources, especially Galactic HI.

In Table~\ref{t:catsum} we break down our sample by membership in
various HIPASS catalogs.  While HICAT and BGC are our primary source
catalogs, due to the concurrent development of the SINGG and HIPASS
projects, preliminary versions of these catalogs had to be used in our
selection.  Likewise, related \HIPASS\ catalogs such as the South
Celestial Cap Catalog (SCCC) of \citet{k02} and the Anomalous Velocity
Cloud Catalog (AVCC) of \citet{p02} were used in our earliest
selections.  

A comparison of our final selection and the published HICAT and BGC
reveals 14 sources not in the published version of the catalogs.  These
made it into our sample for one of three reasons: (1) those located just
to the north of the final HICAT declination cut, $\delta = 2^\circ$ made
it into the version of HICAT used in our selection but were eliminated
from the published version; (2) similarly, some sources near the
detection limit of the cubes did not make it into the final HICAT;
finally (3) sources from earlier selections that were already observed
in our survey were ``grandfathered'' into the SINGG sample.  We
carefully examined the \HIPASS\ data for all targets in our sample that
were neither in the final HICAT nor BGC, in order to check their
reality.  Real sources are those whose angular size is equal to the beam
size, or up to a few times larger, have peak fluxes clearly above the
noise level, and do not correspond to baseline ripples, as determined by
cuts at constant velocity right ascension and declinition through the
data cubes.  Sources that did not meet these criteria were rejected from
our final sample.  The \HI\ properties of the 14 detections neither in
HICAT nor BGC were measured using the standard procedure adopted in BGC.
As was done for BGC creation, special care was taken to split sources
that appear double or which are barely resolved spatially at the 15$'$
resolution of the \HIPASS\ data.  The \HI\ properties of these sources
with new measurements are given in Table~\ref{t:newhi}.  In addition
there is one source in this table, \HIPASS~J1444$+$01, which is also in
HICAT but very close spatially and in velocity to one of the new
measurements, \HIPASS~J1445$+$01.  We adopt our new measurements as an
improved splitting of the \HI\ flux.

\begin{deluxetable*}{l c c c r r r r}
  \tabletypesize{\small}
  \tablewidth{0pt}
  \tablecaption{Additional \HI\ measurements from \HIPASS\ data 
                \label{t:newhi}}
  \tablehead{\colhead{\HIPASS+} &
             \colhead{RA} &
             \colhead{Dec.} &
             \colhead{$S_p$} &
             \colhead{$F_{\rm HI}$} &
             \colhead{\vhel} &
             \colhead{$W_{50}$} &
             \colhead{$W_{20}$} \\
             \colhead{(1)} &
             \colhead{(2)} &
             \colhead{(3)} &
             \colhead{(4)} &
             \colhead{(5)} &
             \colhead{(6)} &
             \colhead{(7)} &
             \colhead{(8)}
}
\startdata
J0249$+$02   &   02 49 06 & $+$02 08 11 & 1.033 &  56.4 & 1104 &  56 &  73 \\
J0400$-$52   &   04 00 33 & $-$52 41 27 & 0.053 &   7.5 &10566 & 298 & 349 \\
J0412$+$02   &   04 12 47 & $+$02 21 20 & 0.069 &  13.6 & 5017 & 393 & 424 \\
J1145$+$02   &   11 45 03 & $+$02 09 57 & 0.163 &   5.6 & 1010 &  30 &  51 \\
J1208$+$02   &   12 08 00 & $+$02 49 30 & 0.435 &  66.6 & 1322 & 200 & 223 \\
J1210$+$02   &   12 10 57 & $+$02 01 49 & 0.127 &  10.0 & 1337 &  80 &  97 \\
J1211$+$02   &   12 11 40 & $+$02 55 30 & 0.085 &   5.2 & 1295 &  88 & 108 \\
J1234$+$02B  &   12 34 20 & $+$02 39 47 & 0.469 & 103.1 & 1737 & 355 & 381 \\
J1234$+$02A  &   12 34 29 & $+$02 12 41 & 0.344 &  77.0 & 1805 & 326 & 348 \\
J1326$+$02A  &   13 26 20 & $+$02 06 24 & 0.119 &  17.1 & 1090 & 152 & 177 \\
J1326$+$02B  &   13 26 20 & $+$02 27 52 & 0.049 &   1.9 & 1026 &  38 &  54 \\
J1328$+$02   &   13 28 12 & $+$02 19 49 & 0.063 &   3.0 & 1023 &  50 &  66 \\
J1444$+$01   &   14 44 28 & $+$01 42 45 & 0.146 &  33.0 & 1569 & 323 & 351 \\
J1445$+$01   &   14 45 00 & $+$01 56 11 & 0.098 &  28.9 & 1727 & 625 & 645 \\
J2000$-$47   &   20 00 58 & $-$47 04 11 & 0.067 &  16.2 & 6551 & 310 & 657 \\
\enddata
\tablecomments{Column descriptions [units]: (1) Source name. (2) and (3)
               Right ascension and declination [J2000]. (4) Peak flux
               density of \HIPASS\ 21cm spectrum [Jy]. (5) Integrated
               \HI\ flux [Jy \kms].  (6) Systemic heliocentric velocity
               of \HI\ measured as the mid-point at the 50\%\ of $S_p$
               level [\kms].  (7) Width of \HI\ profile at 50\%\ of $S_p$
               [\kms]. (8) Width of \HI\ profile at 20\%\ of $S_p$
               [\kms]. }

\end{deluxetable*}

\subsection{Selection criteria}

We selected ``candidate'' targets from the source catalogs using the
following criteria: (a) peak flux density, $S_p \geq 0.05$ Jy; (b)
Galactic latitude, $|b| > 30^\circ$; (c) projected distance from the
center of the LMC, $d_{\rm LMC} > 10^\circ$; (d) projected distance from
the center of the SMC, $d_{\rm SMC} > 5^\circ$; (e) Galactic standard of
rest velocity, $V_{\rm GSR} > 200$ \kms; and (f) \vhel\ not within 100
\kms\ of the following ``bad'' velocities: 586, 1929, 2617, 4279, 4444,
5891, 10155 and 10961 \kms.  Condition a insures that only sources with
adequate $S/N$ are used.  It requires that the {\em peak\/}
signal-to-noise ratio $S/N > 3.8$ in the \HI\ spectra.  As noted in
Sec.~\ref{s:scats}, our selection was from preliminary versions of HICAT
and BGC; hence not all the sources in our final sample meet this
criterion when using the published catalogs (4\%\ of our sample have
$S_p < 0.05$ Jy).  Conditions b-d minimize foreground dust and field
star contamination from the Galaxy and Magellanic Clouds.  Condition e
minimizes contamination from HVCs.  Condition f was included to avoid
radio frequency interference features and Galactic recombination lines
found in some preliminary \HIPASS\ catalogs.  It should be noted that
the final HICAT and BGC have been effectively cleansed of these sources
of interference \citep{hipass1,bgc04}.

Our sample was selected from the candidates defined above based on \HI\ 
mass, \MHI, and distance $D$.  The mass is derived from the integrated
\HI\ flux $\FHI = \int f_\nu\, d\nu$ in Jy km s$^{-1}$ and 
$D$ in Mpc using the formula
\begin{equation}
\MHI = 2.36\times10^5\,\Msun\, D^2\, \FHI
\end{equation}
\citep{r62}. The value of $D$ is derived from \vhel\ corrected for a
model of the local Hubble flow.  Specifically, we employ the multi-pole
attractor model of the $H_0$ key project as discussed by \citet{mould00}
and adopt $H_0 = 70\, {\rm km\, s^{-1}\, Mpc^{-1}}$.  This is the only
distance estimate used during sample selection.  Final distances are
discussed in Sec.\ \ref{s:finalhi}.

When selecting sources, we divided the candidates into \logmhi\ bins and
preferentially selected the nearest objects in each bin to populate our
selection.  This preference allows better morphological information and a
more accurate determination of the \HII\ region luminosity function and
also minimizes confusion in the \HI\ detections.  The distance
preference was not rigorously enforced in order to allow sources we had
already observed to be grandfathered into the sample.  A total of 64
galaxies in our final selection would not meet a strict distance
preference selection.

Our final adopted \MHI\ selection bin width is 0.2 dex.  We found that
using a bin size of 0.4 dex, or greater, results in noticeable biasing
within each bin, in the sense that at the high-mass end, the galaxies
selected tend to be in the lower half of the bin in terms of \logmhi\
and $D$.  The sense of the bias is reversed for the low-mass bins.  The
bias is negligible for a bin width of 0.2 dex. Using a smaller selection
bin size would be meaningless in the face of the $D$ and flux errors.
For $\log(\MHI/\Msun) < 8.0$ and $\log(\MHI/\Msun) > 10.6$ there are
less than 180 candidates per decade of \MHI.  At the low-mass end the
sample is limited to the small volume over which such a low-mass can be
detected, while at the high-mass end the number of sources is limited by
their rarity.  Effectively, we are selecting all \HIPASS\ targets that
meet our candidate constraints in both of these mass ranges.

\subsection{Final \HI\ parameters\/}\label{s:finalhi}

\begin{deluxetable*}{l r r r r c c l}
  \tablewidth{0pt}
  \tablecaption{Final SINGG sample\label{t:sample}}
  \tablehead{\colhead{HIPASS+} &
             \colhead{\vhel} &
             \colhead{\fwhm} &
             \colhead{$D$} &
             \colhead{\logmhi} &
             \colhead{\ebv} &
             \colhead{SR1} &
             \colhead{Catalogs} \\
             \colhead{(1)} & 
             \colhead{(2)} & 
             \colhead{(3)} & 
             \colhead{(4)} & 
             \colhead{(5)} & 
             \colhead{(6)} & 
             \colhead{(7)} & 
             \colhead{(8)} } 
\startdata
J0005$-$28   &     737 &    52 & 10.7        &    8.27 &   0.017 & Y &    HB
          \\
J0008$-$34   &     221 &    30 & 3.3         &    7.17 &   0.012 & N &     B
          \\
J0008$-$59   &    7786 &   353 & 112.1       &   10.68 &   0.012 & N &     H
          \\
J0014$-$23   &     468 &   171 & 7.0         &    9.40 &   0.021 & N &   HBA
          \\
J0019$-$22   &     670 &   121 & 9.8         &    8.55 &   0.019 & Y &    HB
          \\
J0030$-$33   &    1580 &   457 & 22.3        &   10.23 &   0.018 & N &    HB
          \\
J0031$-$22   &     539 &    47 & 7.9         &    8.01 &   0.018 & Y &    HB
          \\
J0031$-$10   &    3573 &   286 & 50.1        &   10.29 &   0.032 & N &    HB
          \\
J0034$-$08   &    1652 &   220 & 23.4        &    9.95 &   0.044 & N &    HB
\enddata
\tablecomments{Column descriptions [units]: (1) Source name. (2)
Systemic heliocentric  velocity from \HI\ profile [\kms].  (3) \HI\
profile width at 50\%\ of the peak flux density [\kms].  (4) Adopted
distance [Mpc]. Sources marked with $^*$ have distances taken from 
Karachentsev, et al.\ (2004); otherwise distances were derived from 
\vhel\ using the model of Mould et al.\ (2000). (5) \HI\ mass [\Msun]. 
(6) Foreground dust reddening
from the maps of Schlegel, Finkbeiner \&\ Davis (1998) [mag]. (7)
Indicates whether target was observed as part of SINGG Release 1.
(8) \HIPASS\ catalogs that contain this source: H - HICAT (Meyer et al.\
2004), B - BGC (Koribalski et al.\ 2004), S - SCCC (Kilborn et
al. 2002), A - AVCC (Putman et al.\ 2002), R - this study
(Table~\ref{t:newhi}). The first catalog listed is the source of the \HI\
parameters for that entry. }
\tablecomments{{\it Sample portion of table.}}
\end{deluxetable*}

\begin{figure}
\plotone{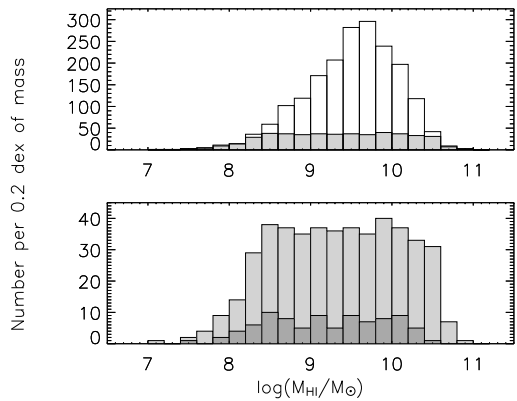}
\caption[]{\HI\ mass histograms.  The top panel shows the candidates
  from HICAT as the plain histogram, and the SINGG selection as the
  shaded histogram. The bottom panel zooms in on the $y$ scale showing
  the total SINGG sample selection in light shading and the SR1 targets
  as the dark shaded histogram.
\label{f:masshist}}
\end{figure}

\begin{figure}
\plotone{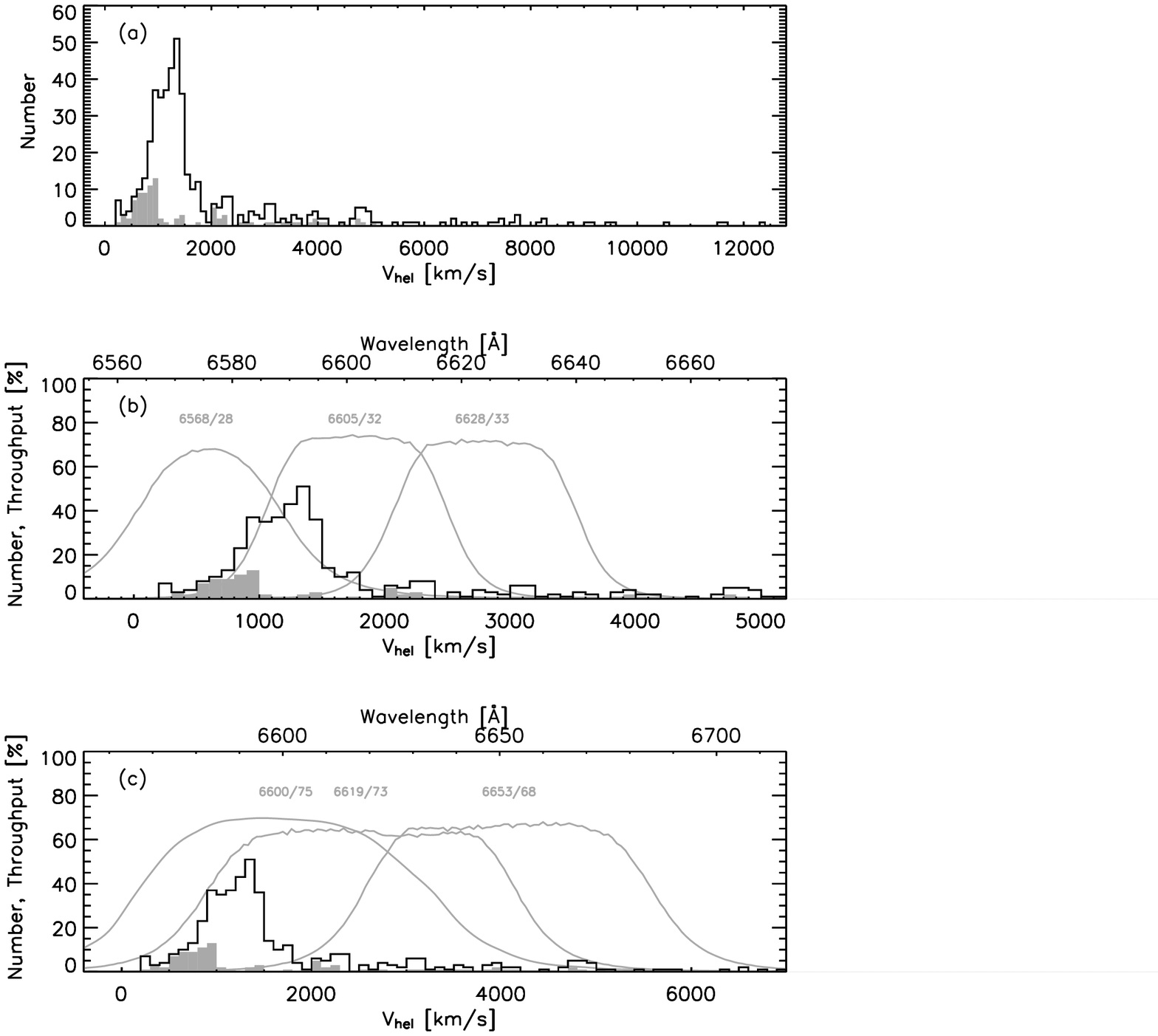}
\caption[]{Heliocentric radial velocity histogram for the SINGG sample.
  All panels show the full SINGG sample as an open histogram, and the
  SR1 targets as the shaded histogram.  Panel (a) shows
  the full velocity range of the sample.  Panel (b) over-plots, in gray,
  SINGG and MCELS \Halpha\ filter throughput curves combined
  with the CCD QE curve on an expanded velocity scale histogram.  Panel
  (c) likewise over-plots the throughput curves for the KPNO filters
  used in run 02.
\label{f:vhist}}
\end{figure}

The full SINGG sample is listed in Table~\ref{t:sample}.
Figure~\ref{f:masshist} compares the \logmhi\ histogram of the full
SINGG sample with the parent distribution of candidate targets.
Figure~\ref{f:vhist} shows the \vhel\ histogram of the full SINGG
sample.  To keep the measurements homogeneous, we took measurements from
HICAT where possible, and used measurements from BGC, or
Table~\ref{t:newhi} for the sources neither in HICAT nor BGC.

Due to small changes in the \HI\ parameters from the preliminary
catalogs used in the sample selection, and the final HICAT and BGC
catalogs used for the adopted measurements, the \logmhi\ histogram of
the sample shown in Fig.~\ref{f:masshist} is not exactly ``flat'' over the
mass range of $\log(\MHI/\Msun) = 8$ to 10.6.  

Hubble flow distances are intrinsically uncertain due to random motions
about the flow, the ``peculiar velocity dispersion''.
Estimates of this range from about 100 to 400 \kms\ \citep[e.g.,
][]{samurai7,sco98,wsdk97,ws98,tbad00} depending on galaxy type and
environment.  Within 7 Mpc the value may be as low as $\sim 70$
\kms\ \citep{mgh05}.  If we adopt 125 \kms\ for the peculiar velocity
dispersion of field spirals \citep{wsdk97,ws98}, then at the median
Hubble flow distance of the full SINGG sample, 18.5 Mpc, we have an
intrinsic distance uncertainty of 10\%\ leading to a 20\%\ luminosity
error.  These uncertainties are much more significant for the nearest
sources in our sample.  We used the Catalog of Neighboring Galaxies
\citep{kkhm04} to improve the distances to the nearest galaxies in our
sample.  We adopt 15 matches between this catalog and our sample
including only galaxies with $D$ based on Cepheid variables (2 cases),
red giant branch measurements (12 cases) or group membership (1 case).
We did not include distances from this catalog based on the brightest
stars or the Hubble flow out of concern for the accuracy of the
distances.  Likewise, we did not use Tully-Fisher relationship distances
since this relationship is usually calibrated with spiral galaxies and
is less reliable for low luminosity, low velocity width galaxies
\citep{msbb00} which dominate our sample in the local volume.  The
\HIPASS\ targets with improved distances are marked in
Table~\ref{t:sample} with an asterisk (*).

\section{Data and analysis}\label{s:data}

\subsection{Observations}

\begin{deluxetable}{c l l c}
  \tablecaption{CTIO 1.5 m observing runs in SR1\label{t:runs}}
  \tablehead{\colhead{Run} &
             \colhead{Dates} &
             \colhead{Filters used} &
             \colhead{Targets} \\
             \colhead{\#} &
             \colhead{~} &
             \colhead{~} &
             \colhead{observed}}
\startdata
01 & 23-27 Oct, 2000 & 6568/28, 6850/95, $R$          & 20 \\
02 & 26-30 Dec, 2000 & 6600/75, 6619/73, 6653/68, $R$ & 25 \\
03 & 13-15 Feb, 2001 & 6568/28,  $R$                  & 21 \\
06 & 12-15 Sep, 2001 & 6568/28, 6605/32, 6628/33, $R$ & 27 \\
\enddata
\end{deluxetable}

The SINGG observations were primarily obtained with the Cerro Tololo
Inter-American Observatory (CTIO) 1.5 m telescope as part of the NOAO
Surveys program.  Additional observations were obtained with the CTIO
Schmidt and 0.9 m telescopes and the Australian National University 2.3 m
at Siding Spring Observatory.  In this paper, we present observations
from four CTIO 1.5 m observing runs consisting of images obtained with
the 2048$\times$2048 CFCCD.  The plate scale of 0.43\as\ pixel$^{-1}$
produces a 14.7\am\ field of view, well matched to the Parkes 64 m beam
width.  Table~\ref{t:runs} presents a brief synopsis of these runs,
whose data comprise SINGG Release 1 (SR1).  The \MHI\ and \vhel\
distributions of the SR1 targets are compared with the full SINGG sample
in Fig.~\ref{f:masshist} and \ref{f:vhist}, respectively.

The images were obtained through narrow-band (NB) filters chosen to
encompass redshifted \Halpha, as well as $R$-band images used for
continuum subtraction.  For three sources (HIPASS J0403$-$01, HIPASS
J0459$-$26, and HIPASS J0507$-$37), continuum observations were obtained
through a narrower filter, 6850/95, which excludes \Halpha\ from its
bandpass.  This was done to test the filter's use in continuum
subtraction or to avoid saturation.  Table~\ref{t:filts} list the
properties of the filters used in this study.  The bandpasses of the NB
filters are plotted in Fig.~\ref{f:vhist}.  These filters include the
primary filters used in this survey, which are $\sim 30$\AA\ wide and
used to observe galaxies with $\vhel < 3300$ \kms, as well as four
broader filters used to extend the velocity coverage of the survey. The
lowest velocity filter used here, 6568/28, was borrowed from the
Magellanic Clouds Emission Line Survey \citep[MCELS;][]{mcels}.  We
purchased additional filters, two of which are used in this study
6605/32, and 6628/33. The remaining filters are from NOAO's collection
at CTIO or KPNO.  The SINGG and MCELS filters were scanned with beams
using a range of incident angles at NOAO's Tucson facility. The scans
were used to synthesize the bandpass through an f/7.5 beam.  Filter
properties are listed in Table~\ref{t:filts}.

\begin{deluxetable*}{l l c r r r r c}
\tabletypesize{\small}
\tablewidth{0pt}
\tablecaption{Filter properties\label{t:filts}}
\tablehead{\colhead{Filter ($F$)} &
\colhead{Owner} &
\colhead{$\max(T_F)$} &
\colhead{$\lambda_{p,F}$} &
\colhead{$\lambda_{m,F}$} &
\colhead{$W_{50,F}$} &
\colhead{$W_{E,F}$} & 
\colhead{$\epsilon_C/C$} \\
\colhead{(1)} &
\colhead{(2)} &
\colhead{(3)} &
\colhead{(4)} &
\colhead{(5)} &
\colhead{(6)} &
\colhead{(7)} &
\colhead{(8)} }
\startdata
6568/28    &  MCELS  & 0.68 & 6575.5 & 6575.5 &   28.1 &   21.2 & 0.042 \\
6605/32    &  SINGG  & 0.74 & 6601.5 & 6601.5 &   32.5 &   25.0 & 0.024 \\
6628/33    &  SINGG  & 0.72 & 6623.7 & 6623.7 &   33.1 &   24.6 & 0.024 \\
6600/75    &  CTIO   & 0.70 & 6600.7 & 6600.8 &   69.4 &   49.4 & 0.043 \\
6619/73    &  KPNO   & 0.65 & 6618.0 & 6618.0 &   73.7 &   49.1 & 0.031 \\
6653/68    &  KPNO   & 0.68 & 6652.2 & 6652.3 &   68.2 &   47.5 & 0.043 \\
6709/71    &  KPNO   & 0.68 & 6708.4 & 6708.4 &   70.6 &   48.8 & 0.043 \\
6850/95    &  MCELS  & 0.72 & 6858.9 & 6859.0 &   94.6 &   70.1 &   $-$ \\
$R$        &  CTIO   & 0.67 & 6507.5 & 6532.4 & 1453.4 &  977.0 &   $-$ \\
\enddata
\tablecomments{Column descriptions [units]: (1) Filter name. (2) Owner
of filter. (3) Peak throughput [dimensionless].  (4) Pivot wavelength
[\AA]. (5) Response weighted mean wavelength [\AA].  (6) Transmission
profile width at 50\% of the peak transmission [\AA].  (7) Response
weighted equivalent width of the filter [\AA].  (8) The adopted 
ratio for the error due to continuum subtraction divided by the 
continuum flux.  Definitions for
$\lambda_{p,F}$, $\lambda_{m,F}$, and $W_{E,F}$ (columns 4,5 and 7)
can be found in appendix~\ref{s:fluxap}. }
\end{deluxetable*}

To perform the observations, the telescope was positioned to place the
\HIPASS\ position near the center of the CCD for each target observed.
Typically, the observations consisted of three 120 s duration $R$
exposures (or 3 $\times$ 200 s with 6850/95) and three NB exposures of
600 s duration.  The observations were obtained at three pointing
centers dithered by 0.5\am\ to 2\am\ to facilitate cosmic ray and bad
column removal.

\subsection{Basic processing}

Basic processing of the images was performed with {\sl
IRAF\/}\footnote{{\sl IRAF\/} is the Image Reduction and Analysis
Facility, and is distributed by the National Optical Astronomy
Observatories, which are operated by the Association of Universities for
Research in Astronomy, Inc., under cooperative agreement with the
National Science Foundation.} using the {\sl QUADPROC\/} package and
consisted of (1) fitting and subtraction of the bias level as recorded
in the overscan section of the images, (2) subtraction of a bias
structure frame typically derived from the average of 15 to 100 zero
frames (CCD readouts of zero duration), and (3) flat-field division.
flat-field frames were obtained employing an illuminated white spot on
the dome as well as during evening and/or morning twilight.  The final
flat-field frames combine the high spatial frequency structure from the
dome flats with the low spatial frequency structure from the sky flats.
They were made by (a) combining the dome flats with cosmic ray
rejection; (b) normalizing the result to unity over the central portion
of the frame; (c) dividing the sky flats with the normalized dome flat;
(d) combining the sky flats, taking care to scale and weight the images
to compensate for the different exposure levels; (e) box median
filtering the result with a box size of 25 to 51 pixels on a side; (f)
normalizing the result; and (g) dividing the result into the normalized
dome flat produced in step b.

\subsection{Red leak correction}

\begin{figure*}
\plottwo{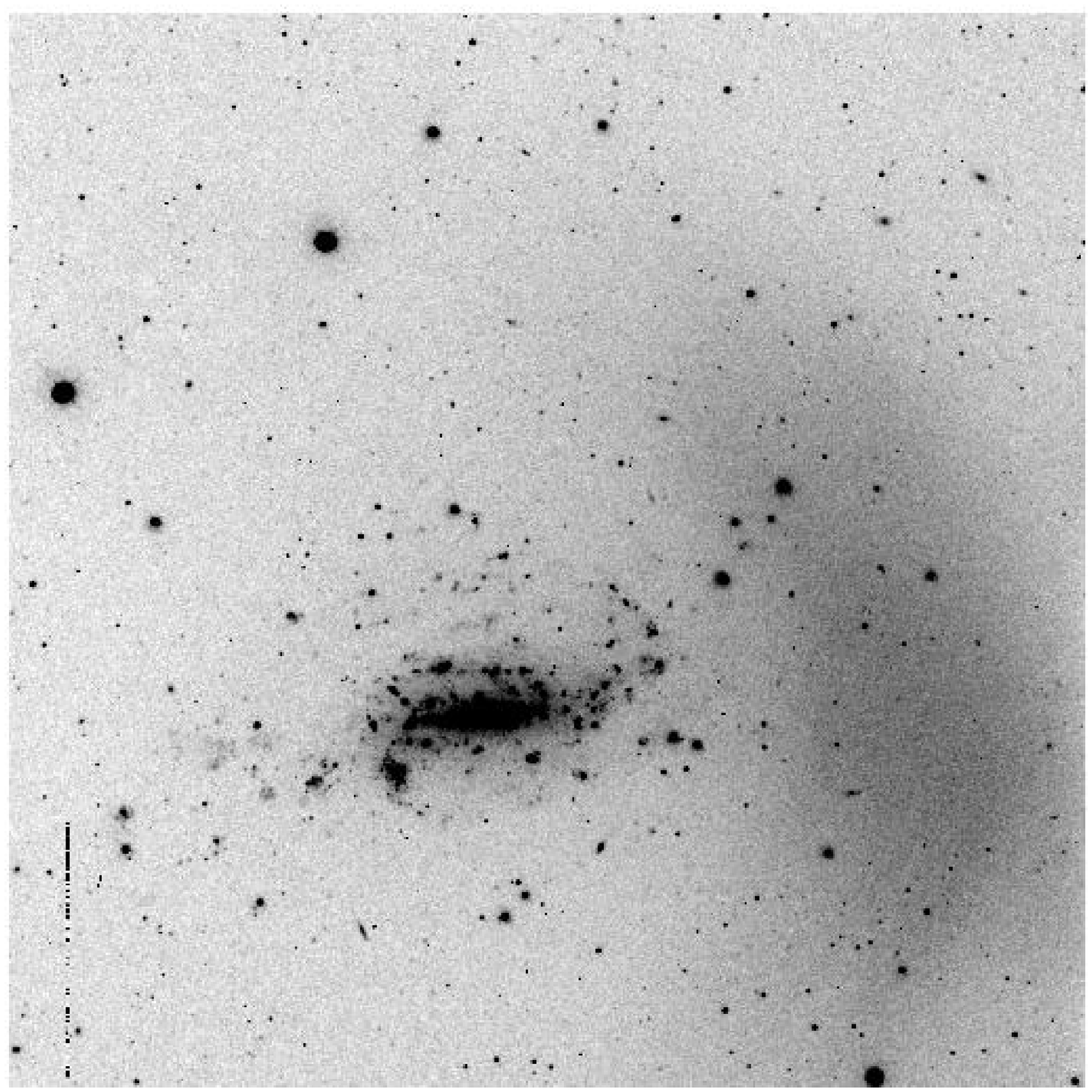}{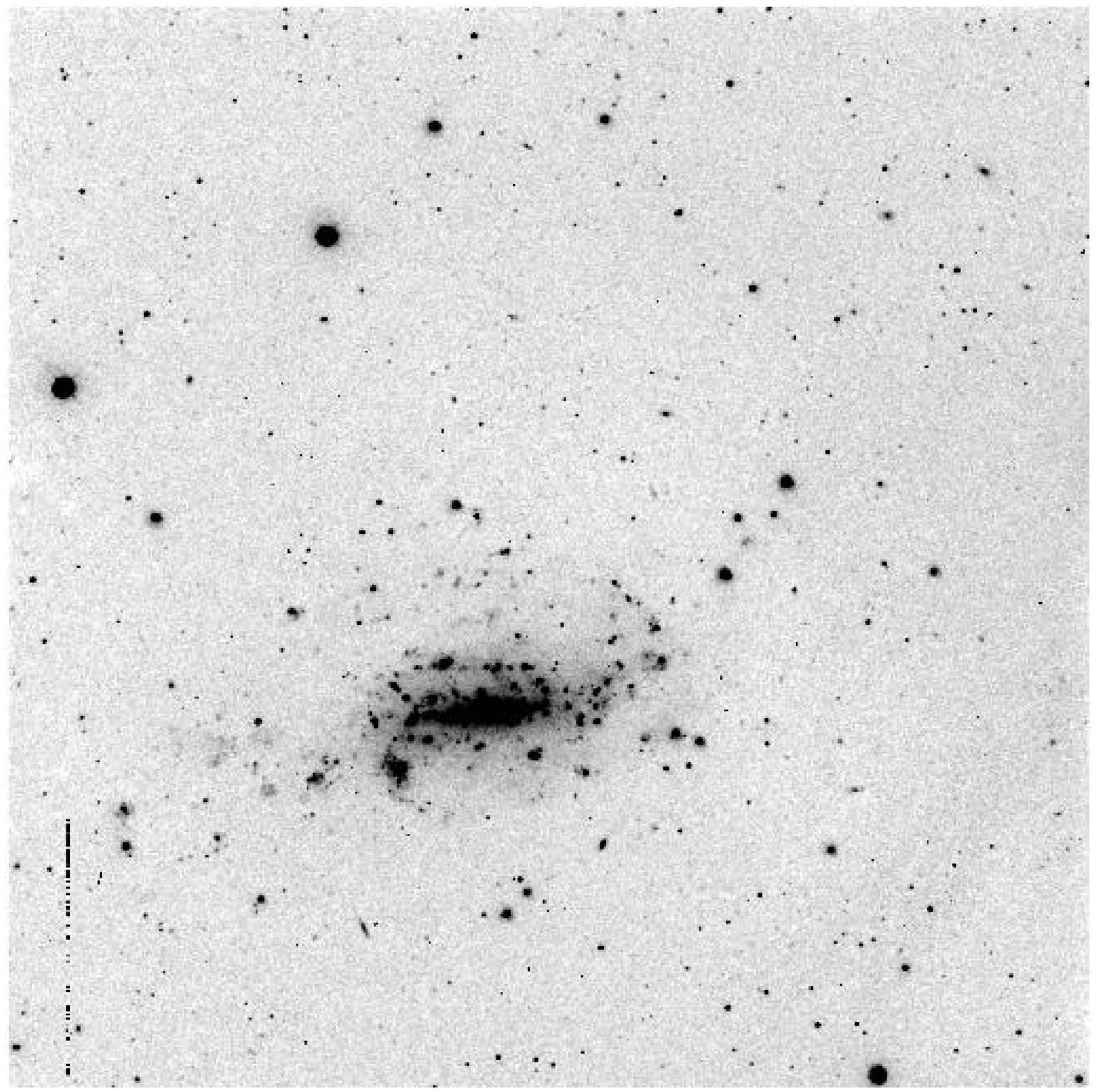}
\caption[]{A single flat-fielded 6568/28 exposure of \HIPASS\ J0459-26
  displayed with an inverse linear stretch showing the ``hump''
  instrumental artifact (left), and after hump removal (right).  These
  images were created using the same stretch after applying a 3$\times$3
  median filter followed by a 4$\times$4 block average in order to
  reject cosmic rays and enhance the appearance of smooth features such
  as the hump.
\label{f:hump}}
\end{figure*}

Examination of the images showed that flat-fielding worked correctly for
most filters; the sky was flat to better than 1\%.  However, this was
not the case for many of the 6568/28 images.  Figure~\ref{f:hump} shows
the nature of the problem: an oblong diffuse emission ``hump'' peaking
on one side of the frame covering $\sim$ 25\%\ of the field of view,
with an intensity up to $\sim 30-40$\%\ of the sky background.  This
feature was intermittent in nature.  For the data presented here, the
hump was only seen in runs 03 and 06.  Run 01 used 6568/28 exclusively
as the NB filter but is not affected, while run 02 did not employ this
filter.  Most, but not all, later observing runs that used this filter
were affected by this feature. Within a run, this feature was
variable in amplitude, although its shape remained stable.  Examination
of individual dithered frames reveal that the count rate of stars is not
affected as they are dithered off and on the hump region.  We attribute
this artifact to a red leak in the filter coating, allowing the filter to
transmit the bright OH sky lines at $\lambda > 6800$\AA.  The
variability in amplitude would then result from the variability of these
lines.

To remove the hump, we created a set of normalized correction
images.  For each affected run, at least 15 object images using the
6568/28 filter were selected, preferably those where the target galaxy
was small, and did not extend into the hump region. Each image was
masked for bad pixels, smoothed with a 7$\times$7 box median to remove
cosmic rays, sky subtracted and then normalized to have a peak in the
hump of 1.0.  The images for each observing run were then combined (with
rejection) to remove stars, galaxies, and other sources, and the
resulting image was again median-smoothed (9$\times$9) to remove any
remaining artifacts of the combine process.  Each affected image was
manually adjusted by subtracting a scaled version of this correction
image.  Typically, the scaling was determined from the intensities of
$\sim$ 2500 pixels surrounding the brightest point of the feature, after
a first pass background sky subtraction.

\subsection{Flux calibration}\label{s:fluxcal}

We used observations of spectrophotometric standards
\citep{hwsghp92,hshwgp94,msba88,oke90} to flux calibrate the data.  The
standards were typically obtained in three sets (at the beginning,
middle and end of each night) of two standards each.  We calibrated
magnitudes in the ABmag system \citep[See ][for a discussion of the
ABmag system and its motivation]{fukugita96}, and \Halpha\ line fluxes
in terms of erg cm$^{-2}$ s$^{-1}$ using synthetic photometry techniques
as detailed in Appendix~\ref{s:fluxap}.

\subsection{Combining images}\label{s:comb}

\begin{figure}
\plotone{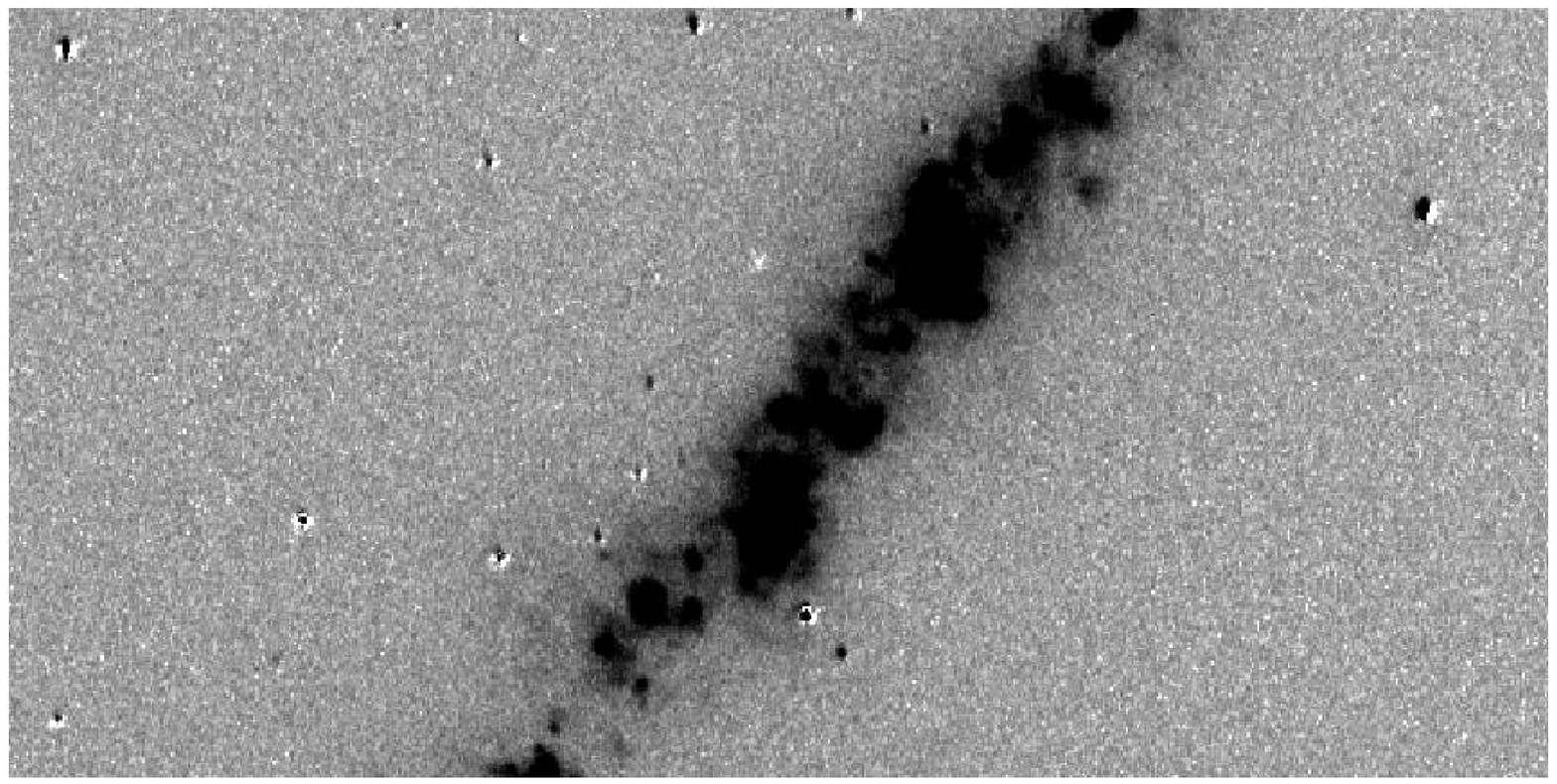}
\plotone{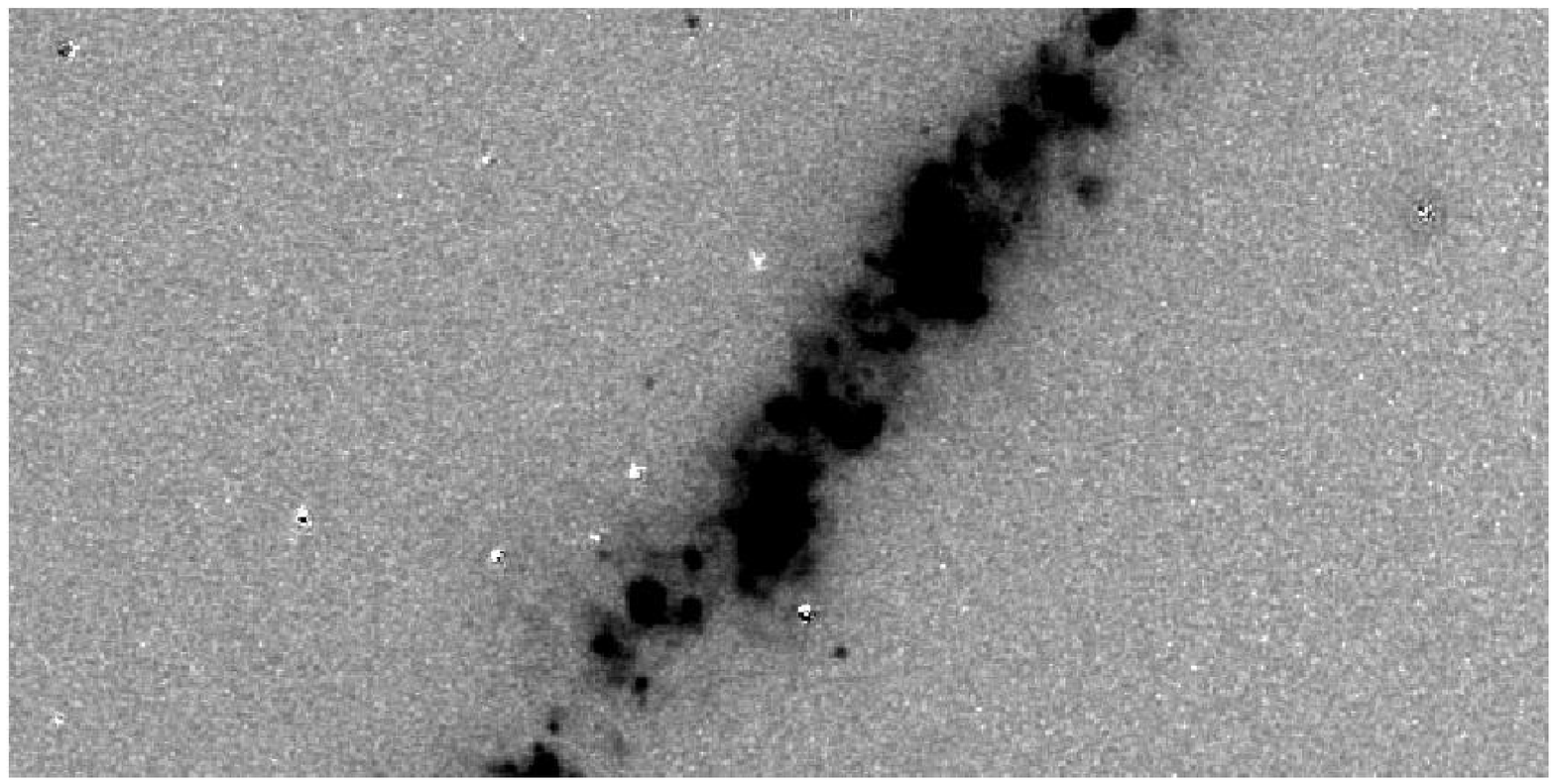}
\caption[]{A section of the R subtracted image of \HIPASS\ J2052-69.  Top:
  ``traditional'' processing;  Bottom ``High-$z$ supernova'' software
  processing.  The two methods are outlined in Sec.\ \ref{s:comb}.
\label{f:alard}}
\end{figure}

In order to align the images and subtract the continuum, we make use of
software kindly provided by the High-$z$ supernova group
\citep{hi_z_sne98} and modified by our team.  As illustrated in
Fig.~\ref{f:alard}, this provides superior final continuum subtracted
images when compared to those of more ``traditional'' processing, which
would typically consist of linearly interpolating all images to a common
origin, combining the images in each filter, and performing a straight
scaled $R$-band image subtraction from the NB image.  Our processing is
somewhat more sophisticated, as follows.

Sources in the individual frames are cataloged using the Source
Extractor (SE) software package \citep{ba96}.  The catalogs include
source positions, fluxes, and structural parameters.  They are used to
align all the frames of each target to a common reference image -
typically the $R$-band frame in the center of the dither pattern.  This
is done by matching the catalogs to derive a linear transformation in
each axis (allowing offset, stretch, rotation and skew).  On the order of
100 matches per frame are typically found.  Registration is done with a
7$\times7$ sinc interpolation kernel to preserve spatial resolution and
the noise characteristics of the frames.  The images in each filter are
then combined in IDL using a modified version of {\sl CR\_REJECT\/}
found in the {\sl ASTROLIB\/} package.  Our modifications remove
sky differences between frames and use the matched catalogs to determine
the multiplicative scaling between frames to bring them to the same flux
scale.  For each filter, the reference image for flux scaling is the one
whose sources have the highest count rate (excluding very short
exposures and saturated images).  The header of this file becomes the
basis of that of the output image.

Continuum subtraction is performed using the algorithm given by
\citet{a00}.  The frame with the best seeing is convolved with a kernel
that matches it to the PSF of the frame with the worst seeing, and the
scaled continuum image is subtracted from the NB frame.  The flux
scaling is implemented by setting the sum of the convolution kernel to
the appropriate scale factor.

\subsection{Astrometric calibration}

SE catalogs were matched to the U.S. Naval Observatory A2.0 database
\citep{usnoa2}.  Typically, on the order of 100 sources were matched
resulting in an rms accuracy of $\sim$0.4\as\ (about one pixel) to the
coordinate system zeropoint.

\subsection{Source identification}\label{ss:sid}

Identification of Emission Line Galaxies (ELGs) was done visually using
color composite images.  These were created using the $R$ image in the
blue channel, the NB image in the green channel and the net \Halpha\
image in the red channel, resulting in emission line sources appearing
red. This assignment is used in all color images presented here.  The
display levels are scaled to the noise level in the frames allowing
sources to be discerned to a consistent significance level in all
images.  We define an ELG to be a discrete source that is noticeably
extended in at least the $R$-band and contains at least one emission
line source.  This phenomenological definition is deliberately broad and
allows an extended galaxy with one unresolved \HII\ region to be
considered an ELG.

The aim is to find any star-forming galaxies associated with the \HI\
source.  However, we can not be certain that the ELGs correspond to the
location of \HI\ within the Parkes beam.  Similarly, we could also
detect background sources in some other emission line redshifted into
the passband of the NB filter (e.g.\ \fion{O}{III}5007\AA, \Hbeta, or
\fion{O}{II}3727\AA\ at $z \sim $ 0.3, 0.4, and 0.8 respectively).
Further spectroscopic and \HI\ imaging follow-up would be required to
unequivocally determine which ELGs are associated with the \HIPASS\
sources.  Despite these concerns, the rich morphology of extended
distributions of \HII\ region in the vast majority of the ELGs is
consistent with them being associated with the \HIPASS\ targets.

We also frequently find emission line sources that are unresolved or
barely resolved in both the $R$ and NB images and projected far from any
apparent host galaxies.  We classified these sources as ``ELDots'' which
is a phenomenological shorthand description for their appearance -
Emission Line Dots.  The nature of the ELDots is not immediately
apparent; they could be outlying \HII\ regions in the targeted galaxy,
or background line emitters.  \citet{rw04} obtained optical spectra of
13 ELDots with the ANU 2.3m telescope and confirmed the detection of
line emission of five in the field of three \HIPASS\ galaxies (\HIPASS
J0209$-$10, \HIPASS J0409$-$56, and \HIPASS J2352$-$52).  For four of
the five ELDots, \Halpha\ was detected at the systemic velocity of the
\HIPASS\ galaxy, while in the fifth case (\HIPASS J2352$-$52) only one
line was detected, at a wavelength outside that expected for \Halpha\ at
the systemic velocity.  The majority of the eight ELDots not detected
spectroscopically were probably fainter than the detection limit of the
observations \citep{rw04}.  Additional ELDots in the SR1 images
presented here are in the process of being cataloged and confirmed
(J.\ Werk et al., in preparation).

\subsection{Sky subtraction}\label{s:ss}

We determine the sky level in an annulus around the galaxy that is set
interactively.  We use color images to define the brightness peak as
well as four points that specify the major and minor axes of the
aperture that encompasses all the apparent emission in both \Halpha\ and
the $R$ band.  In most cases this aperture has a shape and orientation
close to that of the outer $R$ band isophotes.  In cases where a minor
axis outflow is readily apparent in \Halpha, the aperture is made
rounder in shape to accommodate the outflow.  Galaxies with such an
outflow are discussed in Appendix~\ref{s:objnotes}.  For galaxies with a
few small scattered \HII\ regions at large radii, we typically match the
aperture in size and shape to the outer $R$ band isophotes, leaving some
\HII\ regions outside of this aperture.  The semi-major axis size \rsky\
parameterizes the inner size of the sky annulus.  Next, \rsky\ is
tweaked using crude radial surface brightness profiles; the images are
divided into $35\times35$ pixel boxes, the 3$\sigma$ clipped mean level
of each box is plotted as a function of semi-major axis distance, and the
distance at which the mean intensity levels off in both the net \Halpha\
and $R$ band images is selected as the new \rsky. In some cases there
are slight radial gradients in the sky, due to scattered light, and the
mean intensity level does not level off.  In those cases we do not reset
\rsky.  The outer sky radius is set so that the sky annulus has an area
equal to that interior to \rsky.  The exceptions are very large
galaxies, where the available sky area is limited by the CCD boundaries,
and small galaxies, where we set the minimum area to 16 arcmin$^2$. The
sky level is the 3$\sigma$ clipped average of the mean 
level in each box wholly within the sky annulus, rejecting boxes that
have had pixels rejected in the clipping within the box.  The
pixel-to-pixel noise of each image is taken to be the average clipped
rms values within the boxes.  The large scale ($>$ 35 pixels)
uncertainty in the sky is taken as the dispersion in the mean levels in
the boxes used to define the sky; this represents the uncertainty due to
imperfect flat-fielding and scattered light.

\subsection{Image masking}

We use two types of masks, exclusion and inclusion, to indicate how to
use pixels when integrating fluxes.  These masks rely heavily on SE
catalogs as well as ``segmentation images'' produced by SE which
indicate which source each pixel belongs to.

For the $R$ and NB images, the exclusion mask uses the position, SE flag
values, source size, stellarity parameter (star/galaxy classification),
flux, and $R$/NB flux ratio to identify the pixels to exclude.  The SE
segmentation image is displayed, and allows interactive toggling of
which sources are masked or kept.  To make the final exclusion mask,
this mask is grown by convolving it with a circular top hat function
with a radius equal to the seeing width (or a minimum of 1.2\as) so that
the fringes of unrelated stars and galaxies are also rejected.  The net
H$\alpha$ image requires less exclusion masking, because most of the
faint foreground and background sources are adequately removed with
continuum subtraction.  Our algorithm uses the uncertainty in the
continuum scaling ratio to determine which pixels masked in the $R$-band
are likely to have residuals greater than 1.5 times the pixel-pixel sky
noise.  In addition, we exclude pixels corresponding to concave sources
resulting from residuals around bright stars.  The bad pixels are grown
as described above to make the final \Halpha\ exclusion mask.

The inclusion mask is needed primarily to account for \HII\ regions that
are detached from the main body of a galaxy.  In many cases, a simple
aperture that is large enough to include all of a large galaxy's \HII\ 
regions would result in a sky uncertainty that is so big that the
derived total flux would be meaningless.  The inclusion mask is based on
an SE analysis of the net \Halpha\ image.  We use a logic similar
to that adopted to find sources that are most likely foreground,
background, or artifacts, and take all other sources to be part of the
galaxy being measured.  The grow radius of the inclusion mask is twice
the seeing width or a minimum of 2.4\as.

The algorithm for defining the masks is straight forward but not
perfect.  Objects at the edge of frames, satellite trail residuals, and
the wings of bright stars are sometimes mistakenly placed in inclusion
masks, while occasionally portions of the target galaxy, such as
line-free knots, are excluded.  Therefore, each set of masks were
examined by two of us (G.R.M.\ and D.J.H.).  This was done by examining color
images of (a) the entire field, (b) only the pixels included in total $R$
band flux measurements, (c) only the pixels not included in total $R$
flux measurements, (d) only the pixels included in the net \Halpha\
flux, and (e) only the pixels not included in the total \Halpha\ flux.
These images were compared to determine if there were regions that
should or shouldn't be included in the masks.  Mistakenly excluded or
included SE sources were toggled.  In some cases circular
or polygon shaped areas were added as needed to the masks to insure that
the measurements recover as much of the true flux while excluding
obvious contaminating features.

\subsection{Measurements}\label{ss:meas}

The ideal way to measure total \Halpha\ fluxes is to just use a simple
aperture (e.g.\ circular or elliptical) that is large enough to
encompass all \HII\ emission.  In addition to being easy to specify,
this technique has the advantage of including all emission in the
aperture, including that from faint \HII\ regions and Diffuse Ionized
Gas (DIG) that may be below the apparent detection limit of the
observations.  In contrast, measuring \Halpha\ fluxes by summing the
light from HII regions typically underestimates the true flux by 30\% --
50\%\ because of the neglected DIG \citep{fwgh96,hwb01,hwbod04}.
However, as alluded to above, using large apertures may result in very
low S/N due to the sky uncertainty over the very large aperture needed
to contain the outermost \HII\ regions.  We have developed a hybrid
approach that uses the sum of the aperture flux where the S/N level is
reasonable, supplemented with the flux of \HII\ regions outside of this
aperture that are within the inclusion mask described above.  The method
is similar in concept to that employed by \citet{fwgh98}.

\begin{figure*}
\plotone{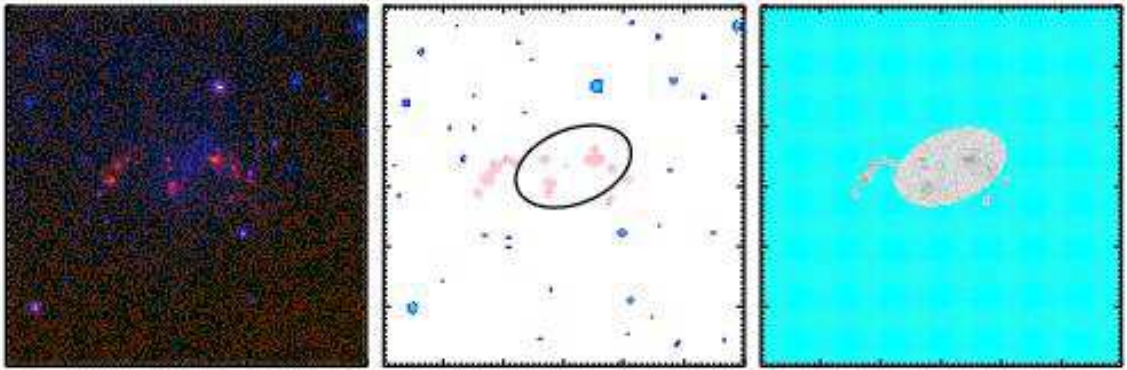}
\caption[]{Steps involved in image masking and total flux measurement.
  The left panel shows a 600 $\times$ 600 pixel subsection of the
  \HIPASS\ J1217$+$00 image in color using our standard assignment: red
  - net \Halpha\ emission; green - NB image (no continuum subtraction);
  and blue $R$ (continuum).  The middle panel indicates pixels masked as
  stars, or stellar residuals in the $R$ and net \Halpha\, marked blue
  and cyan respectively, which comprise the exclusion mask; and pixels
  identified as dominated by \Halpha\ marked pink, the inclusion mask.
  The ellipse indicates indicates the aperture having \rmax.  The right
  hand panel shows a grayscale of the net \Halpha\ image.  Pixels that
  are not used in the total \Halpha\ flux calculation are colored
  cyan. \label{f:exmask}}
\end{figure*}

Surface brightness and curve of growth (enclosed flux) profiles are
extracted for each source using concentric, constant shape elliptical
apertures.  The shape and centers of the apertures are the same as those
set in the sky determination.  The difference in flux between apertures
defines the surface brightness profile.  The curve of growth profiles
are corrected for the excluded pixels in each annulus by adding the
missing area times the mean unmasked intensity in the annulus.  In the
majority of ELGs (96 of 111) the curve of growth plateaus at or very close
\rsky, and we terminate the profiles at a maximum radius $\rmax =
\rsky$.  In some cases the profiles plateau inwards of \rsky, or the
$S/N$ of the enclosed flux is low.  Hence, our adopted algorithm for
determining \rmax\ is to use the smallest of (a) where the curve of
growth flattens, (b) \rsky, or (c) where $S/N = 3$.  Here the noise is
crudely estimated from the large scale sky variation (Sec.\ \ref{s:ss};
as discussed in Sec.~\ref{s:fluxerr} below, this overestimates the error
in the enclosed flux, hence the actual $S/N$ is higher).
Beyond \rmax, we still include the flux of pixels indicated by the
inclusion mask in our total flux measurements.  Figure~\ref{f:exmask}
shows an example of how pixels are masked and which pixels are included
when measuring total \Halpha\ fluxes.

We find some \Halpha\ flux outside of \rmax\ in 30\%\ of the ELGs
studied here.  However, in most cases the fractional \Halpha\ flux
outside of \rmax\ is negligible; it is greater than 0.1, 0.05, 0.01 in
3, 6, and 16 cases respectively.  The most extreme case is \HIPASS\
J1217$+$00 (Fig.~\ref{f:exmask}) where 41\%\ of \FHa\ is beyond \rmax.

The curve of growth is interpolated to determine the effective radius 
$r_e$, the radius along the semi-major axis containing half the flux and
from this the face-on effective surface brightness, defined as
\begin{equation}
S_e = \frac{F}{2 \pi r_e^2}
\end{equation}
where $F$ is the total flux of the target\footnote{the face-on
  correction occurs because $r_e$ is a semi-major axis length, and thus
  $\pi r_e^2$ is the face-on area provided the generally elliptical
  isophotes result from a tilted disk.}.  We are primarily concerned
with the effective surface brightness of \Halpha, \Seha.  We also
calculate the effective surface brightness in the $R$-band which we
convert to the ABmag scale yielding $\mu_e(R)$. Using the same
algorithm, we also calculate $r_{90}$, the radius containing 90\%\ of
the total flux and do not calculate this value if more than 10\%\ of the
flux is beyond \rmax.

The equivalent width we use is that within the \Halpha\ effective radius,
\reha, and is given by 
\begin{equation}
\eweff = \frac{0.5 \FHa}{f_R(\reha)}
\end{equation}
where $f_R(\reha)$ is the $R$-band flux density per wavelength interval
within \reha.  It is derived from the $R$-band aperture
photometry and the standard definition of fluxes in the ABmag system.
We use \eweff\ instead of a total equivalent width since it is directly
comparable to \Seha, which is also measured within \reha.  In addition,
\eweff\ usually has considerably smaller errors due to the smaller
aperture area needed for the measurement.  

For each ELG, two sets of radial profiles are made, one where the
extraction apertures are centered on the brightness peak, the other
where the apertures are centered on the geometric center of the
outermost apparent isophote.  We use the former set to define $r_e$, and
the latter set to define the total flux, $r_{90}$ and \rmax.

\begin{figure*}
\plottwo{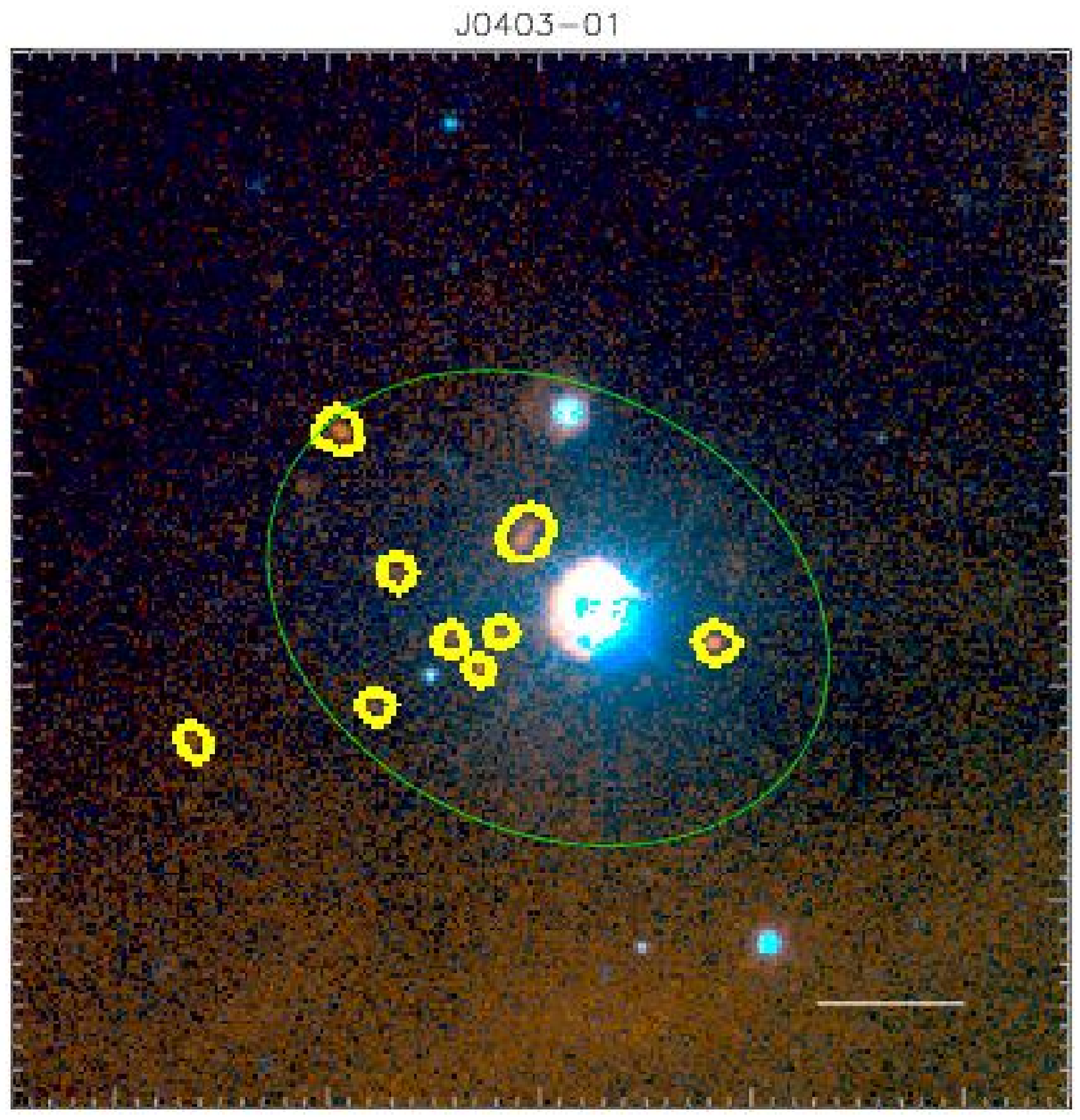}{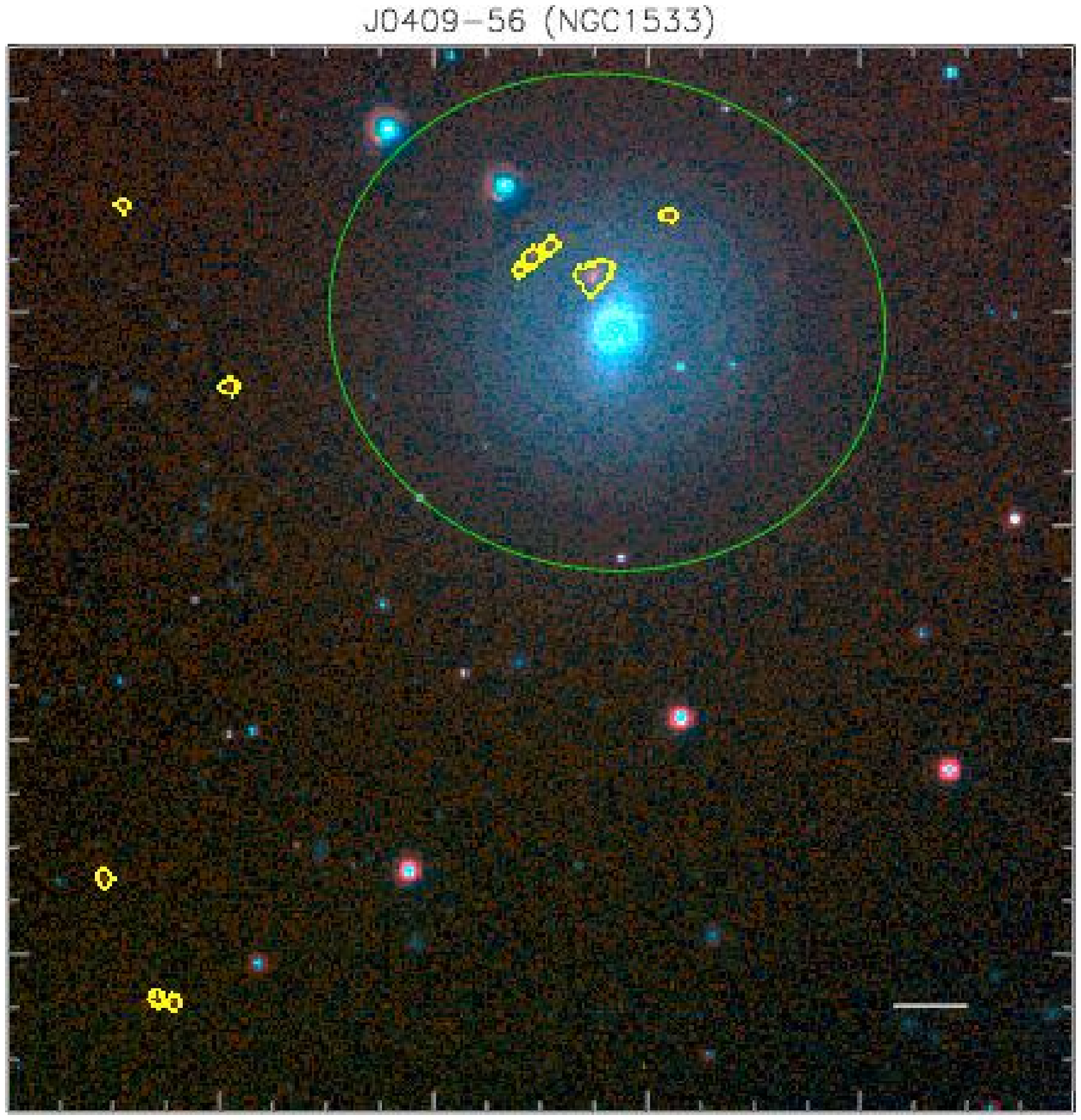}
\caption[]{Color, partial frame images of \HIPASS\ J0403$-$01
  (left) and \HIPASS\ J0409$-$56 (right) with net \Halpha, narrow band
  (not continuum subtracted), and 6850/95 displayed in red, green, and blue,
  respectively.  The apertures used to measure the \Halpha\ flux are
  outlined in yellow, while the aperture used to measure the total
  6850/95 flux is shown in green.  The scale bar (lower right each
  panel) is 30'' long.
  \label{f:spcases}}
\end{figure*}

We found the above method to be sufficient to perform the measurements
in all but two cases, shown in Fig.~\ref{f:spcases}, which we now
detail.  {\bf HIPASS J0403$-$01}: the field of this galaxy is strongly
contaminated with \Halpha\ emitted by Galactic cirrus; in addition there
is a bright star very near the target galaxy.  Because of its presence,
we observed the galaxy with the 6850/95 filter instead of the $R$ band,
so as to minimize saturation.  The galaxy is seen primarily by the
presence of a few \HII\ regions located near the bright star.  If there
is diffuse \Halpha\ it can not be disentangled from the foreground
cirrus.  We therefore measure \FHa\ using the summed flux from the \HII\
regions, measured with small apertures placed around each source.  It is
not clear whether the galaxy is detected in the continuum due to the
glare from the bright star.  We use an elliptical aperture whose center
is offset from the bright star to measure the continuum flux. The center
of the bright star is masked from the aperture, but we were not able to
remove the light from the outskirts of the star.  The measured continuum
flux should be considered an upper limit.  {\bf HIPASS J0409$-$56}
(NGC~1533) is a high-surface brightness SB0 galaxy.  The center of this
galaxy is saturated in the $R$ band so we used images through the
6850/95 filter to obtain the continuum flux.  A few \HII\ regions as
well as the ELDots discussed by \citet{rw04} are visible in the net
\Halpha\ image.  The continuum is so strong relative to \Halpha\ in this
galaxy that the \FHa\ derived using our standard technique is totally
swamped by the continuum subtraction uncertainty.  As for the case of
HIPASS J0403$-$01 we measure \FHa\ through a set of eye-selected small
apertures centered on the \HII\ regions, as well as the ELDots
\citep[since they were shown to be part of the galaxy by ][]{rw04}.  The
continuum flux is measured through a large elliptical aperture, as is
usually the case.  The reader is cautioned that for both these cases the
apertures used to measure \FHa\ and the continuum flux are considerably
different.  Since \FHa\ measures only the light of the noticeable \HII\
regions it may be significantly underestimated in these cases.  We
have not attempted to measure \reha, $r_{90}$, and related quantities, in
them because of the unorthodox nature of the measurement aperture.

\subsection{Flux corrections}\label{s:fcor}

Line fluxes are corrected for the effects of foreground and internal
dust absorption, \fion{N}{II} contamination, and underlying \Halpha\ absorption.
$R$-band fluxes are corrected for foreground and internal dust.

The foreground dust absorption is parameterized by the reddening \ebv\
taken from the \citet{sfd98} maps and listed in Table~\ref{t:sample}.
The extinction at the observed wavelength of \Halpha\ is calculated
using the extinction law of \citet{ccm89}.  For the average $\vhel =
2000$ \kms\ of the full SINGG sample \Halpha\ is at an observed
$\lambda = 6606.6$\AA\ and the foreground Galactic extinction is $A_{\rm
H\alpha,G} = 2.50 E(B-V)$.  For the $R$-band we adopt a foreground dust
absorption of $A_{\rm R,G} = 2.54 E(B-V)$.

To correct for internal dust, we adopt the relationship used by
\citet{hwbod04} between the Balmer decrement ($F_{\rm H\alpha}/F_{\rm
  H\beta}$) derived internal dust absorption $A_{\rm H\alpha,i}$ and the
$R$-band absolute magnitude calculated {\em without\/} any internal dust
absorption correction $M'_R$.  This is based on Balmer line ratios
measured from integrated (drift scan) galaxy spectra of the Nearby
Field Galaxy Survey \citep{jansen00,jffc00}.  After correcting to the AB
magnitude system, the internal dust absorption is given by:
\begin{equation}
\log(A_{\rm H\alpha,i}) = (-0.12\pm 0.048)M'_R + (-2.47 \pm 0.95) 
\label{e:helmboldt_ahai}
\end{equation}

\begin{figure}
\plotone{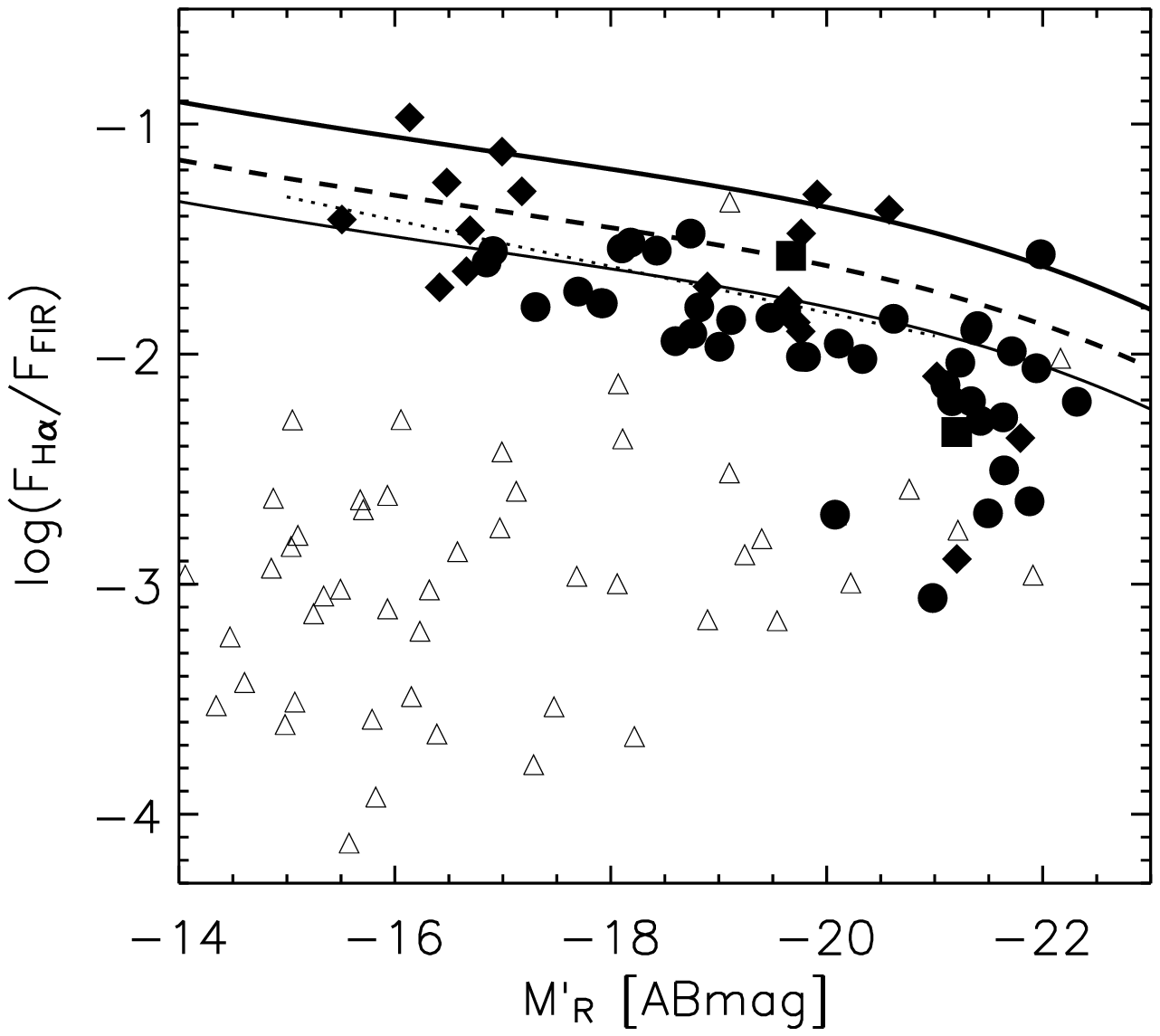}
\caption[]{Comparison of the \Halpha\ to FIR flux ratio, $\FHa/\FFIR$,
  with the $R$-band absolute magnitude (with no internal extinction
  correction) $M'_R$.  IRAS data were used to derive the FIR flux using
  the algorithm of \citet{hkmb88}.  Filled symbols mark IRAS detections
  with the data taken from the IRAS large optical galaxy catalog
  \citep{rice88}, the IRAS Faint Source Catalog \citep{irasfsc}, and the
  IRAS Point Source Catalog marked with squares, diamonds, and circles
  respectively. Triangles correspond to sources which are not in any of
  these catalogs.  We take these to be non-detections by IRAS and place
  them at their $3 \sigma$ lower limits in $\FHa/\FFIR$.  The curves
  represent the application of simple dust reprocessing models on
  stellar population models as discussed in the text, with the main
  difference being in the IMF.  The solid line is for a \citet{s55} IMF
  which has a slope $\alpha = 2.35$, and a mass range of 1 - 100 \Msun;
  the dashed line is for $\alpha = 2.35$ over the mass range of 1 to 30
  \Msun.  The thin solid line and dotted line segment show fits to the
  data having $M'_R > -21$ ABmag: the thick solid line shifted
  vertically and a simple linear fit respectively.
  \label{f:iras}}
\end{figure}

The radiation that dust absorbs is re-emitted in the far-infrared (FIR).
Hence FIR observations can provide a valuable test of the $A_{\rm
H\alpha,i}$ correction.  In Fig.~\ref{f:iras}, we use IRAS 60\micron\
and 100\micron\ flux densities to calculate the ``total'' 40-120\micron\
flux \FFIR\ using the formula given by \citet{hkmb88}. The
IRAS data are taken from three sources as noted in the caption to
Fig.~\ref{f:iras}.  61/113 SR1 ELGs are detected by IRAS, in the
remaining cases we show the ratio at the 3$\sigma$ upper limit to their
\FFIR\ flux.  The IRAS detected ELGs do show a trend of decreasing
$\FHa/\FFIR$ with decreasing $M'_R$, while the non-detections are
consistent with the trend.

To test whether the trend is consistent with eq.~\ref{e:helmboldt_ahai}
we apply a simple model for the dust extinction and re-emission in the
infrared.  In it, a stellar population is enshrouded in dust that obeys
the \citet{c00} dust obscuration law.  The amount of gas phase
extinction is parameterized by $A_{\rm H\alpha,i}$ derived from $M'_R$
using eq.~\ref{e:helmboldt_ahai}.  The flux absorbed by the dust is
re-emitted in the FIR and we assume that 71\%\ of the dust emission is
recovered by \FFIR\ \citep{m99}.  We show curves for two models and two
fits to the data.  For the models, the stellar populations are solar
metallicity 100 Myr duration continuous star formation models from
Starburst99 \citep{starburst99}.  They differ only in their IMF which is
parameterized by a single power law in mass having slope $\alpha = 2.35$
\citep{s55} and a specified mass range, a lower mass limit of 1 \Msun,
and an upper mass limit $m_u$ of 100 \Msun\ (solid line) and 30 \Msun
(dashed line).  The \Halpha\ output in photons per second is taken to be
46\%\ of the ionizing photon output, as expected for case B
recombination of an ionization limited \HII\ region.  Neither of these
models passes through the center of the observed ratios of the detected
sources, although the $m_u = 100$ \Msun\ model nicely defines the upper
envelope.  Adopting $m_u = 30$ \Msun\ results in a better match, but is
still displaced with respect to the data.  We also tested a model with a
steeper $\alpha = -3.3$ and $m_u = 100$ \Msun.  It has the same shape
and falls between the other two models.  We omit showing it so as not to
clutter the figure.

In order to better understand the correlation we make two fits.
In the dot dashed curve we take the $m_u = 100$ \Msun\ model and fit the
best offset in the $y$-axis finding it to be $-0.43$ dex, while the thin
solid line shows a simple linear fit.  For both cases we are only
fitting the data for galaxies with $M'_R > -21$; we use a robust fitting
algorithm and reject outliers.  The rms dispersion in the $\log(F_{\rm
H\alpha}/F_{\rm FIR})$ residuals about the fits are 0.22 dex for the
linear fit, and 0.23 dex for the shifted model.  The fits both yield a
reasonable representation of the data for $M'_R > -21$, while the
galaxies with $M'_R \leq 21$ have an average displacement of $-0.25$ dex
from the offset model fit.

The {\em shape\/} of the model curves is driven by the form of the
$A_{\rm H\alpha,i}$ versus $M'_R$ relationship.  We see that the adopted
model adequately specifies the shape, except for the brightest galaxies.
This can be seen by the fairly good agreement of the linear fit and the
shifted model line.  However, the model does not adequately account for
the {\em zeropoint\/} of the relationship, instead an arbitrary shift is
required.  The zeropoint of the model effectively gives the ratio of the
ionizing to bolometric flux of the stellar populations. As noted above,
adjusting $m_u$ or $\alpha$ can shift the model lines vertically.  An
error in the stellar models themselves can also result in a zeropoint
error.  Recent improvements in the modeling of hot stars using non-LTE
expanding atmospheres with realistic line blanketing \citep{snc02,msh05}
indicate that the ionizing flux output of stars is lower than expected
from the \citet{lcb97} stellar atmosphere models used by Starburst99,
resulting in our model $\FHa/\FFIR$ values being too high.  The use of
improved stellar models should then move the model lines in the correct
direction.  It is also possible that older populations could also
contribute significantly to the dust heating, but not the ionization.
These could result from a star formation history that is declining with
time.  This would be in the correct sense compared to compilations of
the cosmic SFR density evolution \citep[e.g.][]{pgpg05,glaze03}.
Finally, the offset could be due to the inadequacy of the dust model to
account for all star formation.  Then, the fact that our
Balmer-decrement based models don't recover this star formation would
indicate that it is totally hidden by dust.  

Our model adequately models the trend of the \Halpha\ extinction for
galaxies having $M'_R > -21$ ABmag, but is not capable of
self-consistently accounting for the FIR emission.  The SINGG ELGs that
have been detected by IRAS are on average 2.7, 4.8 times brighter in the
FIR than predicted by our model for galaxies less and more luminous than
$M'_R = -21$ ABmag respectively.  As noted above there are a variety of
explanations for this offset.  If the zeropoint offset is removed, then
at the faint end, our dust absorption model predicts the $F_{\rm
H\alpha}/F_{\rm FIR}$ ratio, and by inference the SFR, to
within an a factor of 1.7.  This is sufficient for our purposes - we
wish to determine star formation rates that can be inferred from
\Halpha\ fluxes and quantities that can be inferred from optical
wavelength observations.  Recovery of the star formation that is totally
obscured by dust, is beyond the scope of this survey.  Our adopted dust
absorption model is conservative in that it does not over-predict the
FIR emission.

Since \HII\ regions represent star formation sites, where dust and gas
concentrations are particularly strong, they represent enhanced dust
absorption compared to that seen in the older stellar populations in the
galaxy.  Indeed, it has long been known that internal extinction
estimates of galaxies derived from Balmer lines are larger than those
found by continuum fitting, typically by a factor of $\sim 2$
\citep{fct88,c94}.  Therefore we adopt $A_{R,{\rm i}} = 0.5 A_{\rm
  H\alpha,i}$ to correct $M'_R$ to the total (internal and foreground)
dust corrected absolute magnitude $M_{R}$.

To correct for \fion{N}{II} contamination we adopt the correlation
between the \fion{N}{II} line strength and $M'_R$ given by
\citet{hwbod04} and corrected to the ABmag system
\begin{equation}
\log(w_{6583}) =  (-0.13 \pm 0.035) M'_R + (-3.30 \pm 0.89) \label{e:helmboldt_nii}
\end{equation}
where 
\begin{equation}
w_{6583} \equiv \frac{F_{\rm [NII]6583\AA}}{F_{\rm H\alpha}}
\end{equation}
As before, the correlation is based on the NFGS sample of
\citet{jansen00}.  The correction of the line flux
includes both \fion{N}{II} lines at 6583\AA\ and 6548\AA\ and is
calculated using the filter profile and a simple emission line velocity
profile model as outlined in Appendix \ref{s:fluxap}.

An important source of possible bias results from \Halpha\ emission
being hidden by \Halpha\ absorption.  \citet{mrs85} found a typical
Balmer line absorption EW of 1.9\AA\ in a wide range of extragalactic
\HII\ regions.  \HII\ regions represent active sites of star
formation, and typically have a high equivalent width and small covering
factor over the face of a galaxy.  What we need is a correction for
\Halpha\ absorption appropriate to the integrated spectra of galaxies.
For this we turn to the SDSS, whose fiber spectra typically account for
one third of the flux in nearby galaxies, as shown by B04.  They show
that adopting a uniform \Halpha\ absorption correction corresponding to
$\ew(\Halpha)_{\rm absorption} = 2$\AA\ could cause systematic errors in
SFR determinations with stellar mass.  They note typically the stellar
absorption comprises 2\%\ -- 6\%\ that of the stellar emission in flux.
Therefore we uniformly increase \FHa\ and \ewha\ by 4\%\ to account for
underlying stellar absorption.

A correction to the photometry of the three sources observed
with the 6850/95 narrow band continuum filter was applied in order to
make their magnitudes compatible with those in the $R$ band.  We found
that the fluxes for the two cases where the source were observed with
both filters have identical flux density {\em per wavelength interval\/}
$f_\lambda$ (within the errors) derived from each filter.  Since the two
filters have different pivot wavelengths (Table~\ref{t:filts}) their
flux density {\em per frequency interval\/}, and hence ABmag differs.
For a flat $f_\lambda$ spectrum, the correction to add to $m_{6850/95}$
in ABmag to get the equivalent $m_R$ is $5\log(6858.9/6575.5) = 0.114$
mag, which we apply to all $m_{6850/95}$ measurements.

Finally we note some effects that we have not corrected for.  (1) We
have not corrected our $R$-band fluxes for contamination by \Halpha\ or
other emission lines (\fion{N}{II}, \fion{S}{II} and \fion{O}{I}
typically being the strongest).  Since we find a median $\eweff(\Halpha)
= 16$\AA\ and the width of the $R$ filter is 1450 \AA\
(Table~\ref{t:filts}), then the $R$ fluxes will typically be
underestimated by a few percent.  This in turn means that \eweff\ will
also be underestimated by a few percent.  The galaxies with the highest
\eweff\ would require the largest corrections, up to $\sim 25$\%. (2)
Changes in the NB filter transmission curves due to temperature changes
and filter aging may cause errors in \Halpha\ fluxes.  Neither effect
has been calibrated, but we expect the errors to be limited to the few
percent level.

\section{Global properties database}\label{s:measres}

The results of the image analysis are listed in Tables~\ref{t:ap},
\ref{t:intrinsic} and \ref{t:cor}.  Combined, these represent the
tabulated data of SR1.  In all tables, the first column gives the source
designation used in this study.  If there is only one ELG in the field
the \HIPASS\ designation is used.  If there is more than one, the
\HIPASS\ name is appended with ``:S1'', ``:S2'', etc.\ in order to
distinguish the sources, where the ``S'' stands for SINGG.
Table~\ref{t:ap} defines the apertures used to measure fluxes, presents
the identification of the sources from catalog matching, and provides
morphological information from a variety of literature sources.  The
optical identifications were adopted from HOPCAT, \citep[the \HIPASS\ 
optical catalog of][]{hopcat05}, the BGC, or from the NASA/IPAC
Extragalactic Database (NED\footnote{The NASA/IPAC Extragalactic
  Database (NED) is operated by the Jet Propulsion Laboratory,
  California Institute of Technology, under contract with the National
  Aeronautics and Space Administration.}).  There are four ELGs in SR1
with no previously cataloged optical counterparts: \HIPASS\ J0403$-$01
\citep[also noted by ][]{rw02}; \HIPASS\ J0503-63:S2; \HIPASS\ 
J0504-16:S2, and \HIPASS\ J1131-02:S3.

Table~\ref{t:intrinsic} presents the measured properties of the sources.
These include the $R$ band absolute magnitude $M_R$; the effective and
90\%\ enclosed flux radii in the $R$ band, \reR, $r_{90}(R)$; the
corresponding quantities in net \Halpha: \reha, $r_{90}(\Halpha)$; the
\Halpha\ derived SFR; the face-on star formation rate per unit area,
within \reha, SFA; the face-on $R$ band surface brightness within \reR,
$\mu_{e,0}(R)$; and the \Halpha\ equivalent width within \reha, \eweffo.
These are intrinsic properties, that is corrected for Galactic and
internal extinction and in physically meaningful units.  We also present
\Halpha\ fluxes, \FHa, corrected only for internal extinction and
\fion{N}{II} contamination to allow easy comparison with other work
\citep[e.g.][]{hwbod04,hags1,mmhs97}.  In Table~\ref{t:intrinsic} the
star formation rates given by SFR and SFA have been calculated using the
conversion ${\rm SFR\, [1 \Msun\, yr^{-1}]} = \LHa\, [{\rm erg\,
cm^{-2}\, s^{-1}}] /1.26\times 10^{41}$ calculated by \citet{ktc94}, and
adopted by many other studies \citep[e.g.\
][]{k98,lsitg02,kbsbn04,hopkins04,hwbod04}.  This conversion adopts a
Salpeter (1955) IMF slope with lower and upper mass limits of 0.1 and
100 \Msun.  To compare our results to those that adopt the
\citet{kroupa01} IMF, as do some more recent studies \citep[e.g.\ B04;
]{kauffmann03,tremonti04}, one should divide our SFR estimates by 1.5
(B04).  The errors presented in Table~\ref{t:intrinsic} are derived from
the error model discussed in Sec.~\ref{s:fluxerr}, below.  The
corrections adopted and discussed in Sec.\ \ref{s:fcor} are given in
Table~\ref{t:cor}.

\begin{figure*}
\centerline{\hbox{\includegraphics[width=6.75cm]{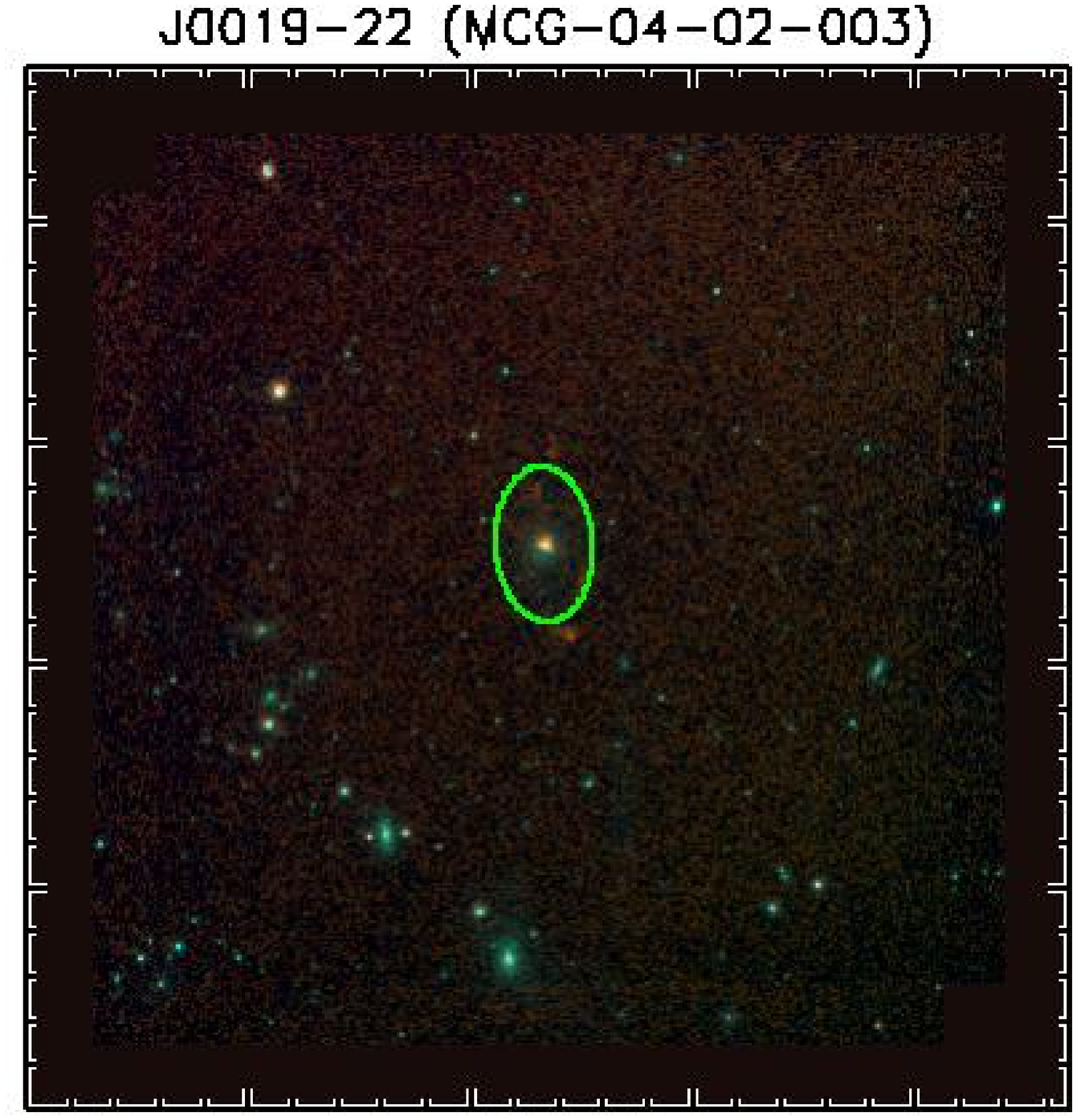}
                  \includegraphics[width=6.75cm]{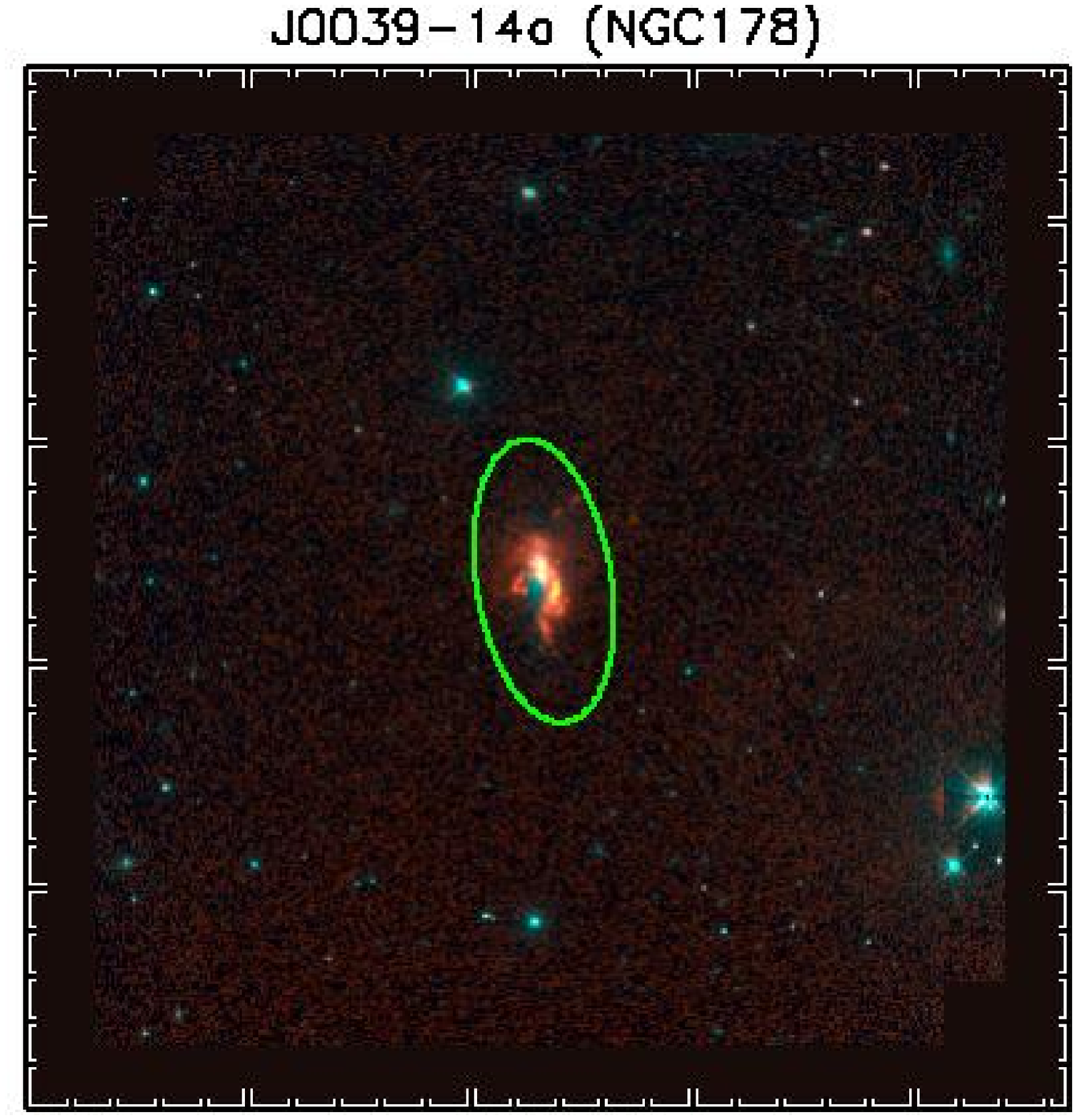}}}
\caption[]{{\it (Sample portion of figure.)\/} Three color images of 
  the target fields, with net \Halpha, narrow band (not continuum
  subtracted), and $R$ displayed in red, green, and blue, respectively.
  North is up, east is to the left, and the tick marks are separated by
  100 pixels (43\as). \HIPASS\ names and optical identifications are
  given above each frame. The elliptical flux measurement aperture is
  shown in green.  For fields with multiple sources they are labeled
  with the SINGG ID (S1, S2, etc.).
\label{f:finder}}
\end{figure*}

Full frame color representations of the images are presented in
Figure~\ref{f:finder}.  The NB image has a larger display range than the
net \Halpha\ image, resulting in the \HII\ regions appearing orange-red
with yellow or white cores.  The paper version of this article shows
only a portion of Fig.~\ref{f:finder}.  All images are available in the
online version of this article.

\subsection{Image quality}\label{s:qual}

\begin{figure}
\plotone{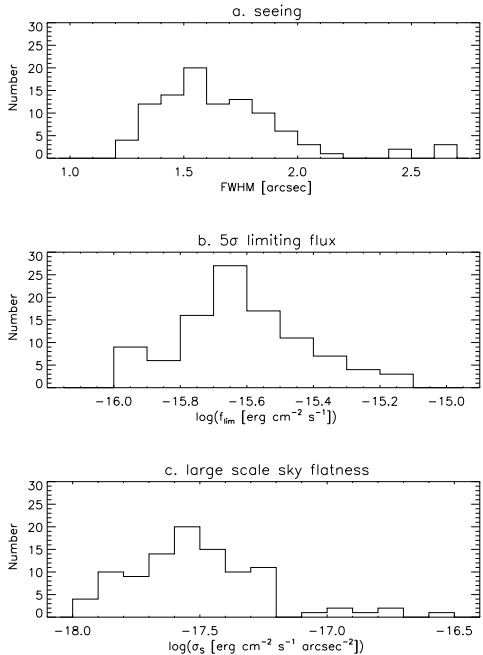}
 \caption[]{Histograms of net \Halpha\ image quality measurements.
  Panel a. (top): FWHM seeing; b.\ (middle) 5$\sigma$ limiting flux; and
  c.\ (bottom) large scale ($\sim 15''$) sky flatness.
  \label{f:imqual}}
\end{figure}

\begin{figure}
\plotone{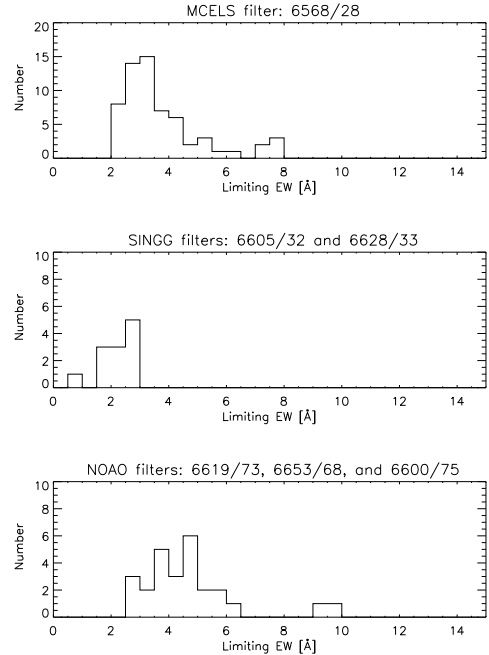}
\caption[]{The scatter in the NB to $R$ (or 6850/95) ratio within a
  frame calibrated to an EW using eq.~\ref{e:sigew}.  $\sigma_{\rm EW}$
  is derived from SE catalogs and represents the typical uncertainty in
  a single foreground or background source.
  \label{f:sigew}}
\end{figure}

The quality of the net \Halpha\ and $R$ band images is specified by the
seeing, the limiting flux, and the flatness of the sky. The limiting EW
is an additional quality measurement that is only applicable to the net
\Halpha\ images.  Statistics on these quantities are compiled in
Table~\ref{t:qastat}, for both the net \Halpha\ images, and where
relevant, the $R$ band images as well.  Histogram plots of these
quantities are shown in Figures~\ref{f:imqual} and \ref{f:sigew} for the
net \Halpha\ images.

The FWHM seeing values (Fig.~\ref{f:imqual}a) are mostly less than 2\as,
with a median of 1.6\as.  The seeing values are slightly worse in net
\Halpha, since our method results in the net image having the larger
seeing of those in $R$ and NB.

The limiting flux, $f_{\rm lim}$ is derived from 
\begin{equation}
f_{\rm lim} = n_\sigma \sigma f_{\rm ap} 
           \sqrt{\pi r_{\rm seeing}^2/a_{\rm pix}}\label{e:flim}
\end{equation}
where $\sigma$ is the pixel-to-pixel rms of the background, $r_{\rm
seeing}$ is the seeing radius (half the FWHM seeing plotted in
Fig~\ref{f:imqual}a), $a_{\rm pix} = 0.19\, {\rm arcsec}^2$ is the pixel
area, $f_{\rm ap}$ is the aperture correction within $r_{\rm seeing}$
(we adopt $f_{\rm ap} = 2.0$), and $n_\sigma = 5$ is the adopted
significance level of the limiting flux.  Defined this way, $f_{\rm
lim}$ is the $n_\sigma$ limiting flux of a point source detection.
Figure~\ref{f:imqual}b plots the histogram of $f_{\rm lim}$.  The median
limiting \Halpha\ flux corresponds to a luminosity $L_{\rm H\alpha}
\sim\ 10^{37}\, {\rm erg\, s^{-1}}$ (neglecting any extinction
corrections) at the median distance of the SINGG sample.  This
corresponds to about half the ionizing output of a single O5V star
(solar metallicity) using the ionizing flux scale of \citet{snc02}.

The sky flatness, $\sigma_S$, is a traditional estimate of the quality
of an image.  It is defined as the {\em large scale\/} variation in the
background.  We measure $\sigma_S$ as the rms of the background
measurements in 35$\times$35 pixel boxes in the sky annulus.  Hence,
this is a measurement of local flatness, rather than a full frame
measurement (except for the largest sources).  Histograms of $\sigma_S$
are displayed in erg cm$^{-2}$ s$^{-1}$ arcsec$^{-2}$ in
Fig.~\ref{f:imqual}c.  Emission line surface brightness is often given
in other units: Rayleighs, defined as ${\cal R} = 10^6/4\pi \, {\rm
photons\, cm^{-2}\, s^{-1}\, sr^{-1}}$ $= 5.67\times 10^{-18}\, {\rm
erg\, cm^{-2}\, s^{-1}\, arcsec^{-2}}$, and emission measure $EM =
2.78\times {\cal R}\,\, {\rm pc\, cm^{-6}}$ for an assumed electron
temperature $T_e = 10^4\, {\rm K}$.  In these units the median large
scale (area $\gtrsim 15'' \times 15''$) rms surface brightness
variations in the net \Halpha\ images corresponds to $0.51 {\cal R}$ and
$EM = 1.4\, {\rm pc\, cm^{-6}}$.  This is about sixty times fainter than
the surface brightness cut used by \citet{fwgh96} to define DIG
emission. In Sec~\ref{s:fluxerr} we parameterize the uncertainty due to
sky subtraction as a function of $\sigma_S$.

The dispersion in the narrow band to $R$ band scaling ratio,  $\sigma_{\rm
rat}$, for background and foreground sources can be used to estimate the
range of intrinsic \ew\ values of sources that are not line emitters,
\sigew.  We define this quantity as
\begin{equation}
\sigew = \sigma_{\rm rat}\frac{U_{\rm NB,line}}{U_R} . \label{e:sigew}
\end{equation}
Here, $U_{\rm NB,line}$ is the unit response to line emission in the NB
frame and $U_R$ is the unit response to the continuum flux density in
the $R$ image.  These quantities are defined in Appendix \ref{s:fluxap}.
The quantity $\sigma_{rat}$ is the dispersion about the mean of the NB
to continuum flux ratio, derived from matched sources in the SE catalogs
of the frames, after applying a three sigma clip to the ratios.
Figure~\ref{f:sigew} shows that \sigew\ is lowest for the two SINGG
filters which have a median $\sigew = 2.4$\AA.  The MCELS 6568/28 has a
significantly higher median $\sigew = 3.2$\AA\ probably because this
filter encompasses Galactic \Halpha.  The NOAO filters have the highest
median $\sigew = 4.5$\AA\ due to their broader bandpass widths.
In Sec~\ref{s:fluxerr} we demonstrate that the mean flux scaling ratio
can be determined to significantly better than \sigew.  However,
measured \ewha\ values approaching \sigew\ should be treated with some
caution because differences between the flux scaling of program sources
versus foreground and background sources could result in systematic
errors approaching \sigew.

\subsection{Quality assurance tests and rejected images}\label{s:qatest}

We subjected the images and our database to a wide range of quality
assurance tests.  As noted in Sec.\ \ref{s:finalhi}, our sample was
checked for possible HVC contamination, and uniform \HI\ properties were
adopted. The reality of all tentative low S/N \Halpha\ detections (in
terms of flux or EW) as well as multiple ELGs was checked by eye,
resulting in the removal of some overly optimistic ELG identifications.
Optical identifications were checked in cases where our identification
did not agree with HOPCAT.  The radial profiles and curves of growth
were checked for the effects of unmasked or improperly masked objects.
We calculated the fraction of the unmasked image covered by the \HIPASS\
half-power beam area for the \HIPASS\ source that was targeted.  We also
checked that the filter used for the observation covered the velocity of
the source.  Color images of all sources were examined to check source
location, large scale sky variations, and other blemishes.  Cases where
the source extends to the edge of the frame or beyond are marked in
Table~\ref{t:intrinsic}.

These tests revealed four sets of observations which we rejected as
non-survey observations.  These include observations of a source
rejected from our final sample (it is part of the Magellanic stream),
two cases of mis-pointing due to \HI\ position errors in earlier versions
of our sample selection, and one observation set that was rejected due
to a very bright sky background (10$\times$ normal due to the proximity
of the gibbous Moon).  In this paper we use these observations only to define
our sky error model in the following subsection.

\subsection{Error model}\label{s:fluxerr}

In measurements of extended sources, typically the largest sources of
random error are sky subtraction, which affects both the $R$ and
\Halpha\ results, and continuum subtraction which affects the \Halpha\
results.  These affect not only the fluxes but also the other
measurements obtained here.  The rest of this subsection details our
error model for these terms.  In addition there is a flux calibration
error.  We have adopted a calibration error of 0.04 mag for data
obtained with the 6568/28 filter and 0.02 mag for data obtained with the
other filters, which was derived from the residuals of the observed
minus intrinsic magnitude versus airmass of the standard stars.  Since
the data presented here span several observing runs and filters, this
error term is considered to be a random error and is added in quadrature
with the other flux uncertainties described below.  

By measuring the sky in an annulus around the source, we can estimate
the sky within \rsky\ more accurately than the large scale sky
fluctuations $\sigma_S$ which we use to characterize the flatness of the
image.  To demonstrate and calibrate this effect we placed apertures, of
a variety of sizes, on ``blank'' portions of our images - that is in
areas away from the target sources.  This allows the sky to be measured
in both the sky annulus and interior to \rsky.  These tests were
restricted to circular apertures; there is no reason to expect the
results to differ for elliptical apertures of equivalent area.

\begin{figure}
\epsscale{0.8}
\plotone{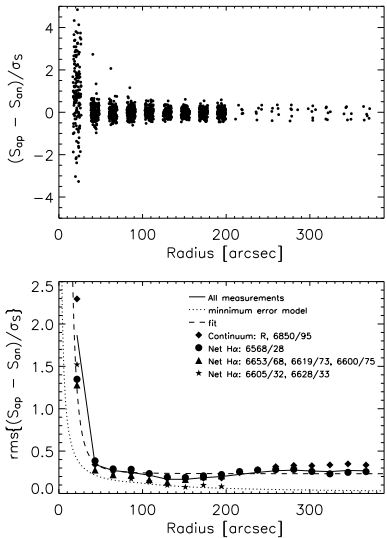}
\caption[]{Derivitation of the sky uncertainty model.  The top panel
  shows the sky level difference between that within a circular aperture
  and that in the surrounding sky annulus, normalized by $\sigma_S$ the
  dispersion in sky measurements within the sky annulus.  The data were
  measured in blank portions of real data frames, and are plotted
  against aperture radius, $r$.  Small random offsets in $r$ are
  employed in this plot to allow the density of the measurements to be
  distinguished.  The measurements for $r > 200\as$ were performed on
  frames containing no ELGs.  The bottom panel shows the rms dispersion
  of the quantity above.  The solid line uses all measurements at each
  $r$ in determining the mean, while the symbols are for different
  subsets of filters as noted in the legend.  The dotted line shows the
  ``least sky error'' model, while the dashed line shows our fit to the
  combined measurements.
  \label{f:skyerr}}
\end{figure}

The results are shown in Fig.~\ref{f:skyerr}.  The top panel plots the
difference in sky levels interior to \rsky\ and that in the annulus
normalized by $\sigma_S$.  For $r_{\rm sky} \geq 50''$, the difference
in sky values is typically less than the large scale sky fluctuations.
The points at $r_{\rm sky} = 50''$ in the top panel of
Fig.~\ref{f:skyerr} have a mean value somewhat offset from zero,
implying that the sky is systematically higher in the measurement
aperture than the sky annulus.  This probably results from a difference
in the sky determination algorithm we had to implement for apertures
this small.  For large \rsky\ we use our standard clipping algorithm
(Sec~\ref{s:ss}) to determine the sky level in both the sky annulus and
interior to \rsky.  However, $\rsky = 50''$ is so small that often too
few 35$\times$35 pixel boxes survive to accurately measure the sky
level.  Hence, in this case we take the sky interior to \rsky\ to be
simply the 3$\sigma$ clipped mean of all the pixels within the aperture.
Since there is no box rejection, the measurement can include the wings
of some stars, and hence may be slightly elevated.

The bottom panel shows the rms of the normalized sky difference
measurements.  We take this quantity to be equivalent to
$\epsilon_S/\sigma_S$ where $\epsilon_S$ is the true sky uncertainty
within the measurement aperture (this approximation
somewhat overestimates $\epsilon_S$ since some of the rms can be
attributed to the uncertainty in the sky level within the sky annulus).  We
show this quantity for cases where we combine all measurements at each
radius to calculate the rms, and when we consider continuum images
separately from net \Halpha\ images which are further subdivided into
logical filter groups.  The dotted line shows a ``least sky error''
model. This would be applicable if the overall sky was flat and residual sky
errors occurred on scale sizes less than the 35 pixel box size used to
make the sky measurements.  For this model 
\begin{equation}
\frac{\epsilon_S}{\sigma_S} = \sqrt{\frac{N_{\rm ap}+N_{\rm an}}{N_{\rm ap}N_{\rm an}}},
\end{equation}
where $N_{\rm ap}$ and $N_{\rm an}$ are the number of measurement
boxes within \rsky\ and the sky annulus respectively.  The
dotted line is drawn assuming perfect packing of the boxes and none 
rejected.  The fact that almost all measurements are above this line
indicates that residual sky errors typically have scale sizes larger
than 35 pixels.  The dashed line shows our fit to the data
\begin{equation}
\frac{\epsilon_S}{\sigma_S} \approx 0.23 + \left(\frac{22.5\as}{r}\right)^3.
\end{equation}
This is an ``eye'' fit to the data adopted for convenience of
calculation, and is not meant to provide insight to the origins of the
residual sky errors.  When applying this model to the elliptical
apertures used in the actual galaxy measurements we replace $r$ with the
equivalent radius, $\sqrt{ab}$, where $a$ and $b$ are the semi-major and
semi-minor axes dimensions of the flux measurement aperture.  To
determine the total flux error due to the sky, we multiply the model by
$\sigma_S$ in units of count rate per pixel and the aperture area in
pixels and calibrate to yield the total flux error due to sky in the
appropriate units.  We adopt a maximum $\epsilon_S/\sigma_S = 2$ to
avoid the model blowing up at small $r$.

To translate this to uncertainties in $r_e$, $r_{90}$ and $S_e$ we
derive what the curve of growth would be if the sky level was changed by
adding or subtracting $\epsilon_S$.  This results in two additional
curves of growth.  The $r_e$, $r_{90}$ and $S_e$ values are found as
before resulting in two additional estimates of these quantities.  We
then find the maximum difference in these quantities between three
estimates - that derived from the nominal curve of growth and those
derived from the additional curves of growth.  We take the error to be
one half this maximum difference.

The random uncertainty on \Halpha\ flux measurements due to continuum
subtraction is set by how well the adopted continuum scaling ratio is
determined.  Since many foreground and background sources are used to
determine this ratio, we expect the accuracy to be better than the
source to source rms in the flux ratio $\sigma_{\rm rat}$ (defined in
Sec~\ref{ss:meas}).  Since the NB filters and continuum filters have
similar mean wavelengths, and to first order the spectral properties of
foreground and background sources should not vary significantly from
field to field at the high latitudes of our survey, then we take the
fractional error due to continuum subtraction, $\epsilon_C/C$ to be the
field to field dispersion in the continuum scaling ratio normalized by
the mean continuum ratio.  The adopted values of $\epsilon_C/C$ are
given in Table~\ref{t:filts}.  They range from 0.024 to 0.043, about one
third of $\sigma_{\rm rat}$.  We made sufficient observations to
determine the normalized rms for four NB filters.  For two other NB
filters (6628/33 and 6600/75) we have not made enough observations to
determine an accurate rms (we require at least four), and so adopt the
fractional continuum error from similar filters.  The \Halpha\ flux
error within an aperture is then determined by multiplying the continuum
count rate $C$ by $\epsilon_C/C$ to get the count rate uncertainty.
This is then multiplied by the flux scaling coefficient to get the
\Halpha\ flux uncertainty.  The errors on $r_e$, $r_{90}$ and $S_e$ due
to continuum subtraction are found in a method analogous to the sky
error.  The errors due to continuum subtraction and that due to sky
subtraction are added in quadrature to yield the total random error on
the \Halpha\ flux.  We derive the uncertainty on \eweffo\ by propagating
the flux errors in \Halpha\ and $R$ band within \reha.

\subsection{Tests of the error model}

To test the internal accuracy of our error model we use repeat
measurements: We repeated observations in three cases albeit with
slightly different filters.  In each case, one of the two measurements
was superior, and that was adopted in the measurements given in
Table~\ref{t:intrinsic}.  Nevertheless, the other set was of sufficient
quality to test our error model.  We now briefly discuss the results and
note which observation was chosen for our results.  Measurements given
in these comparisons have not been corrected for internal extinction.
\bhipass\ {\bf J0507$-$7} (NGC~1808) was observed with both the $R$ and
6850/95 filters in run 01 as a test of the accuracy of narrow band
continuum subtraction.  Using an extraction aperture $\rmax = 6'$ we
measure the quantities [$r_e$, $r_{90}$, $m$, $\mu_e$] for the $R$ and
6850/95 observations of [$70.0 \pm 0.8$, $213 \pm 7$, $9.451 \pm 0.025$,
$20.592 \pm 0.023$] and [$67.1 \pm 1.7$, $197 \pm 15$, $9.496 \pm
0.036$, $20.547 \pm 0.033$] respectively (in units of [arcsec, arcsec, ,
ABmag, ABmag arcsec$^{-2}$]).  Hence, the difference between the
observations are [$2.8 \pm 1.8$, $17 \pm 17$, $0.044 \pm 0.043$, $0.045
\pm 0.40$] - all within about 1.5$\sigma$ of zero. The $R$ band
observations are centered better on the galaxy than the 6850/95 images.
They also have higher S/N and this clearly shows in the smaller errors
above, hence we adopt the $R$ image for our final measurements.
\bhipass\ {\bf J0409$-$56} (NGC~1533) was observed with the 6850/95
filters in run 01 after realizing that $R$ images were saturated in the
nucleus.  For this reason we adopt the 6850/95 results for our published
measurements. Using a broad annulus from $r = 9''$ to $240''$ (in order
to avoid the saturated nucleus), we measure $m_R = 10.476 \pm 0.021$
ABmag and $m_{6850/95} = 10.504 \pm 0.066$ ABmag, yielding a magnitude
difference of $0.028 \pm 0.033$, or zero within errors.  Because of the
saturation in $R$, we have not compared half or 90\%\ enclosed light
quantities. \bhipass\ {\bf J0943-05b} (UGCA~175) was observed with two
different NB filters on separate nights of run 02.  The first set of
observations with the $6600/75$ filter have an elongated PSF due to poor
tracking.  The second set of images obtained with the $6619/73$ filter
have a superior PSF and results from it are used as our adopted
measurements.  Using our adopted extraction aperture $\rmax = 1.66'$ we
measure [\reha, $r_{90}(H\alpha)$, $\log(F_{\rm H\alpha})$,
$\log(\Seha)$] = [$33.8 \pm 2.9$, $67.3 \pm 3.6$, $-12.33 \pm 0.08$,
$-16.19 \pm 0.13$] and [$33.3 \pm 2.5$, $66.2 \pm 1.8$, $-12.36 \pm
0.08$, $-16.21 \pm 0.13$] with the 6600/75 and 6619/73 images
respectively (in units of [arcsec, arcsec, $\log({\rm erg\, cm^{-2}\,
s^{-1}})$, $\log({\rm erg\, cm^{-2}\, s^{-1}\, arcsec^{-2}})$]).  Hence
the difference between the filters for these quantities are [$0.4 \pm
3.9$, $1.1 \pm 4.0$, $0.03 \pm 0.12$, $0.02 \pm 0.18$]; again the
results agree within the errors.

\begin{figure}
\epsscale{0.6}
\plotone{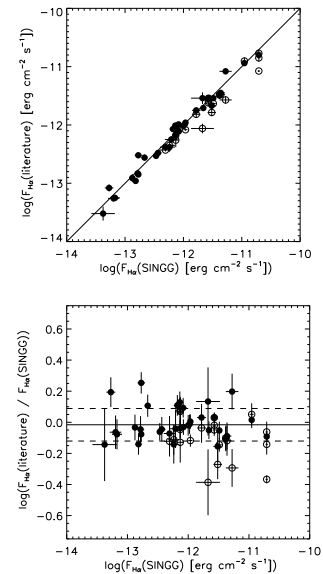}
\caption[]{Comparison of our total \Halpha\ fluxes with values from the
  literature (open circles) and from the HUGS group (Kennicutt et al.\
  2006, in preparation, filled circles).
  The top panel plots fluxes from other sources against ours.  The solid
  line is the unity relationship.  The bottom panel plots the ratio of
  fluxes from other sources to our own, compared to the SINGG flux.  The
  solid line marks the mean using only the HUGS data.  The dashed lines
  indicate the $\pm 1\sigma = 0.104$ dex dispersion about the mean HUGS
  value.  The sources for the published \FHa\ values are \citet{gmp03};
  \citet{hwb01}; \citet{hhg93}; \citet{hvwg94}; \citet{he04};
  \citet{mmhs97}; \citet{martin98}; \citet{mk06}, and \citet{rd94}.  
  \label{f:litplot}}
\end{figure}

As an external check of our fluxes, Fig.~\ref{f:litplot} compares our
total log(\FHa) measurements with a variety of published measurements as
well as with measurements from 11HUGS (11 Mpc \Halpha\ and UV Galaxy
Survey, Kennicutt et al.\ 2006, in preparation).  11HUGS has completed
an H$\alpha$ and R-band imaging survey of an approximately
volume-limited sample of $\sim$350 spiral and irregular galaxies within
a distance of 11Mpc.  The comparisons in Fig.~\ref{f:litplot} are made
as close to ``raw'' values as possible in order to reduce the possible
sources of error.  We correct the \FHa\ for \fion{N}{II} contamination,
because NB filter transmission curves vary strongly from survey to
survey, but almost always transmit some \fion{N}{ii}.  No internal
extinction, nor \Halpha\ absorption corrections were applied.  Likewise
we have not attempted to exactly match apertures with the literature or
HUGS measurements.  The errors are taken from the publications, where
available, otherwise we adopt a mean error of 0.063 dex, derived from
the SINGG \FHa\ used in the plot.  The bottom panel compares the
logarithmic ratio of the published \FHa\ fluxes to the SINGG value
plotted against the SINGG flux.  Hence the errors are the $x$ and $y$
errors in the top panel added in quadrature.

The weighted mean $\log(F_{\rm H\alpha}({\rm literature\/})/F_{\rm
H\alpha}({\rm SINGG\/}))= -0.030$ with a dispersion of 0.12 dex, when
using all 56 measurements.  Concentrating on just the 34 HUGS
measurements yields a weighted mean $\log(F_{\rm H\alpha}({\rm
literature\/})/F_{\rm H\alpha}({\rm SINGG\/}))= -0.016$ and a scatter of
0.10 dex.  We conclude that the SINGG \Halpha\ fluxes agree well with
other measurements - to within 33\%\ on average.  The agreement is a bit
better, to within 27\%\ for galaxies in common with the recent HUGS
survey.  For an average error of 16\%\ from SINGG and 12\%\ from HUGS we
expect a scatter of 0.08 dex about the mean.  While our error model can
account for much of the measured variance an additional $\sim 11$\%\
flux uncertainty (added in quadrature) in both the SINGG and HUGS fluxes
would be required for a full accounting.  Possible sources of additional
error include aperture placement, flux calibration (particularly in the
filter transmission curves and flux standards), and the \fion{N}{II}
correction.

\section{Results}\label{s:results}

\subsection{\Halpha\ detectability of \HI-selected galaxies}\label{s:detect}

\begin{figure}
\epsscale{1.0}
\plotone{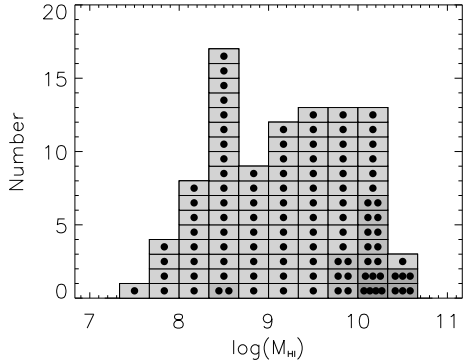}
\caption[]{The \HI\ mass histogram of the 93 SR1 targets.  Each
  rectangle represents one \HIPASS\ target, while each dot within a
  rectangle represents an Emission Line Galaxy (ELG).  \label{f:detect}}
\end{figure}

Figure~\ref{f:detect} shows the \HI\ mass histogram of the 93 SR1
targets.  In this histogram, each box represents a single \HIPASS\
source.  Each dot within a box indicates a discrete \Halpha\ emitting
galaxy as defined in Sec.\ \ref{ss:sid}.  Thus some \HIPASS\ sources
contain multiple ELGs, while all SR1 targets contain at least one ELG.
This does not mean that all \HI\ rich galaxies are also star forming.
Later (non SR1) SINGG observations have uncovered at least one \HIPASS\
galaxy that is undetected in \Halpha\ despite deep \Halpha\
observations. The present study shows that high-mass star formation is
highly correlated with the presence of \HI, and that \HI\ rich but
non-star-forming galaxies are rare.

The high detectability of \HI\ sources in \Halpha\ is remarkable.
Recently \citet{hopcat05} showed that there are no ``dark'' (optically
invisible) \HI\ galaxies among the 3692 HICAT sources with low
foreground Galactic extinction, bolstering earlier claims that starless
galaxies are rare \citep{zbss97,rw02}.  The dearth of dark \HI\ galaxies
may be due to the fact that when there is sufficient \HI\ for a gas
cloud to be self gravitating, it is gravitationally unstable until newly
formed stars and supernovae heat the ISM enough to arrest further star
formation.  Thus an \HI\ cloud that is massive enough to be
self-gravitating is likely to have already formed at least some stars,
and hence should be visible.  Star formation should set in at a lower
\HI\ mass if there is already some matter (e.g.\ dark matter) available
to bind the ISM.  Low-mass \HI\ clouds that are not self-gravitating
would have low column density and would be susceptible to ionization by
the UV background \citep{zbss97}.  \HI\ is therefore either associated
with stars or destroyed.  The theory behind this scenario is studied in
detail by \citet{tw05} who conclude that galaxies with baryonic masses
$\gtrsim 5\times 10^6$ \Msun\ should be unstable to star formation and
hence not be dark.

Our results allow a stronger statement - gas bearing dormant galaxies
are rare.  That is, if a galaxy has an ISM with $\MHI \gtrsim 3\times
10^7 \Msun$, then it almost always has {\em recently\/} (within 10 Myr)
formed high-mass stars.  The gravitational instability in the ISM is not
halted globally by feedback from evolved stellar populations.  Instead,
new stars continue to form, including the massive stars that ionize
\HII\ regions.

\subsection{Range of Properties}

\begin{figure*}
\epsscale{0.8}
\plotone{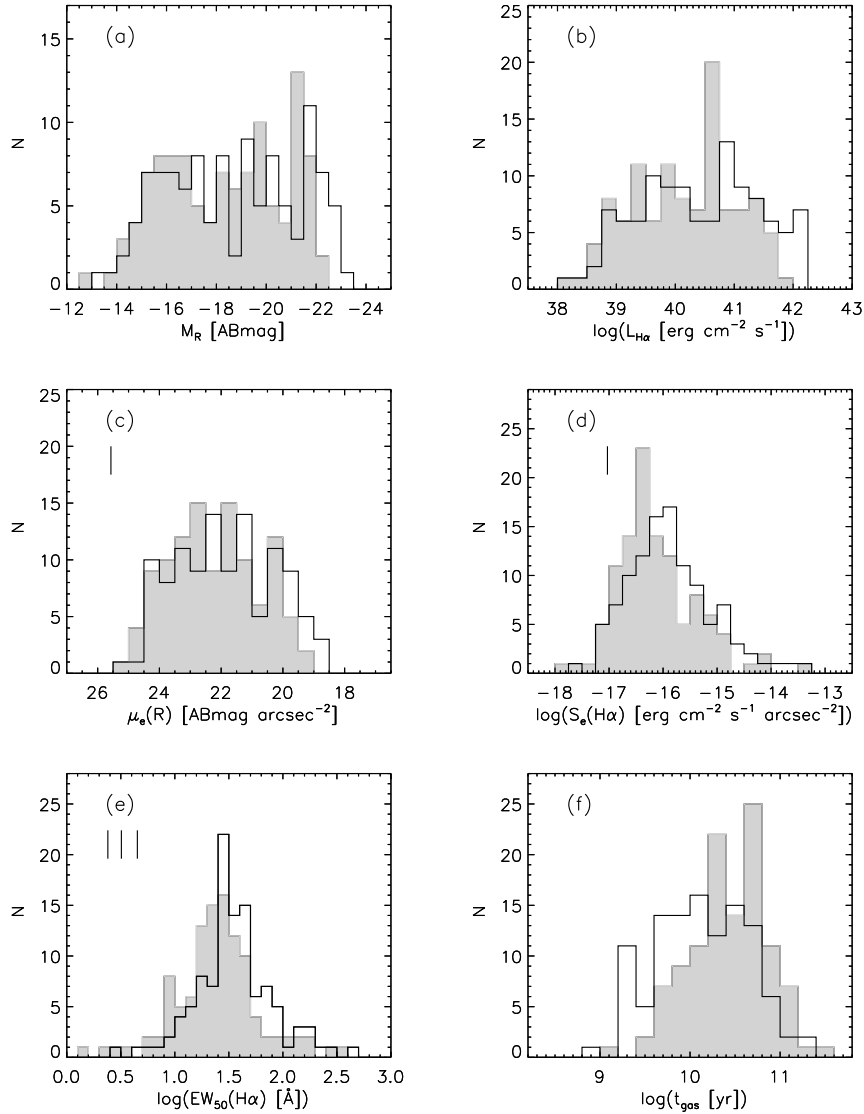}
\caption[]{Histogram of ELG observed properties.  In all panels, the
  gray shaded histogram shows the distribution of quantities with no
  internal dust extinction correction, while the black line shows the
  quantities with internal dust corrections.  Panel (a) shows the
  $R$-band absolute magnitude $M_R$ distribution.  Panel (b) shows the
  \Halpha\ luminosity \LHa\ distribution.  Panel (c) shows the $R$-band
  effective surface brightness $\mu_e(R)$ distribution.  The tick mark
  shows the average $3\sigma_S$ where $\sigma_S$ is the large scale sky
  variation.  Panel (d) shows the \Halpha\ effective surface brightness
  \Seha\ distribution. The tick mark shows the average $3\sigma_S$.
  Panel (e) shows the effective \Halpha\ equivalent width \ew\
  distribution.  The tick marks indicate the median $\sigma_{\rm EW}$
  (eq.~\ref{e:sigew}) or the NB SINGG filters, the MCELS 6568/28 filter, and
  the 75\AA\ wide NOAO filters.  Panel (f) shows the distribution of the
  gas cycling timescale \tgas.
  \label{f:range}}
\end{figure*}

The SINGG ELGs cover a wide range of properties, as shown by the
histograms in Fig.~\ref{f:range}.  These show the distribution of the
properties before (shaded histogram) and after internal dust absorption
correction (solid line).  We caution the reader that these are measured
distributions of the detected ELGs, and do not necessarily easily
transform into true volume averaged number densities.  While we do make
some comparisons with other samples, the aim is to show the diversity of
the ELGs, rather than to quantify differences with other samples.

Figure~\ref{f:range}a shows the histogram of $R$ absolute magnitudes,
which is a crude measure of the stellar content of the sources.  The
distribution is broad, covering four orders of magnitude in luminosity,
with no strong peaks.  We find ELGs ranging from $M_{R,0} = -13.1$
(corresponding to \HIPASS\ J1131$-$02:S3, a barely extended anonymous
ELG) to $M_{R,0} = -23.1$ (\HIPASS\ J2202$-$20:S1 = NGC~7184); that is
from well in the dwarf galaxy regime to nearly two magnitudes brighter
than the knee in the $R$-band luminosity function $M_*(R) = -21.5$
\citep[found from interpolating the SDSS luminosity functions of ][]{blanton03}.

The \Halpha\ luminosity, \LHa, is our basic measurement of the star
formation rate.  The \LHa\ distribution, shown in Fig.~\ref{f:range}b,
covers about four orders of magnitude in luminosity and has no strong
peaks.  It ranges from $\log(\LHa) = 38.2\,\, {\rm erg\, cm^{-2}\,
s^{-1}}$ (\HIPASS\ J0043$-$22 = IC~1574) to $\log(\LHa) = 42.25\,\, {\rm
erg\, cm^{-2}\, s^{-1}}$ (\HIPASS\ J0224-24 = NGC~922), corresponding to
a star formation rate of 0.0012 to 14 $\Msun\, {\rm yr^{-1}}$.  None of
the SR1 ELGs has a star formation rate approaching that of an
ultraluminous infrared galaxy, of $\sim 150$ \Msun\ yr$^{-1}$.  The
ionizing output of the weakest ELG corresponds to ionization by 7 O5V
stars \citep{snc02}.

The $R$-band face-on effective surface brightness, $\mu_e(R)$, gives the
integrated surface density of stars.  The distribution, shown in
Fig.~\ref{f:range}c, spans about 3.5 orders of magnitude in intensity
(surface brightness), ranging from $\mu_e(R) = 25.2\, {\rm ABmag\,
arcsec^{-2}}$ (\HIPASS\ J1106$-$14, an LSB dwarf irregular galaxy) to
18.6 ${\rm ABmag\, arcsec^{-2}}$ (\HIPASS\ J0209$-$10:S2 = NGC~838, a
starburst galaxy in a compact group).  The distribution is broad with a
sharp drop at the low surface brightness end.  The edge is near the
detection limit of our data, so may represent a bias.  If the ELGs
contain a lower surface brightness component, we would not be able to
detect it.

\begin{figure}
\plotone{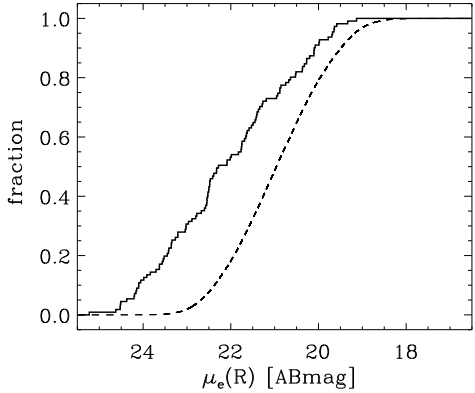}
\caption[]{Cumulative histograms of $R$-band face-on effective surface
  brightness, $\mu_e(R)$, of the SINGG ELGs (thick solid line) and the
  sample of about 28,000 SDSS spectroscopic targets cataloged by
  \citet[][thin dashed line]{blanton05}.  We derived $\mu_e(R)$
  from their cataloged quantities as described in the text.
  \label{f:sbcum}}
\end{figure}

The ELGs typically have lower surface brightness than the low redshift
galaxies targeted for spectroscopy by the Sloan Digital Sky Survey
(SDSS).  This is illustrated in Fig.~\ref{f:sbcum} which compares the
cumulative histograms in $\mu_e(R)$ for the SINGG ELGs and a sample of
$2.8\times 10^4$ low redshift SDSS galaxies cataloged by
\citet{blanton05}. The latter catalog includes SDSS spectroscopic sample
targets weeded of portions of larger galaxies that were incorrectly
identified as separate sources.  From their published catalog we
calculated $\mu_e$ in the SDSS $r'$ and and $i'$ passbands using the
Petrosian flux and half light radii.  We then interpolated these to the
effective wavelength of the Harris $R$ filter to obtain $\mu_e(R)$.
Both the SINGG and \citet{blanton05} samples have been corrected for
Galactic extinction but not internal extinction in this plot. The
inter-quartile range of the Blanton et al sample is 21.75 to 20.16 ABmag
arcsec$^{-2}$ significantly narrower and brighter than that of the SINGG
ELGs: 23.30 to 20.91 ABmag arcsec$^{-2}$.  \citet{blanton05} note that
the deficit of the lowest surface brightness galaxies ($\mu_e(r') >
23.5$ mag arcsec$^{-2}$) in their catalog is largely a result of their
software for selection of sources for spectroscopy.  \citet{kniazev04}
demonstrate that significantly lower intensity sources can indeed be
found in the SDSS images.

The \Halpha\ effective surface brightness indicates the intensity of
star formation, that is the rate of star formation per unit area.  This
is the key observable quantity to test any model where the energetic
output of star formation balances the hydro-static pressure of the disk
ISM \citep[e.g.][]{k89}.  \citet{heckman05} argues that the most
physically important distinguishing characteristic of starburst galaxies
is their very high star formation intensities.  The observed
distribution, shown in Fig.~\ref{f:range}d, spans 4.4 orders of
magnitude, ranging from $\log(\Seha) = -17.69\,\, {\rm erg\, cm^{-2}\,
s^{-1}\, arcsec^{-2}}$ to $\log(\Seha) = -13.31\,\, {\rm erg\, cm^{-2}\,
s^{-1}\, arcsec^{-2}}$.  This corresponds to a range in star formation
intensity, \SSFR, from $8\times 10^{-5}$ to 2.0 \Msun\ kpc$^{-2}$
yr$^{-1}$. The least intense detected star-formation occurs in \HIPASS\
J1106$-$14, while the most intense star formation occurs in \HIPASS\
J1339$-$31A (NGC~5253), a well known starburst dwarf galaxy \citep[or
blue compact dwarf; e.g.\ ][]{c97}.  The low surface brightness end of
the distribution corresponds to the approximate detection limit of the
data, indicating that there may be lower surface brightness emission
that we are missing.

\eweff\ indicates the star formation rate compared to the past average.
Figure~\ref{f:range}e shows that for the cases where this is defined it
ranges from 2.8\AA\ for \HIPASS\ J0514$-$61:S1 (or ESO119-G048 an SBa
galaxy)\footnote{It is possible that \HIPASS\ J0409$-$56 has a lower
\eweff, but in this case we can not accurately measure \eweff\ due to
the strength of its continuum} to 451 \AA\ (\HIPASS\ J1339$-$31A), for
the sources detected in \Halpha.  While the lowest \eweff\ measurements
are likely to be highly uncertain due to continuum subtraction, the
distribution is peaked, centered at $\eweff \approx 24$\AA\ well beyond
the detection limits of the data.  Using the models of \citet{ktc94} and
the adopted IMF this corresponds to a birthrate parameter $b \approx
0.2$, where $b$ is the ratio of current star formation to the past
average.

Figure~\ref{f:range}f plots the histogram of gas cycling time \tgas,
which we define to be:
\begin{equation}
\tgas \approx 2.3 \left(\frac{\MHI}{SFR} \right).
\end{equation}
Here the factor 2.3 corrects the \HI\ mass for helium content and the
expected mean molecular content of galaxies.  The latter was derived
from the optically-selected sample of galaxies observed by
\citet{yaklr96} which has $\langle \MHtwo/\MHI \rangle = -0.06$ with a
dispersion of 0.58 dex.  We approximate this as equal masses in
molecular and neutral components.  \tgas\ estimates how long star
formation at its present rate would take to process the observed neutral
and inferred molecular phases of the ISM.  Hence \tgas\ is an estimate
of the future potential of star formation.  \tgas\ ranges from 0.7 Gyr
(\HIPASS\ J1339$-$31A, again) to 220 Gyr (\HIPASS\ J0409$-$56), that is,
from starburst like timescales to many times the Hubble time $t_H =
13.5$ Gyr ($H_0 = 70\, {\rm km\, s^{-1}\, Mpc^{-1}}$, $\Omega_M = 0.3$,
$\Omega_\lambda = 0.7$).  Figure~\ref{f:range}f shows that the \tgas\
distribution is broad, with 41\%\ of the sample having $\tgas < t_H$.

Figure~\ref{f:range} shows that our adopted internal dust absorption
corrections have a modest impact on the observed distributions.  In
general, the dust correction spreads out the histograms.

The SINGG ELGs exhibit diverse morphologies.  They include spirals
(e.g.\ \HIPASS\ J1954$-$58 = IC~4901) and later type systems (e.g.\ 
\HIPASS\ J0459$-$26 = NGC~1744), but also residual star formation in Sa
and S0 systems (e.g.\ \HIPASS\ J0409$-$56).  Irregular galaxies are well
represented in the sample from low surface brightness dwarf irregulars
with just a few \HII\ regions (e.g.\ \HIPASS\ J0310$-$39 = ESO300-G016)
to high-surface brightness windy blue compact dwarf (e.g.\ \HIPASS\ 
J1339$-$31A).  The sample also includes interacting systems (e.g.\ 
\HIPASS\ J0209$-$10 = four members of HCG 16) and mergers (e.g.\ 
\HIPASS\ J0355$-$42 = NGC~1487).  The \Halpha\ images often enhance
structures that are relatively subtle in broad band images thus
revealing information on the dynamics of the system.  These include
small scale inner rings, large outer rings (\HIPASS\ J0403$-$43:S1 =
NGC~1512, for example, has both), bars (e.g.\ \HIPASS\ J0430$-$01 =
UGC~3070) and spiral arms (e.g.\ \HIPASS\ J0512$-$39 = UGCA106).  In other
cases the structures that are apparent in the $R$-band are less obvious
in \Halpha\ (e.g.\ \HIPASS\ J2334$-$36 = IC~5332 shows a grand-design
spiral structure in $R$ and an apparent random \HII\ region distribution
in \Halpha).

While most of our images reveal only a single ELG, multiple ELGs were
found in 13 pointings.  In the most extreme case, \HIPASS\ J0209$-$10
(Hickson Compact Group 16) four ELGs were detected in a single frame.
Thus the total number of ELGs in SR1 is 111, significantly larger than the
number of fields observe.  While in some cases the companions would have
been recognized immediately at any optical wavelength (e.g.\ the two
large spirals in \HIPASS\ J2149$-$60), in many cases the companion is
compact and has low luminosity, and hence could easily be mistaken for
background sources (e.g.\ \HIPASS\ J0342$-$13:S2, and the dwarf member,
S3, of the \HIPASS\ J2149$-$60 system).  This result demonstrates the
value of \Halpha\ imaging for identifying interacting companions with an
unobtrusive appearance.  Comments on the morphologies of all multiple
ELGs can be found in Appendix~\ref{s:objnotes}.

\begin{figure}
\plotone{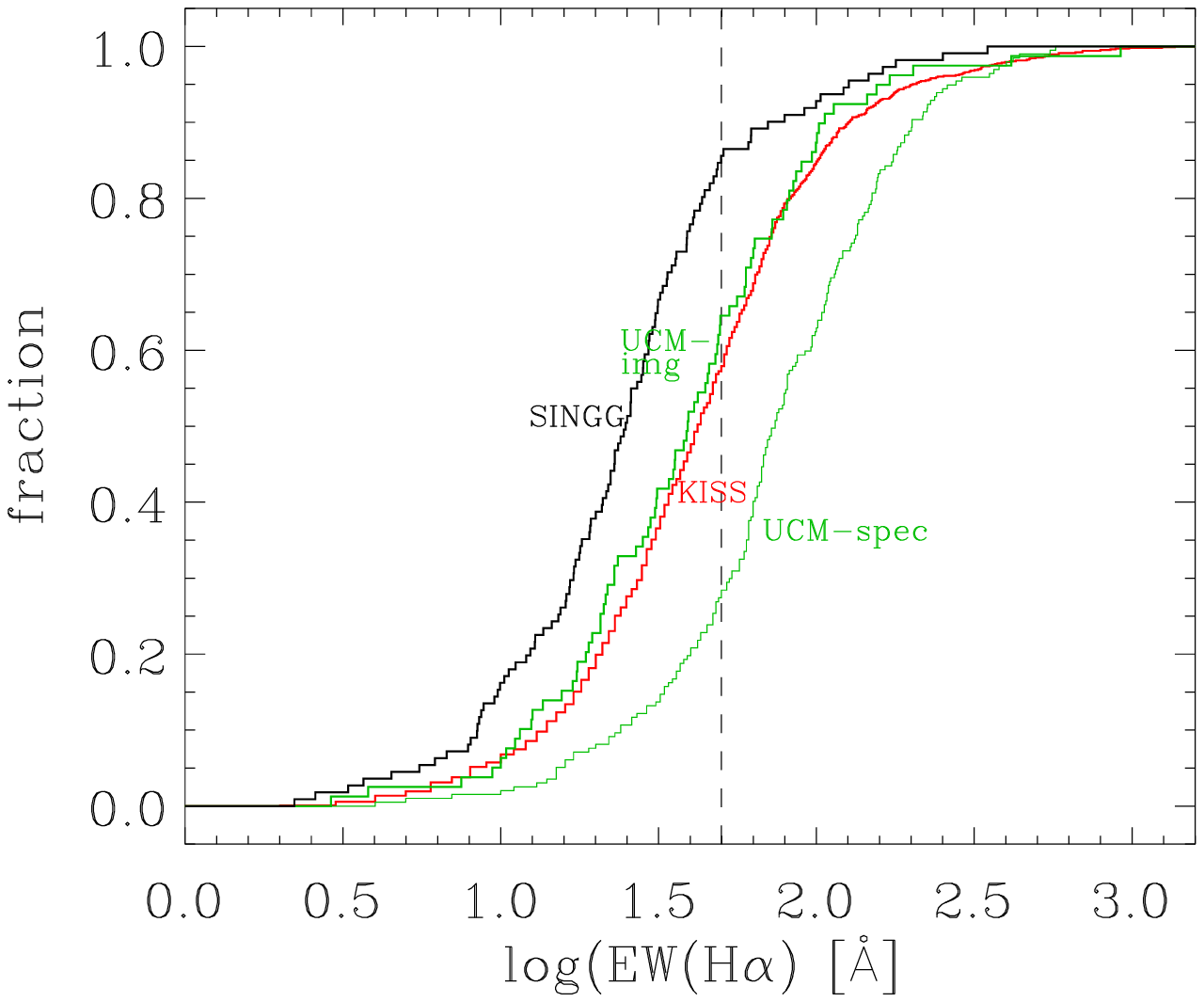}
\caption[]{Cumulative histograms of \ewha\ for the SINGG SR1 ELGs (black
  line), and those from the KISS (red line) and UCM prism (green lines)
  surveys.  For the SINGG galaxies, we plot \eweff.  For the KISS survey
  we plot the prism \ewha\ values from \citet{kiss4} while the UCM
  sample results for spectroscopically confirmed ELGs are shown from
  both spectroscopic data \citep[UCM-spec,][]{gzrav96} as well as NB
  imaging data \citep[UCM-img,][]{pzgag03}.  The dashed line marks the
  ``traditional'' starburst definition cut at $\ewha \geq 50$\AA.
  \label{f:ewcum}}
\end{figure}

The wide range of star formation properties observed in our sample
supports our contention that the SINGG survey is not strongly biased
toward any particular type of star forming galaxy.  This is not
generally the case in star formation surveys.  This is illustrated in
Fig.~\ref{f:ewcum} which shows the cumulative histogram of $EW(\Halpha)$
for SINGG compared to two prism based emission line surveys: KISS
\citep{kiss4} and UCM \citep{gzrav96,pzgag03}.  The prism-selected
sources are weighted considerably more to high \ewha\ systems.  This
difference can not be attributed totally to how the \ewha\ measurements
are made.  \ewha\ measurements for the UCM survey were made from
long-slit spectroscopic data \citep{gzrav96} as well as NB \Halpha\ 
imaging \citep{pzgag03}.  The latter study was done to recover the
``total'' \Halpha\ flux including that beyond the spectroscopic slit used
by \citep{gzrav96}.  As shown in
Fig.~\ref{f:ewcum}, the \ewha\ distribution in both UCM cases is skewed
toward higher values than the SINGG sample.  Taking the ``traditional''
definition of starbursts as having $\ewha \geq 50$\AA, then 14\%, 42\%,
35\%, and 72\%\ of the SINGG, KISS, UCM imaging, and UCM spectroscopic
surveys, respectively, are starbursts.  Rather than SINGG missing out on
starbursts, it is more likely that the prism surveys are missing low
\ewha\ systems.  Because we are not dealing with volume averaged
densities in this comparison, it is premature to say how these
differences translate into the relative biases of the surveys.  That
issue will be addressed further in paper II \citep{singg2}.

\section{Conclusions}\label{s:conc}

The Survey for Ionization in Neutral Gas Galaxies (SINGG), is providing
a view of star formation in the local universe that is not hampered by
the strong stellar luminosity based selection biases found in many other
surveys.  Our first results are based on observations of 93 of the total
468 \HIPASS\ targets.  These observations comprise the first release of
SINGG data: SR1.  All of these 93 targets contain \Halpha\ Emission Line
Galaxies (ELGs).  The high detectability of star formation in \HI\ rich
galaxies confirms that \HI\ is an important indicator of the presence of
star formation.  The detected galaxies cover a wide range of
morphologies, including LSB spirals and irregulars, normal spirals,
strong starburst activity with minor axis wind features, and residual
star formation in early type disk systems.  The ELGs we find have a
$\mu_e(R)$ distribution extending to fainter intensities typically
targeted for SDSS spectroscopy, while the \eweff\ distribution appears
to be less biased toward starbursts than are prism surveys.

Multiple ELGs were found in 13 systems bringing the total number of ELGs
imaged to 111.  In many cases, the relationship between the companion
and the primary source was not obvious from previous optical images.
This illustrates how \Halpha\ follow-up imaging is a valuable tool for
identifying star forming companions to \HI-selected galaxies.

This introduction to SINGG shows the potential for using a homogeneous
\HI-selected sample to explore star formation in the local universe.
Other papers in this series will discuss the contribution of \HI\
galaxies to the local cosmic star formation rate density \citep[][paper
II]{singg2}; the correlations between the global star formation
properties of galaxies (Meurer et al. 2006); the \HII\ region luminosity
function and demographics of the diffuse ionized gas (DIG; Oey et al.\
in preparation); and the compact emission line sources projected far from their
apparent hosts (the ELDots, J.\ Werk et al.\ in preparation).  The SR1
data, both images and a database, are made available at {\sf
http://sungg.pha.jhu.edu/\/} for the benefit of other researchers and
the public and as part of our commitment to the NOAO Surveys Program.

\acknowledgments

Partial financial support for the work presented here was obtained
through grants from NASA including HST-GO-08201, HST-GO-08113, NAG5-8279
(ADP program), and NAG5-13083 (LTSA program) to G.R.\ Meurer; a Space
Telescope Science Institute Director's Discretionary Research Fund grant
to H.C.\ Ferguson, NSF 0307386 and NASA NAG5-8426 to R.C.\ Kennicutt; an
Australia Research Council Discovery Project - ARCD DP 0208618; NSF
grant AST-0448893 to M.S.\ Oey; and NNG04GE47G (LTSA) to T.\ Heckman.
The observations were made possible by a generous allocation of time
from the Survey Program of the National Optical Astronomy Observatory
(NOAO), which is operated by the Association of Universities for
Research in Astronomy (AURA), Inc., under a cooperative agreement with
the National Science Foundation.  NOAO also supported this work with
assistance in the purchase of two filters, support in scanning the
filters at Tucson, as well as in excellent visitor support and
hospitality at CTIO, La Serena and Tucson. At CTIO we received the
expert assistance at the 1.5 m telescope from its operators: Alberto
Alvarez, Mauricio Fern\'{a}ndez, Roger Leighton, Hernan Tirado, and
Patricio Ugarte, and especially for the observer support work of Edgardo
Cosgrove and Arturo G\'{o}mez.  Our smooth operations are largely due to
their skill and experience.  We thank the NOAO Tucson filter experts -
James DeVeny, Daryl Willmarth, Bill Blinkert, Bruice Bohannan and Heidi
Yarborough, for scanning the filters and carefully performing other
acceptance tests.  We thank Arna Karick and Shay Holmes who assisted in
obtaining the observations presented here.  We are very thankful for
data reduction assistance from Cheryl Pavlovsky (STScI) and Sanae
Akiyama (U.\ Arizona) for performing initial CCD data reductions.  The
High-$z$ Supernovae group assisted our efforts by providing access to
their data reduction scripts.  GRM is grateful for hospitality received
during visits to the Australia Telescope National Facility Headquarters,
the University of Melbourne, The University of Queensland, and the
Australian National University's Mount Stromlo Observatory, were parts
of this work were accomplished.  We are thankful for useful
conversations and correspondence with John Blakeslee, Holland Ford, Joe
Helmboldt, Christy Tremonti, Linda Smith, Michael Blanton, and John
Moustakas.  We are grateful to the anonymous referee for comments and
suggestions that have improved this paper.  We thank Capella Meurer for
suggestion of the SINGG acronym.  This research has made use of the
NASA/IPAC Extragalactic Database (NED) which is operated by the Jet
Propulsion Laboratory, California Institute of Technology, under
contract with the National Aeronautics and Space Administration.

{\it Facilities:} \facility{CTIO:1.5m}

\begin{appendix}

\section{Flux calibration relations}\label{s:fluxap}

Here we present the formalism for converting observed count rates to
calibrated magnitudes and integrated \Halpha\ line fluxes \FHa. These
relationships are easily derived using the principles of synthetic
photometry \citep{synphot}.  We denote the count rate as $C_F(X)$ where
the $F$ subscript, used throughout this section, denotes the filter
dependence. The airmass, $X$, dependence of the calibration is derived
from the standard airmass equation:
\begin{equation}
m_{\rm true,F} - m_{\rm obs,F} = A X + B_F \label{e:ext}
\end{equation}
where $m_{\rm true,F}$ is the true magnitude above the atmosphere and
$m_{\rm obs,F} = -2.5 \log(C_F(X))$ is the observed magnitude.  Because
the filters used in this study all have similar central wavelengths, we
simultaneous fit a single extinction term $A$ (in units of mag
airmass$^{-1}$) for all filters and individual zeropoints $B_F$ for each
filter.  Typically a single night's worth of standard star observations
were used in each fitting, although in periods of fine and stable
weather we have been able to combine the data from several nights in a
single fit.

Calibration is to spectro-photometric standards, and we use flux
calibrated spectra of these stars to derive the true magnitude of the
stars through the relevant filters.  There are a variety of ways to
define the true magnitude from a flux calibrated spectrum $f_\lambda$.
For deriving the formulas here, the STmag system is most convenient
\begin{equation}
m_{\rm ST} = -2.5 \log\langle f_\lambda \rangle - 21.1
\end{equation}
where $\langle f_\lambda \rangle$ is the bandpass averaged flux density 
(defined in eq.~\ref{e:mflam} below) and is
in units of erg cm$^{-2}$ s$^{-1}$ \AA$^{-1}$.  The magnitudes we quote
here are in the more familiar ABmag system which is related to the
STmag system by
\begin{equation}
m_{\rm AB} = m_{\rm ST} + 5 \log\left(\frac{5500}{\lambda_p}\right)
\end{equation}
and $\lambda_p$ is the pivot wavelength in \AA\ of the filter given by
\begin{equation}
\lambda_p = \sqrt{\frac{\int \lambda\, T_F(\lambda)\, d\lambda}{\int \lambda^{-1}\, T_F(\lambda)\, d\lambda}}.
\end{equation}

Denoting the total system throughput as a function of wavelength
$T_F(\lambda)$ then the mean flux density in the band is
\begin{equation}
\langle f_\lambda \rangle = \frac{\int \lambda\, T_F(\lambda)\, f_\lambda\,
  d\lambda}{\lambda_{m,F}\,W_{E,F}}.\label{e:mflam}
\end{equation}
Here $f_\lambda$ is the spectrum of the source in erg cm$^{-2}$
s$^{-1}$ \AA$^{-1}$ and $\lambda_{m,F}$ and $W_{E,F}$ are the response
weighted mean wavelength, and equivalent width of the passband given
by 
\begin{equation}
\lambda_{m,F} = \frac{\int{\lambda\, T_F(\lambda)\, d\lambda}}{\int{T_F(\lambda)\, d\lambda}}
\end{equation}
and 
\begin{equation}
W_{E,F} = \int{T_F(\lambda)\, d\lambda}.
\end{equation}
Ideally $T_F(\lambda)$ should be the product of the CCD response, the
throughput of all the optical elements (filters, primary and secondary
mirrors), as well as the atmospheric transmission as a function of
wavelength and airmass.  For our purposes the standard extinction
equation~\ref{e:ext} is sufficient to remove the atmospheric response.
Since there is very little wavelength variation in the mirror coatings,
we take $T_F(\lambda)$ to be the product of the filter and CCD
responses.  The unit response of a given observation is given by
\begin{equation}
U_{F}(X) = \frac{\langle f_\lambda \rangle}{C_F(X)}.
\end{equation}
The airmass dependence is given by 
\begin{equation}
\log\left(\frac{U_F(0)}{U_F(X)}\right) = - 0.4 A X.
\end{equation}
where the unit response above the atmosphere is given by 
\begin{equation}
\log U_F(0) = -0.4(21.1 + B_{F,ST})
\end{equation}
and $B_{F,ST}$ is the zeropoint from eq.~\ref{e:ext} in the STmag
system.  

The unit response to line emission is defined to be 
\begin{equation}
U_{\rm F,line}(X) = \frac{C_{\rm F,line}(X)}{F_{\rm line}}
\end{equation}
where $F_{\rm line}$ is the integrated line flux, in erg cm$^{-2}$ s$^{-1}$
and $C_{\rm F,line}(X)$ is the count rate after continuum subtraction.  
$U_{\rm F,line}(X)$ is given by
\begin{equation}
U_{\rm F,line}(X) = U_{F}(X)\, \lambda_{m,F}\,W_{E,F} \, \frac{\int f_{\rm \lambda,line}\, d\lambda}{\int \lambda\, f_{\rm \lambda,line}\, T_F(\lambda) d\lambda}
\label{e:uline}
\end{equation}
where $f_{\rm \lambda, line}$ is the emission line spectrum. For a
single line this is the line profile, for multiple lines in the filter
bandpass this is the summed profiles of all the lines.  We experimented
with various models for the line profile including $\delta$
function, Gaussian, and square function line profiles.  Our adopted model is a
Gaussian having the same \vhel\ and \fwhm\ as the integrated \HI\
profile:
\begin{equation}
G(\lambda_0,\vhel,\fwhm, \lambda) = e^{-0.5x^2/\sigma^2}
\end{equation}
where the peak amplitude is 1.0, $\lambda_0$ is the rest wavelength of
the line and $x$ and $\sigma$ are given by the usual relationships $x =
c(\lambda - \lambda_0)/\lambda_0 - V_{\rm hel}$ and $\fwhm =
\sqrt{8\ln(2)}\sigma$, where $\sigma$ is the Gaussian dispersion of the
line.  This model is meant to give a first approximation to the
integrated \Halpha\ velocity profile.  While we do not know the \Halpha\ 
velocity profile of the targets, we do know their \HI\ profiles which
are often Gaussian in shape in dwarf galaxies to double horn profiles
for large spirals.  As long as the profiles avoid the steep edges of the
bandpass, we find that profile shape does not make a significant
difference to the value calculated for $U_{\rm F,line}$.  Square
profiles give $U_{\rm F,line}$ values that are very similar to the
Gaussians of the same \fwhm, as do $\delta$ functions centered at \vhel.
We did not test double horn profiles mainly because of the difficulty in
modeling them.  In addition, generally we do not expect the \Halpha\ 
profiles to have as much power at high relative velocities as do double
horn profiles for two reasons.  First, the horns results from the nearly
flat rotation curves of most disk galaxies at large radii, often
extending significantly further than the \Halpha\ distribution
\citep{k89,mk01}.  Second, the dip between the horns need not indicate
the lack of ISM at systemic velocity but rather may indicate the ISM at
the galaxy's center is not primarily neutral.

The filters used in this study are not sufficiently narrow to exclude
the [{\sc Nii}] lines at rest $\lambda = 6548.05$\AA\ and 6583.45\AA.  Quantum mechanics
sets the flux ratio of these two lines to $F_{6548}/F_{6583} = 0.338$.
Calling $w_{6583} = F_{\rm [NII]6583}/F_{\rm H\alpha}$ then 
then the fraction of the total line count rate due to \Halpha\ is 
\begin{equation}
\frac{C_{F,\Halpha}(X)}{C_{\rm F,line}(X)} = \frac{1}{1 + w_{6583}K_{\rm [NII]}}
\label{e:hafrac}
\end{equation}
where 
\begin{equation}
K_{\rm [NII]} = \frac{1.0031\int \lambda G(6583,\vhel,\fwhm, \lambda)
  T_F(\lambda) d\lambda + 0.337\int \lambda G(6548,\vhel,\fwhm, \lambda)
  T_F(\lambda) d\lambda}{\int \lambda G(6563,\vhel,\fwhm, \lambda)
  T_F(\lambda) d\lambda}\, . \label{e:knii}
\end{equation}
For a given $w_{6583}$, then it is a matter of using eq.~\ref{e:knii}
and \ref{e:hafrac} to determine the count rate from \Halpha\ alone,
and then using a $f_{\rm \lambda,line} = G(6563,\vhel,\fwhm, \lambda)$
in eq.~\ref{e:uline} to get the unit response to \Halpha\ line emission.
We estimate $w_{6583}$ from the $R$-band absolute magnitude of the line
using the empirical relation of \citet{hwbod04} and given in our
eq.~\ref{e:helmboldt_nii}. 

\section{Notes on individual \HIPASS\ targets}\label{s:objnotes}

Here we present notes on individual \HIPASS\ targets.  We concentrate on
two classes of targets: (1) cases where the measurements were difficult
to perform; and (2) ``interesting'' targets including all those with
multiple ELGs, cases where strong outflows are seen, resolved galaxies
(near enough to break up into stars), and objects with peculiar or
striking morphological features such as rings, or a dominant bulge or
nucleus.  The sources are listed by their \HIPASS/SINGG designation
(with NGC, IC, UGC, or ESO designations in parenthesis).  In the
descriptions we use the following abbreviations: AGN - active galactic
nucleus; ELG - emission line galaxy; HSB - high surface
brightness; LSB - low surface brightness; MSB - moderate surface
brightness; DIG - diffuse ionized gas; BCD - blue compact dwarf Sy -
Seyfert; and the cardinal directions N,S,E,W.  \pino

{\bf\bhipass~J0005$-$28} (ESO409-IG015): HSB BCD with a detached \HII\ 
region located at $r = 61''$ to NW along the optical major axis.

{\bf\bhipass\ J0019$-$22}: A possible polar ring (otherwise a faint outer
disk ring) of faint \HII\ regions encloses a somewhat off-center
elliptical core, featureless in $R$, but containing a central compact HSB
\HII\ complex.

{\bf\bhipass\ J0039$-$14A} (NGC~178): This galaxy has a very peculiar
morphology, suggestive of a merger.  In the $R$-band, the galaxy is
predominantly aligned NS, with two tails extending S.  The HSB core
is double, with components separated by 9.5\as, with the S component
being considerably brighter in \Halpha.  \HII\ arms to W and E are
suggestive of a polar ring, while minor axis fans of extra-planar DIG to
north of central components have no obvious power sources.  Faint
detached \HII\ regions exist to NW of galaxy.

{\bf\bhipass\ J0135$-$41} (NGC~625): A well known amorphous / BCD galaxy
\citep{sb79,mmhs97}.  In net \Halpha\ we see a HSB core, containing a
few knots as well as LSB extra-planar features including a nearly
complete loop, rising 82\as\ from the major axis or the N side.  This
feature was not seen in the images of \citet{mmhs97} but is consistent
with the \HI\ kinematics \citep{cmsc04}.

{\bf\bhipass\ J0145$-$43} (ESO245-G005): A resolved LSB IBm/SBm containing
bubbly \HII\ regions especially at the bar ends.  This galaxy was
imaged in \HI\ by \citet{ccf00} while \citet{miller96} present 
NB imaging in \Halpha\ and \fion{O}{III}.

{\bf\bhipass\ J0209$-$10}: There are four ELGs in the field (the most in
SR1) - the four bright members of Hickson Compact Group 16. Earlier
\Halpha\ images were presented by \citet{vi98}, while spectra were
presented by \citet{rcccz96} and \citet{dc99} who found a high incidence
of AGN characteristics.  All four galaxies have prominent nuclear HSB
\Halpha, and at least three have a minor axis outflow.  S1 (NGC~839) is
an inclined disk galaxy with a LINER $+$ Sy2 nuclear spectrum which is
prominent in \Halpha, while DIG in a minor axis extends out to $r =
31''$.  S2 (NGC~838) is a moderately inclined disk with a lumpy nuclear
region having a starburst spectrum.  Its compact nucleus is surrounded
by a HSB \Halpha\ bright ring with a diffuse wind emanating out the
minor axis to 53\as.  S3 (NGC~835) and S4 (NGC~833) are closely
interacting.  S3 is nearly face-on with a double ring morphology ($r
\approx 10'', 43''$) in \Halpha\ and a LINER $+$ starburst nuclear
spectrum.  In the $R$-band a tidal arm extends to the E.  S4 is a
lopsided moderately inclined barred galaxy.  Its nucleus has an Sy2 $+$
LINER spectrum and is embedded to one side of the bar.  DIG extends out
from the disk at an angle intermediate between major and minor axes,
merging with the DIG from S3.

{\bf\bhipass\ J0216$-$11C} (NGC~873): A sharp edged HSB spiral well
covered in \HII\ regions, somewhat more extended in the $R$-band than
\Halpha. Its nucleus is off-center compared to the outer isophotes.

{\bf\bhipass\ J0221$-$05}: S1 (NGC~895) is an SBc having a weak \Halpha\ 
emitting nucleus, two tight \HII\ region rich arms emerge from the bar
with two flocculent armlets between the primary arms. At $r \approx
115''$ the arms merge to form an outer pseudo-ring off-center toward E
compared to the nucleus.  The displacement may result from interaction
with S2 \citep[NGC~895A as assigned by][]{zsfw93}, a very compact nearly
circular BCD with LSB \Halpha\ extensions along its minor axis.

{\bf\bhipass\ J0224$-$24}: This collisional ring galaxy system
\citep{lt76,hw93} is similar to the Cartwheel galaxy \citep[see
e.g.][]{fh77,higdon95}.  The primary S1 (NGC~922) has a ``C'' shaped
morphology and is the ELG with the highest \Halpha\ luminosity ($\LHa =
1.8 \times 10^{42}\,\, {\rm erg\, s^{-1}}$) in SR1, while we identify
the interloper S2 as the compact \Halpha\ bright galaxy 8.36\am\ away to
the WNW.  \citet{ngc922} present a more detailed analysis of this system
using a variety of observations including the SINGG SR1 data.

{\bf\bhipass\ J0256$-$54} (ESO154-G023): A nearly resolved LSB Sd or Sm whose
brightest \HII\ region, (located NE of center) has bipolar bubbles
aligned near the minor axis with an overall extent of 64\as.

{\bf\bhipass\ J0317$-$22} (ESO481-G017): Face-on spiral with \Halpha\ 
concentrated in a parallelogram shaped ring 28\as\ in diameter, with a
weak sparse \HII\ region distribution beyond.

{\bf\bhipass\ J0317$-$41} (NGC~1291): This large face-on SB0 has a
distribution of faint \HII\ regions that traces tightly wound arms or pseudo
ring at $r = 4.7'$.  A swirling pattern of DIG, dominates
the central \Halpha\ morphology.  This structure is devoid of \HII\
regions and increases in surface brightness toward the nuclear region,
reminiscent of the DIG in the bulge of M31
\citep{ciardullo88}.  The measurements of this galaxy are difficult
because the nucleus is saturated in the $R$-band (hence the nucleus is
masked from our image) and the galaxy's sparse population of \HII\
regions extends to the edge of the frame or beyond.

{\bf\bhipass\ J0320$-$52} (NGC~1311): Edge-on Sd or Sm with \HII\ knots
along the major axis and DIG streamers along the minor axis.

{\bf\bhipass\ J0342$-$13}: This system is dominated by S1 (NGC~1421), a
highly inclined SBb galaxy whose brightest two arms, bar and nucleus are
well covered with HSB \HII\ regions.  Its dwarf companion S2 could
easily be mistaken for a background galaxy, except for its two HSB
nuclear \HII\ regions.

{\bf\bhipass\ J0355$-$42} (NGC~1487): A well known merger \citep{vv59}.
Our $R$ image shows two nuclei separated by 10.3\as\ on a nearly NS
line, with a third similar hotspot (or nucleus) located 26\as\ to
the E of the northern, while two long tidal tails stretch out to
$\sim 75$\as.  The \Halpha\ image is dominated by the HSB core
surrounding the three hotspots, while fainter \HII\ regions can be seen
all along the tidal arms.

{\bf\bhipass\ J0359$-$45}:  S1 (The Horologium Dwarf) is a face-on LSB
dwarf with sparse population of dozens of faint \HII\ regions.  S2
(ESO249-G035) is an edge-on LSB to MSB disk galaxy. 

{\bf\bhipass\ J0403$-$01}: A very difficult galaxy to measure due to the
supposition of a bright (saturated) star and pervading Galactic cirrus
emission in \Halpha.  This galaxy is discussed in Sec.~\ref{ss:meas}
while Fig.~\ref{f:spcases}a shows an expanded image of the galaxy.

{\bf\bhipass\ J0403$-$43}: This is the well known starburst pair
NGC~1512/NGC~1510 (S1/S2), strongly interacting galaxies sharing a
common \HI\ envelope \citep{hvmgp79}.  S1 is a large SBb with a central
starburst ring surrounding its nucleus and embedded in a bar that is
otherwise devoid of \Halpha.  An outer ring, well populated with \HII\ 
regions, circles the bar, while a weak distribution of \HII\ regions
extends to the edges of the frame.  S2 is an amorphous/BCD galaxy that
dominates over S1 in total \Halpha\ flux.  Its \Halpha\ morphology is
strongly concentrated toward two central knots, separated by 4\as, with
radial filaments including an apparent jet toward the SW.

{\bf\bhipass\ J0409$-$56} (NGC~1533): This is an HSB face-on SB0, having bar
length of 40\as\ with a few faint \HII\ regions over the optical face,
as well as the more distant ELDots discussed by \citet{rw04}. This is
the second difficult to measure galaxy discussed in Sec.~\ref{ss:meas}
and displayed in Fig.~\ref{f:spcases}b.

{\bf\bhipass\ J0430$-$01} (UGC3070): Sm galaxy with parallelogram outer ring
having dimensions 36\as$\times$64\as\ enclosing a central bar.

{\bf\bhipass\ J0441$-$02} (NGC~1637): Three armed asymmetric spiral having
a sharp change in the \HII\ regions properties.  Interior to $r \approx
1.5\am$ there is a modest density of bright \HII\ regions, exterior to
this there is a sparse distribution of faint \HII\ regions out to $r\sim
3.6$\am.

{\bf\bhipass\ J0454$-$53} (NGC~1705): An amorphous / BCD galaxy well
known for its off-center super star cluster and strong galactic wind.
Our \Halpha\ image shows minor axis arcs, not seen by \citet{mfdc92},
which reach out to 90\as\ to the S, and 94\as\ to the NNW.

{\bf\bhipass\ J0456$-$42} (ESO252-IG001): Sm/Im galaxy superimposed on
an edge-on background galaxy. Contains a curious near linear \Halpha\
arc through center along minor axis.

{\bf\bhipass\ J0503$-$63}: S1 (ESO085$-$G034) is an inclined Sa with
faint tightly wound spiral arms more apparent in \Halpha. Its \HII\  
distribution is brighter on the E side toward S2, its compact dwarf
companion, which has two bright \HII\ regions, and has not been
previously cataloged (according to NED).

{\bf\bhipass\ J0504$-$16}: This system contains two LSB galaxies.  S1 is
a large face-on SBcd with \HII\ regions over its face and two long outer
arms.  The longest stretches SW toward S2, a small LSB galaxy, not
previously cataloged (according to NED), with a few faint \HII\ regions
on its SE side.

{\bf\bhipass\ J0506$-$31} (NGC~1800): Another well studied amorphous /
BCD with extra-planar \Halpha\ and a HSB core region
\citep[e.g.][]{hvwg94,mmhs97}.  Most of the star formation and
extra-planar \Halpha\ is located in the E half of the galaxy.

{\bf\bhipass\ J0507$-$37} (NGC~1808): A well studied starburst with an
embedded Sy2 nucleus \citep{vv85}.  The starburst corresponds to the
lumpy core $\sim 22\as$ across with intense \Halpha\ emission, embedded
in an oval shaped bar 270\as\ long.  \HII\ regions trace a somewhat smaller
and tighter integral symbol shaped bar 192\as\ long.  Spiral arms emerge
from the larger bar to form a nearly complete figure-eight pseudo-ring
containing a few faint \HII\ regions.  There are numerous dust lanes in
the central region, and an edge-darkened spray of diffuse dust
obscuration emanating from the core toward the NE projecting out to at
least 40\as, previously noted by \citet{gw74} \citep[cf.][]{vv85}, is
indicative of an edge-darkened dust entrained outflow.

{\bf\bhipass\ J0514$-$61}: There are three ELGs in this system.  S1
(ESO119-G048) is an SBa, which has a 143\as\ long oval bar resembling a
strongly nucleated HSB elliptical galaxy.  Two weak arms start at the
bar's ends to form a faint pseudo-ring.  Sparsely distributed \HII\
regions populate the region from the bar minor-axis to the faint outer
arms.  S2 (ESO119-G044) is a face-on Sbc having a fairly random
distribution of \HII\ regions covers its optical face.  S3 is a compact
HSB SBab with strong line emission along the bar and much of the tight
spiral arms that emerge from it.

{\bf\bhipass\ J1054$-$18}: S1 (ESO569-G020) is a moderate to LSB spiral
containing a small bar, invisible in \Halpha, and flocculent \HII\ 
region rich arms.  S2 (ESO569-G021) is a small disk galaxy with a compact
nucleus, and an HSB \Halpha\ ring (12$''$ diameter).

{\bf\bhipass\ J1105$-$00} (NGC~3521): This moderately inclined bulge
dominated Sb galaxy has a HSB nucleus that is saturated in $R$, and
masked out in our net \Halpha\ image.  Hence our \FHa\ and \Seha\ 
measurements are underestimated.  However examination of the NB images
suggests that nucleus does not significantly contribute to the total
\FHa.  There is an apparent \Halpha\ ring at $r \approx 25''$, while the
disk beyond that to $r=116''$ is well covered by \HII\ regions and DIG.

{\bf\bhipass\ J1109$-$23} (IC~2627): This is a face-on grand-design Sc galaxy
that is somewhat lopsided on large scales.  At its heart is a very
compact HSB ring of \HII\ emission at $r = 1.8''$ surrounding the
nucleus.  The two arms are well delineated in bright \HII\ regions. 

{\bf\bhipass\ J1131$-$02}: S1 (UGC06510) is a face-on SBc with
flocculent \Halpha\ rich arms and a small 16\as\ long bar containing a
strong nucleus in $R$.  S2 is an edge on disk with strong line emission
along much of its length.  S3 is a small source, not previously
cataloged (according to NED), located between S1 and S2 that is similar
to an ELDot except that it is clearly two faint connected line emitting
knots separated by 3.5\as.

{\bf\bhipass\ J1303$-$17c} (UGCA320,DDO161): This is a partly resolved,
edge-on low surface brightness disk, with nearly rectangular bulge or
bar 1\am\ across containing numerous clusters or knots and a well
defined nucleus.  \HII\ regions at the edge of the bulge have DIG
extending out the minor axis especially on the N side.  This source was
imaged in \HI\ by \citet{ccf00}.

{\bf\bhipass\ J1337$-$29} (NGC~5236): The well studied Messier 83 is a
large face-on SBb.  The thick bar is 199\as\ long and contains numerous
dust lanes.  The bar dust lanes terminate in a central, knotty very HSB
region (in both $R$-band and \Halpha) 14\as\ across - the central
starburst.  Numerous bright \HII\ regions have a high covering factor,
especially along the two arms, out to $R \sim 290''$.  There the
\Halpha\ distribution is largely truncated, as pointed out by
\citet{k89} and \citet{mk01}, while the UV light profile shows no
truncation \citep{thilker05}.  However, a few fainter \HII\ regions can
be seen out to the edge of our frame.

{\bf\bhipass\ J1339$-$31A} (NGC~5253): Like the other amorphous / BCD
galaxies, we see smooth elliptical outer isophotes and a knotty core
which has been imaged extensively by HST \citep[e.g.][]{c97}.  This
source has the most extreme star formation properties, in terms of
\SSFR\ and \eweff\ of any of the SR1 galaxies.  At large radii
the \Halpha\ morphology is bubbly along the minor axis.  A
well known dust lane darkens the SE minor axis.

{\bf\bhipass\ J2149$-$60}: A spectacular system consisting of a binary
spiral pair with a compact dwarf in between.  S1 (NGC~7125) is a
moderately inclined Sb with an \Halpha\ bright inner ring, a nucleus
devoid of \Halpha\ and thin MSB outer arms.  S2 (NGC~7126) is a low
inclination SBbc with a small \Halpha\ bright bar, two main arms and
many armlets all rich in \HII\ regions.  S3, located between them
is a small almost featureless LSB galaxy containing one \HII\ region and
some DIG.  All three sources correspond to \HI\ detections in the VLA
map of \citet{ncst97}.  A fourth \HI\ source identified by them (their
145G17B) is not apparent in \Halpha.

{\bf\bhipass\ J2202$-$20}: S1 (NGC~7184) is a dusty inclined SBbc with an
inner ring enclosing a foreshortened bar which contains a compact
\Halpha\ bright nucleus.  Two symmetric arms, well traced by \HII\ 
regions emerge from the bar, become flocculent in their \HII\ 
distribution, and regain distinction at the outermost radii.  S2 is a
small featureless edge-on disk with MSB line emission along its length.

{\bf\bhipass\ J2334$-$36} (IC~5332): This is a large angular extent
face-on Sc galaxies with two arm morphology in $R$ all the way to the
compact bulge, but a flocculent distribution of bubbly \HII\ regions.

{\bf\bhipass\ J2343$-$31} (UGCA442): A partly-resolved edge-on LSB
galaxy showing several \HII\ regions along the major axis having 
loop morphologies.  This galaxy was imaged in \HI\ by \citet{ccf00} and
HST WFPC2 by \citet{k03} and \citet{mould05}.

{\bf\bhipass\ J2352$-$52} (ESO149-G003): This is an edge-on LSB to MSB
disk, flared at large radii, having minor axis \Halpha\ filaments
emanating from the central region despite the lack of a central HSB
core.

\end{appendix}



\clearpage

\begin{deluxetable}{l l l c c c c c c}
  \tablecaption{Source identification and measurement 
   apertures for SR1\label{t:ap}}
  \tabletypesize{\small}
  \tablewidth{0pt}
  \tablehead{\colhead{HIPASS$+$} &
             \colhead{Optical ID} &
             \colhead{Morph} &
             \colhead{RA} &
             \colhead{Dec} &
             \colhead{Filters} &
             \colhead{\rmax} &
             \colhead{$a/b$} &
             \colhead{$\theta$} \\
             \colhead{(1)} &
             \colhead{(2)} &
             \colhead{(3)} &
             \colhead{(4)} &
             \colhead{(5)} &
             \colhead{(6)} &
             \colhead{(7)} &
             \colhead{(8)} &
             \colhead{(9)} }
\startdata
\object[ESO409-IG015]{J0005$-$28}                     & ESO409-IG015          &                & 
00 05 31.7 & -28 05 53 & 6568/28;$R$ &  65 &  1.72 &  141 \\
\object[MCG-04-02-003]{J0019$-$22}                    & MCG-04-02-003         & \verb#.I..9*P# & 
00 19 11.5 & -22 40 06 & 6568/28;$R$ &  75 &  1.60 &    3 \\
\object[ESO473-G024]{J0031$-$22}                      & ESO473-G024           & \verb#.IBS9..# & 
00 31 22.2 & -22 46 02 & 6568/28;$R$ &  72 &  1.41 &   26 \\
\object[NGC178]{J0039$-$14a}                          & NGC178                & \verb#.SBS9..# & 
00 39 08.2 & -14 10 29 & 6605/32;$R$ & 139 &  2.12 &    9 \\
\object[IC1574]{J0043$-$22}                           & IC1574                & \verb#.IBS9..# & 
00 42 27.0 & -22 06 19 & 6568/28;$R$ &  71 &  1.73 &  172 \\
\object[NGC625]{J0135$-$41}                           & NGC625                & \verb#.SBS9$/# & 
01 35 06.2 & -41 26 04 & 6568/28;$R$ & 238 &  2.37 &   94 \\
\enddata
\tablecomments{Column descriptions [units]: (1) Source name. (2) Optical
identification, from HOPCAT (Doyle et al.\ 2005), the BGC
(Koribalski et al.\ 2004) or NED. For sources from NED with prefixes 2MASX 
(2 Micron All Sky Survey Extended object) and APMUKS(BJ) (Automated 
Plate Measurement United Kingdom Schmidt (B$_{\rm J}$); Maddox et al. 1990) 
we use the abbreviated prefixes ``2M' and ``A-', respectively. 
(3) Morphology code from the RC3 (de Vaucouleurs et al.\ 1991). 
(4) And (5) right ascension and declination
[J2000]. (6) Filters used in the observation [narrow band ;
continuum]. (7) Semi-major axis length of elliptical aperture used to
measure total flux [\as]. (8) Axial ratio of aperture used to measure
total flux.  (9) Position angle of the major axis, measured from north
toward the east, of the aperture used to measure total flux.  }
\tablecomments{{\it Sample portion of table.}}
\end{deluxetable}

\clearpage
\begin{landscape}
\begin{deluxetable}{l c c c c c c c c c c}
  \tablecaption{Intrinsic quantities for SR1\label{t:intrinsic}}
  \tabletypesize{\tiny}
  \tablewidth{0pt}
  \tablehead{\colhead{Designation} &
             \colhead{$M_{R}$} &
             \colhead{$r_e(R)$} &
             \colhead{$r_{90}(R)$} &
             \colhead{$\mu_{e,0}(R)$} &
             \colhead{$\log(\FHa)$} &
             \colhead{$r_e({\rm H\alpha})$} &
             \colhead{$r_{90}({\rm H\alpha})$} &
             \colhead{$\log({\rm SFR})$} &
             \colhead{$\log({\rm SFA})$} &
             \colhead{$EW_{50,0}$} \\
             \colhead{(1)} &
             \colhead{(2)} &
             \colhead{(3)} &
             \colhead{(4)} &
             \colhead{(5)} &
             \colhead{(6)} &
             \colhead{(7)} &
             \colhead{(8)} &
             \colhead{(9)} &
             \colhead{(10)} &
             \colhead{(11)} }
\startdata
J0005$-$28     &      $-15.51\pm 0.03\,\,\,$ & $ 0.773\pm 0.013$ & $ 2.115\pm 0.054$ & $22.33\pm 0.03$ &         $-12.32\pm 0.02\,\,\,$ &   $ 0.453\pm 0.002$ &   $ 0.860\pm 0.026$ &                      $-2.24$ & $ -1.55\pm0.02$ & $292.8\pm  3.2$ \\
J0019$-$22     &     $-15.50\pm 0.03^{\dag}$ & $ 0.912\pm 0.016$ & $ 2.934\pm 0.023$ & $22.70\pm 0.03$ &       $-13.14\pm 0.08^{\ddag}$ &   $ 0.595\pm 0.191$ &                 $-$ &                      $-3.13$ & $ -2.68\pm0.18$ &  $ 27.9\pm 3.4$ \\
J0031$-$22     &      $-14.06\pm 0.06\,\,\,$ & $ 0.909\pm 0.029$ & $ 2.035\pm 0.088$ & $24.17\pm 0.03$ &         $-13.08\pm 0.04\,\,\,$ &   $ 0.682\pm 0.019$ &   $ 1.868\pm 0.159$ &                      $-3.29$ & $ -2.95\pm0.03$ & $ 78.7\pm  4.3$ \\
J0039$-$14a    &     $-19.01\pm 0.02^{\dag}$ & $ 2.304\pm 0.013$ & $ 6.038\pm 0.218$ & $21.00\pm 0.02$ &        $-12.06\pm 0.03^{\dag}$ &   $ 2.141\pm 0.011$ &   $ 5.159\pm 0.167$ &                      $-1.24$ & $ -1.90\pm0.03$ &  $ 56.8\pm 3.4$ \\
J0043$-$22     &      $-14.47\pm 0.03\,\,\,$ & $ 0.738\pm 0.005$ & $ 1.333\pm 0.008$ & $23.30\pm 0.02$ &         $-13.34\pm 0.20\,\,\,$ &   $ 0.622\pm 0.082$ &   $ 1.257\pm 0.121$ &                      $-3.96$ & $ -3.55\pm0.11$ & $  9.5\pm  3.6$ \\
J0135$-$41     &      $-17.30\pm 0.02\,\,\,$ & $ 1.309\pm 0.012$ & $ 3.109\pm 0.053$ & $21.61\pm 0.02$ &         $-11.33\pm 0.07\,\,\,$ &   $ 0.436\pm 0.021$ &   $ 1.404\pm 0.525$ &                      $-2.02$ & $ -1.30\pm0.03$ & $222.9\pm  7.3$ \\
J0145$-$43     &      $-16.14\pm 0.09\,\,\,$ & $ 1.869\pm 0.075$ & $ 3.554\pm 0.175$ & $23.60\pm 0.02$ &         $-11.93\pm 0.05\,\,\,$ &   $ 2.181\pm 0.049$ &   $ 3.244\pm 0.234$ &                      $-2.59$ & $ -3.27\pm0.05$ & $ 30.6\pm  4.4$ \\
J0156$-$68     &      $-16.23\pm 0.03\,\,\,$ & $ 1.589\pm 0.020$ & $ 3.384\pm 0.061$ & $23.13\pm 0.02$ &         $-13.32\pm 0.05\,\,\,$ &   $ 1.515\pm 0.031$ &   $ 2.413\pm 0.205$ &                      $-2.71$ & $ -3.07\pm0.05$ & $ 27.3\pm  2.9$ \\
J0209$-$10:S1  &     $-20.98\pm 0.02^{\dag}$ & $ 3.338\pm 0.030$ & $10.708\pm 0.254$ & $19.59\pm 0.02$ &        $-12.46\pm 0.08^{\dag}$ &   $ 1.488\pm 0.191$ &   $ 7.446\pm 1.519$ &                      $-0.62$ & $ -0.97\pm0.04$ &  $ 83.4\pm 4.9$ \\
J0209$-$10:S2  &      $-21.16\pm 0.02\,\,\,$ & $ 2.348\pm 0.013$ & $ 8.627\pm 0.118$ & $18.62\pm 0.02$ &         $-11.83\pm 0.02\,\,\,$ &   $ 1.500\pm 0.026$ &   $ 4.527\pm 0.264$ &                      $ 0.04$ & $ -0.32\pm0.01$ & $183.0\pm  4.9$ \\
J0209$-$10:S3  &      $-21.88\pm 0.02\,\,\,$ & $ 4.829\pm 0.050$ & $19.870\pm 0.290$ & $19.34\pm 0.03$ &         $-12.12\pm 0.08\,\,\,$ &   $ 3.239\pm 0.148$ &   $17.297\pm 2.324$ &                      $-0.16$ & $ -1.18\pm0.04$ & $ 39.9\pm  5.2$ \\
J0209$-$10:S4  &      $-21.22\pm 0.02\,\,\,$ & $ 3.574\pm 0.007$ & $ 8.476\pm 0.033$ & $19.46\pm 0.02$ &         $-12.88\pm 0.25\,\,\,$ &   $ 2.990\pm 0.241$ &   $ 7.640\pm 1.399$ &                      $-1.01$ & $ -1.96\pm0.19$ & $ 11.1\pm  4.9$ \\
\enddata
\tablecomments{Column descriptions [units]: (1) Source name. (2) $R$
band absolute magnitude [ABmag].  (3) Effective radius in the $R$ band
[kpc]. (4) The radius enclsoing 90\%\ of the $R$ band flux [kpc]. 
(5) Extinction corrected (Galactic and internal), face-on $R$ 
band effective surface brightness [ABmag arcsec$^-2$]. (6) Logarithm 
of total H$\alpha$ flux corrected for Galactic extinction and 
[{\sc Nii}] contamination [erg cm$^-2$ s$^{-1}$].  (7)
Effective radius in H$\alpha$ [kpc]. (8) The radius enclsoing 90\%\ of the 
H$\alpha$ band flux [kpc]. (9) Logarithm of star formation rate
[\Msun\ year$^{-1}$]. The errors for this quantity are the same as 
those in column (5).  (10) Logarithm of star formation rate per unit
face-on area [\Msun\ year$^{-1}$ kpc$^{-2}$].  (11) Extinction corrected
equivalent width within $r_e({\rm H\alpha})$ [\AA].  Cases where there is flux
outside of the region measured with elliptical apertures are indicated
in columns (2) and (6) with $^\dag$ ; $^\ddag$ is used when the flux
exterior to the elliptical apertures is $>$ 10\% of the total flux ; and
the two cases where the H$\alpha$ flux was measured with just small
apertures centered on the obvious \HII\ regions instead of the standard 
method are marked $^{\#}$. Cases which were measured with this technique, 
have invalid $r_e$, $r_{90}$ in which case columns
(3), (4), (5), (6), (7), (8) and (10) are marked with ``$-$'', as 
appropriate. Cases where the source extends close to or beyond the 
frame boundary are marked $^\S$ in columns (2) and (6).}
\tablecomments{{\it Sample portion of table.}}
\end{deluxetable}

\begin{deluxetable}{l c c c c c c}
  \tablecaption{Corrections used in flux measurements\label{t:cor}}
  \tabletypesize{\small}
  \tablehead{\colhead{HIPASS+} &
             \colhead{$A_{\rm H\alpha,G}$} &
             \colhead{$M'_R$} &
             \colhead{$A_{\rm H\alpha,i}$} &
             \colhead{$A_{\rm R,i}$} &
             \colhead{$w_{6583}$} &
             \colhead{$k_{\rm [NII]}$} \\ 
             \colhead{(1)} &
             \colhead{(2)} &
             \colhead{(3)} &
             \colhead{(4)} &
             \colhead{(5)} &
             \colhead{(6)} &
             \colhead{(7)} }
\startdata
J0005$-$28     & 0.04 & $-15.51$ & 0.24 & 0.12 & 0.052 &  0.352\\
J0019$-$22     & 0.05 & $-15.50$ & 0.24 & 0.12 & 0.052 &  0.353\\
J0031$-$22     & 0.05 & $-14.06$ & 0.16 & 0.08 & 0.034 &  0.381\\
J0039$-$14a    & 0.05 & $-19.01$ & 0.64 & 0.32 & 0.148 &  0.797\\
J0043$-$22     & 0.04 & $-14.47$ & 0.18 & 0.09 & 0.038 &  0.525\\
J0135$-$41     & 0.04 & $-17.30$ & 0.40 & 0.20 & 0.089 &  0.488\\
J0145$-$43     & 0.04 & $-16.14$ & 0.29 & 0.14 & 0.063 &  0.492\\
J0156$-$68     & 0.07 & $-16.23$ & 0.30 & 0.15 & 0.065 &  0.893\\
J0209$-$10:S1  & 0.06 & $-20.98$ & 1.10 & 0.55 & 0.267 &  1.335\\
J0209$-$10:S2  & 0.06 & $-21.16$ & 1.16 & 0.58 & 0.283 &  1.335\\
J0209$-$10:S3  & 0.06 & $-21.88$ & 1.41 & 0.71 & 0.351 &  1.335\\
J0209$-$10:S4  & 0.06 & $-21.22$ & 1.18 & 0.59 & 0.288 &  1.335\\
\enddata
\tablecomments{Column descriptions [units]: (1) Source name. (2)
Absorption of \Halpha\ by Galactic foreground dust [mag]. (3) $R$ band
absolute magnitude without internal dust absorption correction
[ABmag].  (4) Absorption of \Halpha\ by dust internal to the source,
estimated from $M'_R$ using eq.~\ref{e:helmboldt_ahai} [mag]. (5) Absorption in the
$R$ band by dust internal to the source [mag].  (6) Line flux ratio
$F({\rm [NII]6583\AA}\/)/F({\rm H\alpha}\/)$ estimated from $M'_R$ using
eq.~\ref{e:helmboldt_nii}.  (7) Filter and velocity profile dependent
correction coefficient for contamination of the \Halpha\ flux by the
\fion{N}{II} lines (see eqs.~\ref{e:hafrac} and \ref{e:knii}).}
\tablecomments{{\it Sample portion of table.}}
\end{deluxetable}

\begin{deluxetable}{l c c c c }
  \tablecaption{Median image quality statistics\label{t:qastat}}
  \tabletypesize{\small}
  \tablehead{\colhead{Image type} &
             \colhead{Seeing} &
             \colhead{Limiting mag or flux} &
             \colhead{Limiting surface brightness} &
             \colhead{Limiting EW}}
\startdata
$R$ or continuum & 1.57\as & 22.73 ABmag & 26.95 ABmag arcsec$^{-2}$ & \nodata\ \\
NB             & 1.56\as & 20.83 ABmag & 25.76 ABmag arcsec$^{-2}$ & \nodata\ \\
Net \Halpha\   & 1.61\as & $2.6\times 10^{-16}\, {\rm erg\, cm^{-2}\, s^{-1}}$ & 
  $2.9\times 10^{-18}\, {\rm erg\, cm^{-2}\, s^{-1}\, {arcsec}^{-2}}$ & 3.3\AA\ \\
\enddata
\end{deluxetable}
\clearpage
\end{landscape}

\end{document}